\documentclass[a4paper,11pt]{article}
\pdfoutput=1 

\usepackage{jheppub} 
\usepackage{subfigure}
\usepackage{reviewphysics}

\usepackage[T1]{fontenc} 

\title{\boldmath Diboson Production in Proton-Proton Collisions at $\sqrt{s}=7$ TeV}

\author[a,1]{M. Schott,\note{Corresponding author.}}
\author[b]{J. Zhu}

\affiliation[a]{Johannes Gutenberg-University, Mainz, Germany}
\affiliation[b]{University of Michigan, Ann Arbor, MI, United States of America, 48103}

\emailAdd{mschott@cern.ch}
\emailAdd{junjie@umich.edu}

\abstract{This review article summarizes results on the production cross section measurements of electroweak boson pairs ($WW$, $WZ$, $ZZ$, $W\gamma$ and $Z\gamma$) at the Large Hadron Collider (LHC) in $pp$ collisions at a center-of-mass energy of $\sqrt{s}=7$ \TeV. The two general-purpose detectors at the LHC, ATLAS and CMS, recorded an integrated luminosity of $\approx5\,\ifb$ in 2011, which offered the possibility to study the properties of diboson production to high precision. These measurements test predictions of the Standard Model (SM) in a new energy regime and are crucial for the understanding and the measurement of the SM Higgs boson and other new particles. In this review, special emphasis is drawn on the combination of results from both experiments and a common interpretation with respect to state-of-the-art SM predictions.}

\begin{document} 
\maketitle
\flushbottom

\section{Introduction}

The Standard Model (SM) of particle physics is a quantum field theory based on the $SU(3)_C \otimes SU(2)_L \otimes U(1)_Y$ gauge symmetry group and
 describes the strong, weak and electromagnetic interactions among elementary particles\cite{Nakamura:2010zzi}. As a direct consequence of the non-Abelian gauge symmetry of the electroweak sector, i.e.  the $SU(2)_L \otimes U(1)_Y$ gauge group, the SM predicts self-coupling vertices of the gauge bosons.  At the Large Hadron Collider (LHC), these vertices lead to the production of diboson final states. Moreover, the discovery of the Higgs boson in the year 2012 \cite{Aad:2012tfa}, \cite{Chatrchyan:2012ufa} was due to its diboson decay channels. The production of pairs of vector bosons is therefore not only an important background for studies of the newly-discovered Higgs boson, but also provides a unique opportunity to test the electroweak sector of the SM. 
 

These precision tests have already a long history in particle physics. The LEP experiments performed precise measurements of the $e^+ e^- \rightarrow W^+ W^-$ and $e^+ e^- \rightarrow ZZ$ cross sections as a function of center-of-mass energy\cite{Schael:2004tq, Abbiendi:2003mk, Achard:2004ji, Abdallah:2007ae}. 
The clean experimental signature and nature of the purely electroweak calculations allowed for a precise study of the gauge-group nature of the SM. 
Limits on possible extensions and deviations from the SM predictions were also drawn. Some of those limits are still the most stringent ones available.
Since the $e^+ e^- \rightarrow WW$ cross section depends crucially on the mass of the $W$ boson ($m_W$), the cross section dependence on the center-of-mass energy allowed for an indirect determination of $m_W$.

The Tevatron experiments allowed studies of all possible diboson final states $W^+W^-$, $W^\pm Z$, $ZZ$, $W\gamma$ and $Z\gamma$ \cite{Neubauer:2011zz, Abulencia:2007tu, Abazov:2007ai}. 
In contrast to the LEP collider, the bulk part of the production process is governed by the Quantum Chromodynamic (QCD) processes. Nevertheless,  
precise measurements of these production cross sections allow again tests of the predicted gauge-boson self-interactions and 
searches for physics beyond the SM. 

With the start of the Large Hadron Collider at CERN, a completely new energy regime became accessible, and the corresponding production cross sections increased by more than a factor of five compared to the Tevatron. 
The increase in available statistics and the higher center of mass energies allow for an improved study of perturbative Quantum Chromodynamic predictions, which are by now available to next-to-next-to-leading order in the strong coupling constant $\alpha_s$ for some processes\cite{Cascioli:2014yka}. In addition, for the first time, limits on new physics scenarios that modify the triple gauge coupling vertices could be improved compared to the LEP experiments. 

This review article summarizes the results of diboson production 
cross section measurements at the LHC with a center-of-mass energy of $\sqrt{s}=7\,\TeV$, based on data collected by the ATLAS and CMS experiments in 2011. We present the experimental signatures and the differences between the measurement strategies in detail. 
Special focus is drawn on combinations of various results from both experiments and their interpretations within the SM. The results published by the ATLAS and CMS collaborations form the basis for this review. The combinations of these results and further 
derived information have been conducted with great care, but are solely based on the private work of the authors of this article 
and do not reassemble any official joint ATLAS and CMS effort. 


This article is structured as follows. In Sect. \ref{sec:Theory}, we
briefly summarize the most important features of the electroweak
sector of the SM, the theoretical methodology for the predictions of 
diboson production in $pp$ collisions, and the description of new physics scenarios regarding diboson production.
The LHC collider, the ATLAS and CMS experiments, as well as further experimental aspects are discussed in 
Sect. \ref{sec:ExperimentalAspects}. Measurements that are sensitive to the $WWZ$ and $WW\gamma$ vertices, i.e.measurements of the 
$W^+W^-$, the $W^\pm Z$ and the $W^\pm\gamma$ final states, are presented in Sect. \ref{sec:wwvvertex}. 
The $ZZ$ and $Z\gamma$ final states are sensitive to $ZZZ$, $ZZ\gamma$ and $Z\gamma\gamma$ 
vertices that are not allowed in the SM; they are discussed in Sect. \ref{sec:zzgammavertex}. 
The results on the cross section measurements and the limits on anomalous triple gauge couplings (aTGCs) 
are interpreted in Sect. \ref{sec:interpretation}. In addition, the sensitivity to quartic gauge couplings (QGCs) is briefly discussed. Section \ref{sec:conclusion} summarizes all the results and gives an outlook for future measurements at the LHC.

\section{\label{sec:Theory}Dibosons in the Standard Model}

\subsection{\label{sec:EWSector}The Electroweak Sector}

\begin{figure}[h]
\begin{center}
\resizebox{0.95\textwidth}{!}{\includegraphics{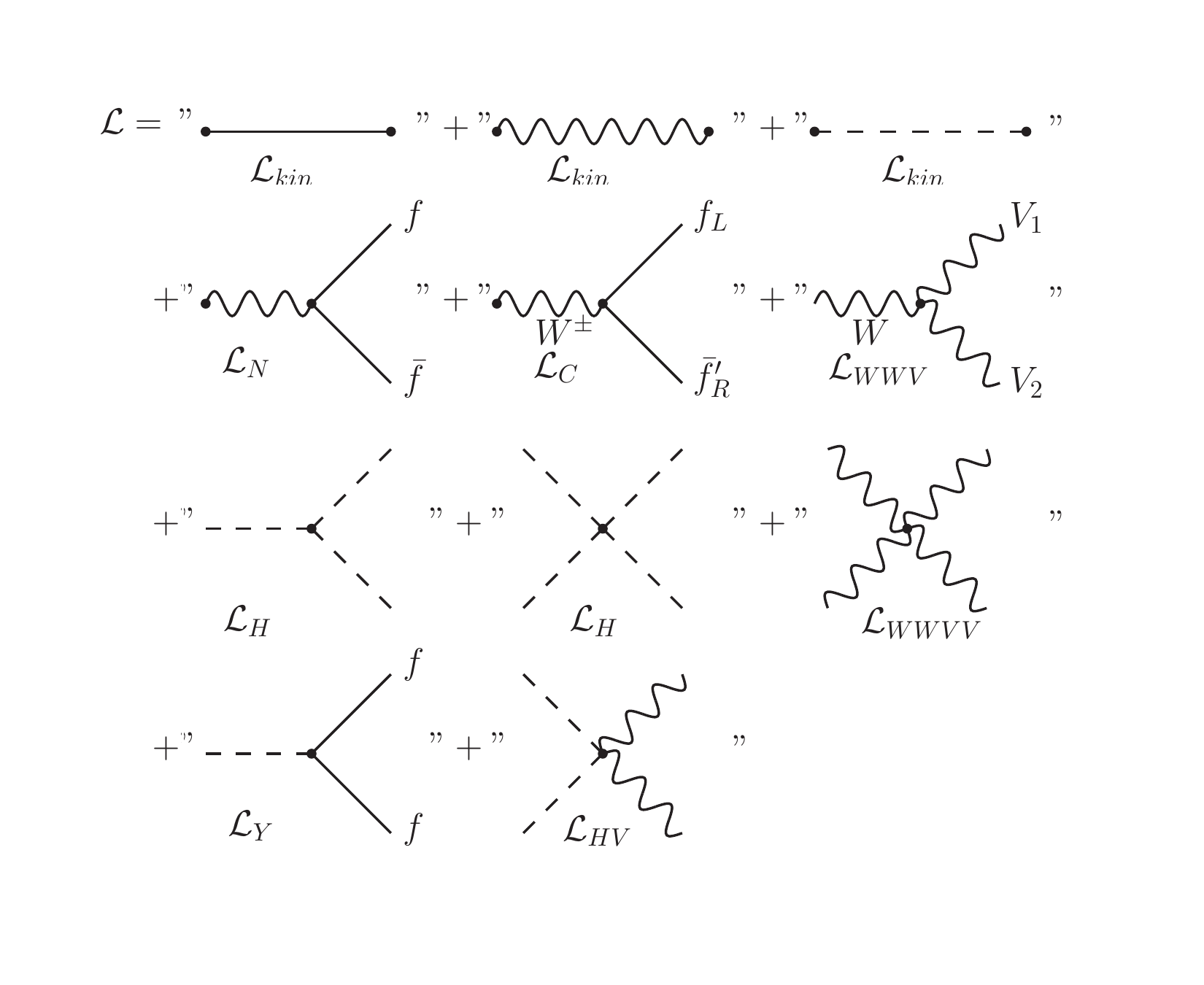}}
\caption{\label{fig:TSMLagrangian} Feynman diagrams of tree-level interactions of the Lagrangian describing the electroweak sector of the SM.}
\end{center}
\end{figure}

The Lagrangian of the electroweak sector of the SM can be written as
\begin{equation}
\label{EQN:SMLagrangian}
\LC_{EW} = \LC_{Kin} + \LC_{N} + \LC_{C} + \LC_{WWV} + \LC_{WWVV} + \LC_{H} + \LC_{HV} + \LC_{Y},
\end{equation}
where the different terms are schematically shown in
Fig. \ref{fig:TSMLagrangian} as tree-level Feynman diagrams\cite{Nakamura:2010zzi}. The free
movements of all fermions and bosons are described by the kinematic
term $\LC_{Kin}$. 
The neutral current interactions,
i.e. the exchange of photon and the \Zboson boson, are summarized in
$\LC_{N}$. The $W$ boson interaction to left-handed particles and
right-handed anti-particles is represented by the charged current term $\LC_{C}$. Since the
weak interaction is based on an $SU(2)_L$ group structure, three-point
and four-point interactions of the electroweak gauge bosons
appear, denoted by $\LC_{WWV}$ and $\LC_{WWVV}$. The self-interactions
of the Higgs boson and the interaction of the Higgs boson with the
electroweak gauge bosons are represented by the terms $\LC_{H}$ and
$\LC_{HV}$, respectively. The Yukawa couplings between the Higgs field
and the fermions is denoted by $\LC_Y$.

The production of single photons and \Zboson bosons within the SM is given by
\begin{equation}
\label{EQN:LN}
\LC_N = e J_{\mu}^{em}A^\mu + \frac{g}{\cos\theta _W} \left( J_\mu^3 - \sin^2\theta_W J_\mu^{em}\right) Z^\mu,
\end{equation}
where $J_\mu^{em}$ and $J_\mu^3$ denote the electromagnetic and weak
current, $A^\mu$ is the photon field, and $Z^\mu$ is the $Z$ boson. Each current contains the sum over all fermionic
fields weighted by their respective charges. In addition, the weak
current $J_\mu^3$ involves only left-handed particles and right-handed
anti-particles. The relative strength of the electromagnetic and weak
interactions is described by the weak mixing angle, $\theta_W$.

The charged current term,
\begin{equation}
\label{EQN:LC}
\LC_C = - \frac{g}{\sqrt{2}} \left[\bar u_i \gamma^\mu \frac{1-\gamma^5}{2} M_{ij}^{CKM} d_j + \bar v_i \gamma ^\mu \frac{1-\gamma^5}{2} e_i \right] W_\mu^+ +  h.c.
\end{equation}
describes the production of charged bosons via fermions,
i.e. $f\bar f' \rightarrow W^\pm$. For simplicity, only the terms
of the first generation are shown and the quark and lepton spinor
fields are labeled by $u_i, d_j$ and $\nu_i, e_i$, respectively. The
mixing of quark generations is described by the CKM matrix
$M_{ij}^{CKM}$. The $(1-\gamma^5)/2$ operator is introduced to describe the exclusive interaction of left-handed particles
and right-handed anti-particles with the \Wboson boson.

It should be noted that the terms $L_N$ and $L_C$ can lead to
diboson final states at hadron colliders through $t$- and $u$-
exchange of a fermion. However, as a direct consequence of the $SU(2)$
gauge group of the weak interaction, the production of vector boson
pairs through the $s$-channel process is also allowed. The corresponding term in
the Lagrangian reads as
\begin{equation}
\label{EQN:WWV}
\LC_{WWV} = - i g \left[ W^+_{\mu\nu}W^{-\mu} (A^\nu \sin\theta_W - Z^\nu \cos\theta_W) + W_\nu^- W_\mu^+ (A^{\mu\nu} \sin\theta_W - Z^{\mu\nu}) \right]. 
\end{equation}
This term leads to TGCs at the tree level. It is thus interesting to experimentally demonstrate that 
gauge bosons couple not only to electric charge but also to weak isospin.
It is apparent from Eqn. \ref{EQN:WWV} that the SM allows only
$\gamma WW$ and $ZWW$ couplings. Various models of physics beyond the
SM predict aTGCs but also new vertices like $ZZZ$, $ZZ\gamma$ and $Z\gamma\gamma$. 

The last contribution to diboson final states comes from the decay of the Higgs boson, described by
\begin{equation}
\label{EQN:HVV}
\LC_{HVV} = \left(g m_W H + \frac{g^2}{4} H^2 \right) \left( W_\mu^+ W^{\nu -} + \frac{1}{2 \cos^2 \theta_W} Z_\mu Z^\mu \right).
\end{equation}
However, for a SM Higgs boson with a mass close to 125 GeV, the decay into a diboson pair must involve at least one off-shell vector
boson. This production mode is therefore significantly suppressed if we only consider on-shell vector bosons in the final state.

The vector boson fusion or four-point self-interaction as described
by $L_{WWVV}$, is not discussed in this review, as the expected
production cross sections at $7\,\TeV$ $pp$ collisions are
negligible. However it should be noted that this channel has never been
experimentally observed and measurements of these processes at 14
\TeV\ would be a crucial test of the SM predictions.


\subsection{\label{sec:DiBosonProd}Diboson Production at the LHC}

Several introductory articles on the production of vector bosons in
hadron collisions are available. Before discussing the experimental
results, we summarize here the essential aspects of the theoretical
predictions of the diboson production at the LHC along the lines of the following publications.
 \cite{Neubauer:2011zz, Campbell:2011bn, Nunnemann:2007qs, Schott:2014}.

The diboson production at hadron 
colliders is significantly different than the production mechanism at lepton colliders due to the complicated internal structure of protons. 
The proton structure can be described phenomenologically by parton density functions (PDFs), written as
$f_{A,q} (x, Q^2)$ for parton-type $q$ in proton $A$ with a
relative momentum of $x=p_q/p_A$ in the direction of the proton's
motion for an energy scale $Q^2$ of the scattering process. Here,
$p_q$ and $p_A$ denote the momenta of the parton and proton,
respectively. For the production of vector boson pairs ($VV$), the
energy scale is often set to the invariant mass of the two vector
bosons, i.e. $Q^2=M^2_{VV}$.

\begin{figure*}
\begin{minipage}{0.44\textwidth}
\resizebox{1.0\textwidth}{!}{\includegraphics{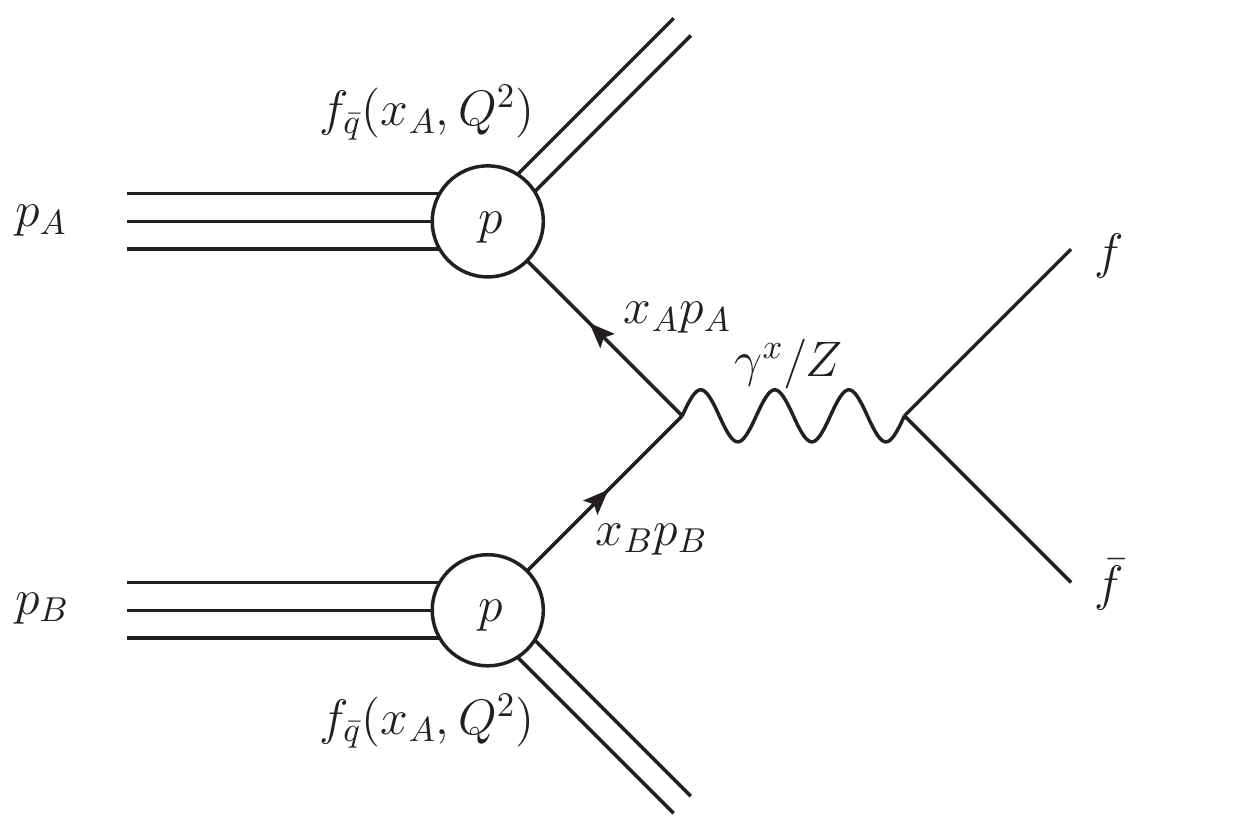}}
\caption{\label{Fig:FakTheorem}Illustration of the factorization theorem used to calculate the production cross section in $pp$ collisions.}
\end{minipage}
\hspace{0.2cm}
\begin{minipage}{0.52\textwidth}
\resizebox{0.33\textwidth}{!}{\includegraphics{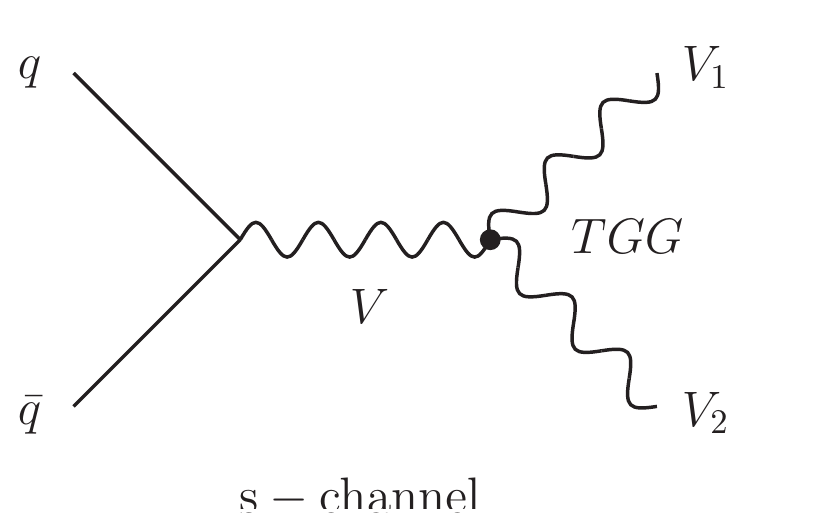}}
\resizebox{0.32\textwidth}{!}{\includegraphics{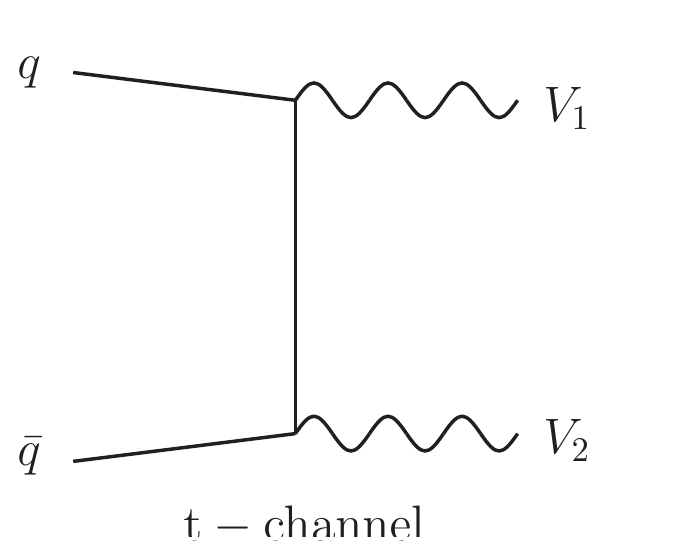}}
\resizebox{0.32\textwidth}{!}{\includegraphics{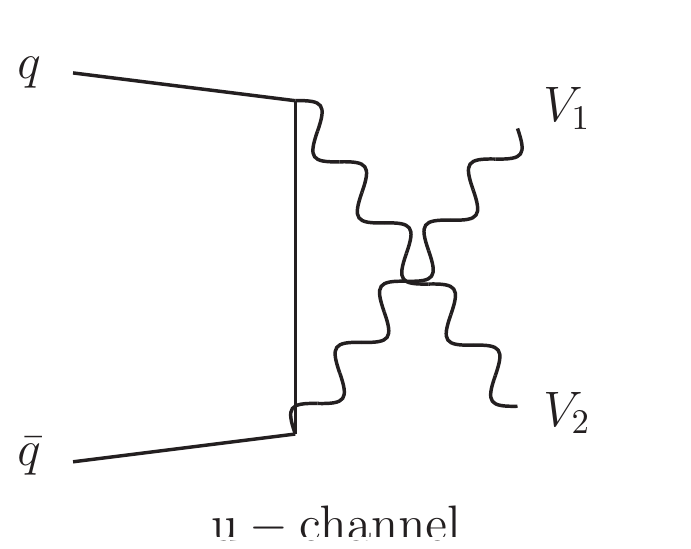}}
\resizebox{0.33\textwidth}{!}{\includegraphics{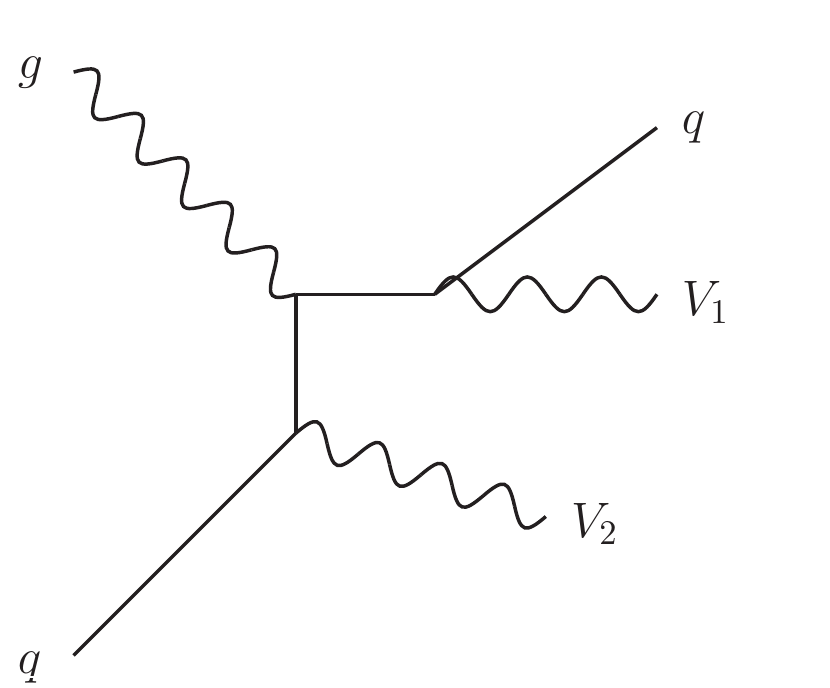}}
\resizebox{0.32\textwidth}{!}{\includegraphics{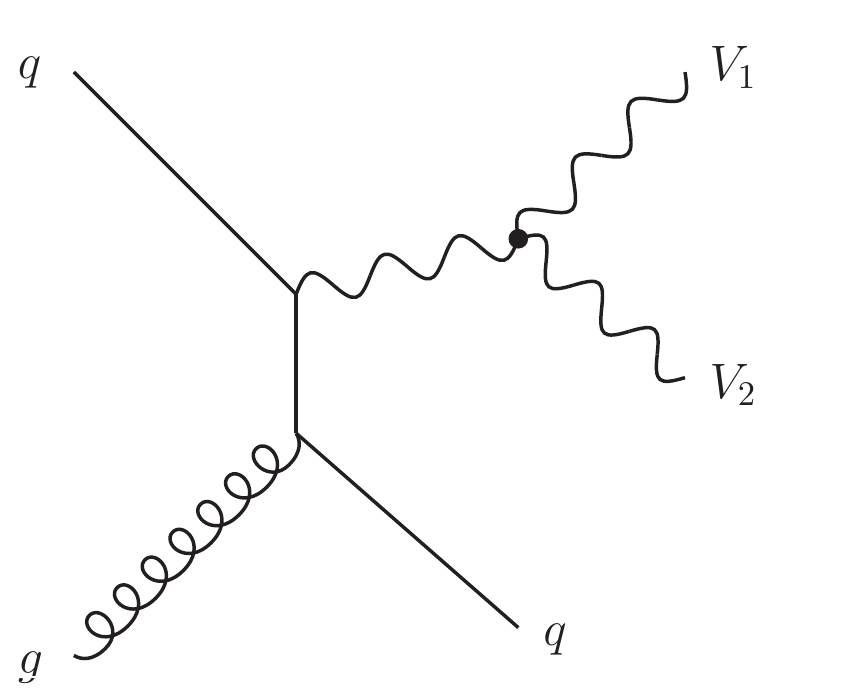}}
\resizebox{0.32\textwidth}{!}{\includegraphics{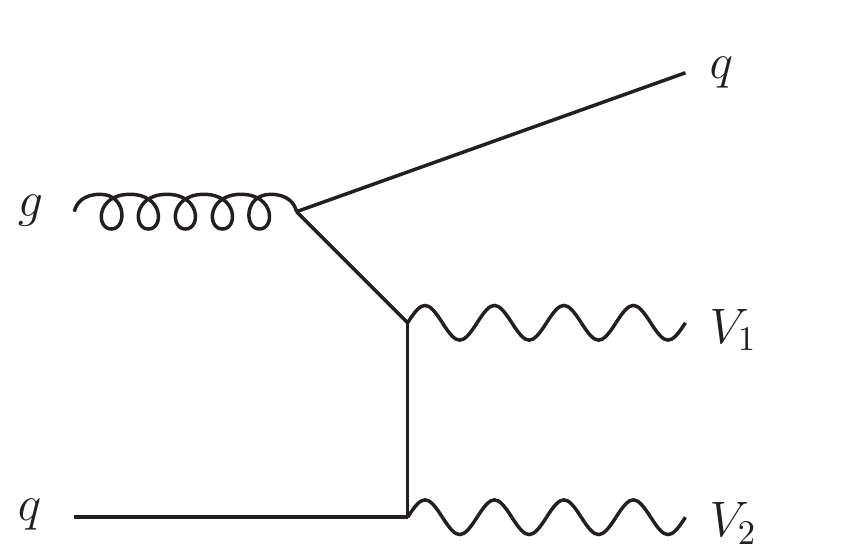}}
\caption{\label{Fig:LONLOVV} LO (top) and NLO (bottom) Feynman diagrams for diboson production.}\vspace{0.25cm}
\end{minipage}
\end{figure*}

\begin{figure*}
\begin{center}
\resizebox{.45\textwidth}{!}{\includegraphics{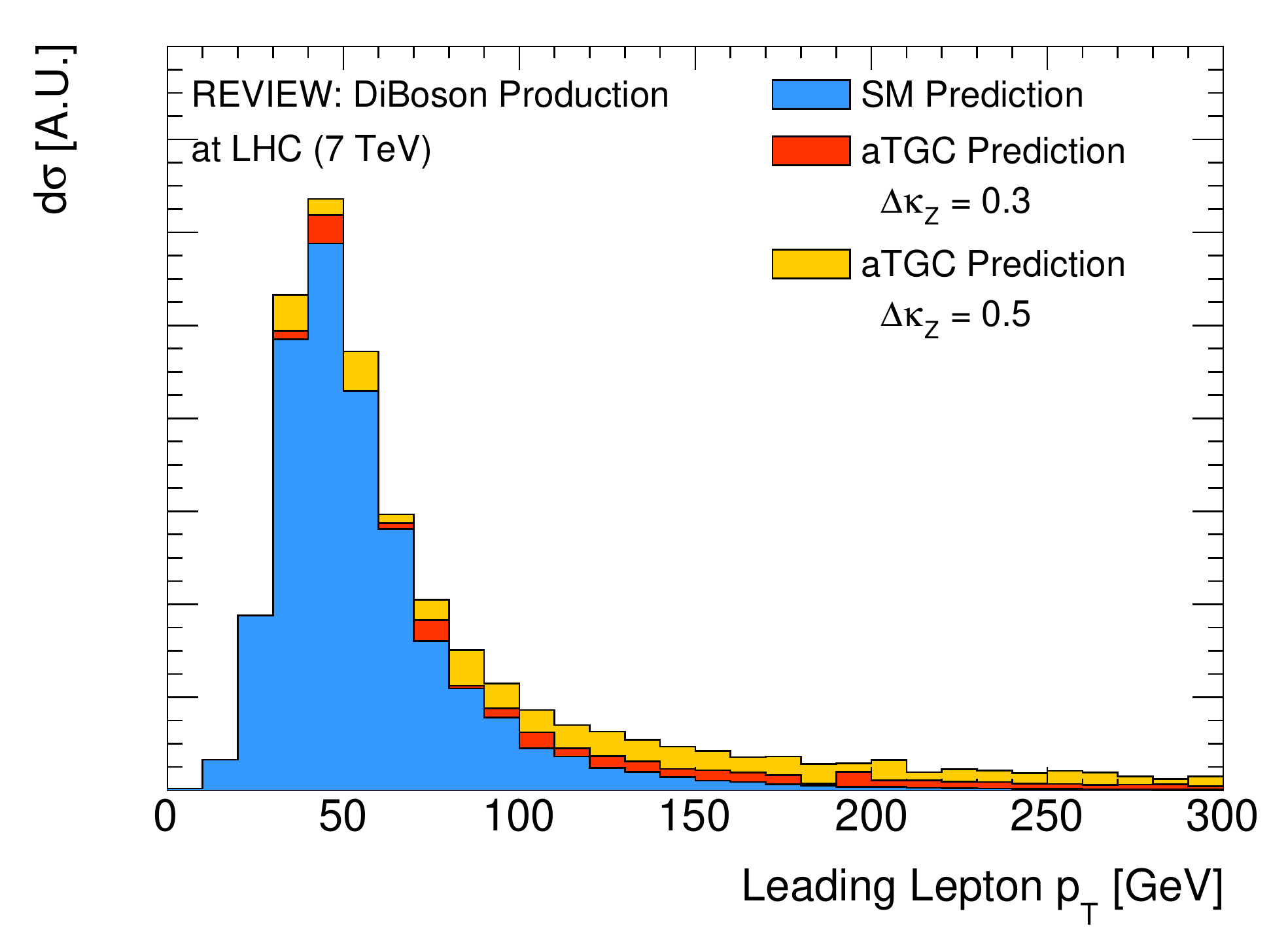}}
\resizebox{.45\textwidth}{!}{\includegraphics{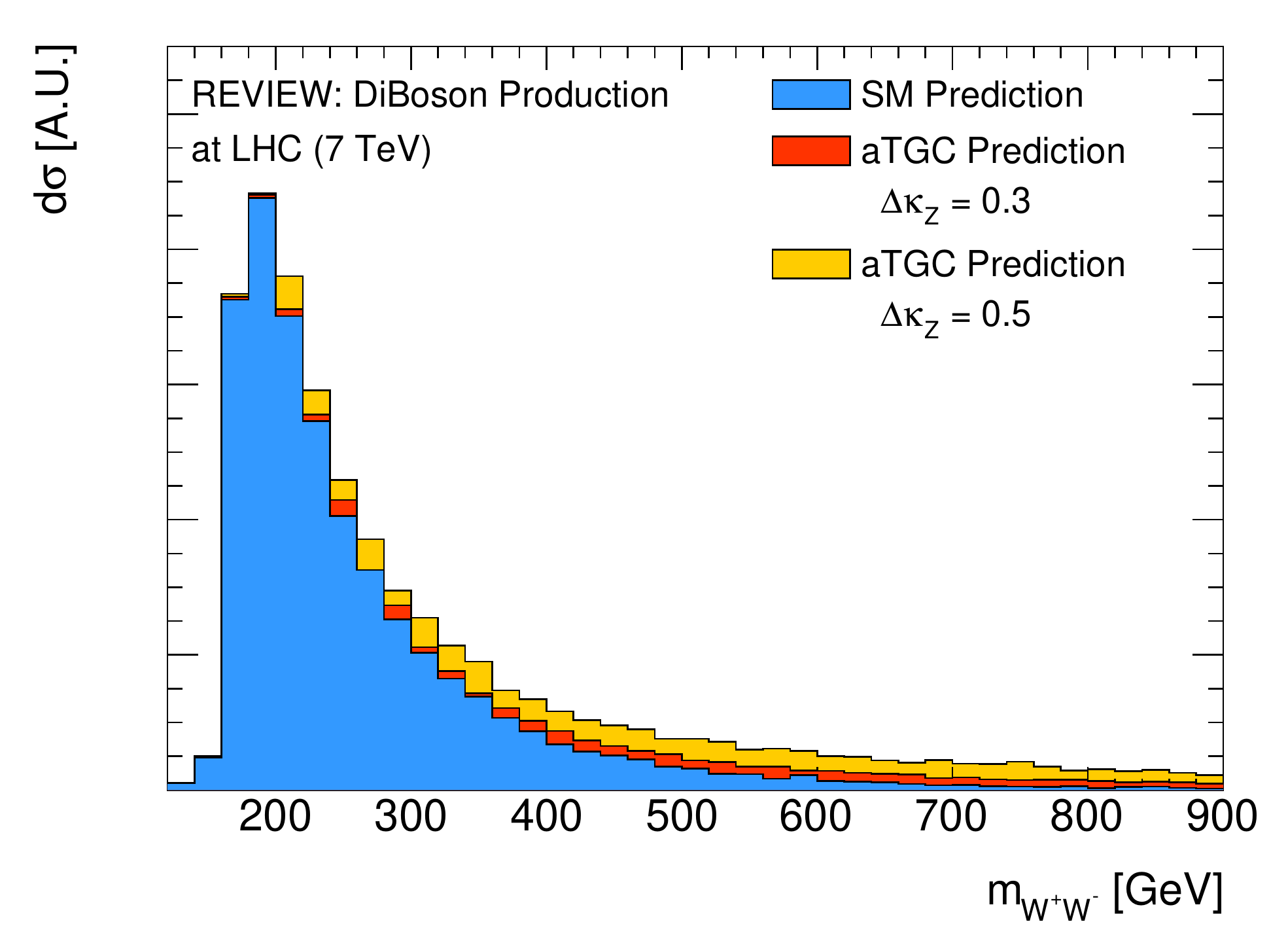}}
\caption{\label{Fig:SChannelDependence}Example of an enhanced cross section as a function of the leading lepton transverse momentum (left) and the invariant mass of the $W^+W^-$ system (right) with two aTGC scenarios.}
\end{center}
\end{figure*}

The factorization theorem states that the production cross section in
$pp$ collisions can be expressed by combining PDFs with a
fundamental partonic cross section $\hat \sigma _{q\bar q \rightarrow VV}$, illustrated in Fig. \ref{Fig:FakTheorem}:
\begin{equation}
\label{EQN:FacTheorem}
\sigma_{p p \rightarrow VV} = \sum _q \int dx_A dx_B f_{q_B} (x_A, M^2_{VV}) f_{q_A} (x_B, M^2_{VV}) \hat \sigma _{q\bar q \rightarrow VV},
\end{equation} 
where $\hat \sigma_{q\bar q' \rightarrow VV}$ is the cross section of the inclusive hard
scattering process of two partons leading to two vector bosons in the
final state. The sum runs over all quark flavors, and the integration
is performed over the momentum fractions of the two colliding partons
$x_A$ and $x_B$. The factorization theorem holds not only for
inclusive hard-scattering processes but also for perturbative QCD corrections.
 
The partonic cross section is governed by the $\LC_N$, $\LC_C$, and
$\LC_{WWV}$ terms of the SM Lagrangian given in
Eqn. \ref{EQN:SMLagrangian}.  The corresponding leading-order (LO) and 
some next-to-leading-order (NLO) Feynman diagrams are shown in Fig. \ref{Fig:LONLOVV} 
for different production channels. It should
be noted that $W\gamma$ and $Z\gamma$ final states can also be 
realized by initial and/or final state radiation processes of the participating fermions.

Of special importance is the $s$-channel as it involves TGCs that are predicted within the SM. Any new
physics model that involves new or alternative interactions between
the SM electroweak gauge bosons may change these TGCs and hence change the corresponding observables.  
An enhancement of the $s$-channel contribution to the full diboson
production cross section is predicted to increase with rising
parton collision energy $\hat s$, i.e. with the invariant mass of
diboson system $M_{VV}$. This dependence is shown in
Fig. \ref{Fig:SChannelDependence} where an aTGC scenario is assumed for the $WW(Z/\gamma)$ production
vertex. Regions in the phase space that correspond to high center-of-mass energies of the interacting partons therefore provide a high
sensitivity to new physics scenarios.

Figure \ref{Fig:LONLOVV} shows some examples of Feynman diagrams due to the
NLO QCD corrections. In particular, the quark-gluon fusion in the
initial state leads to a significant increase in the production cross
section at a $pp$ collision energy of $7\,\TeV$\footnote{E.g. the $WW$
  cross section increases by $\approx 25\%$.}. The gluon-gluon fusion
with fermions in the loop contribute approximately $3\%$. The NLO QCD
corrections have been first calculated in \cite{Ohnemus:1990za, Ohnemus:1991kk, Ohnemus:1992jn, Baur:1993ir, Baur:1994aj, Baur:1995uv, Baur:1997kz, Campbell:1999ah, Dixon:1999di}. 
It should be noted that usually additional jets in the $W^+W^-$
diboson production are experimentally vetoed to reduce the background
from the pair production of top quarks.  Hence, a significant
contribution of the NLO QCD corrections to the $W^+W^-$ process are not directly studied experimentally.

The most up-to-date QCD calculations include off-shell effects in
gluon fusion processes, subsequent decay, and effects of massive
quarks in the loop \cite{Glover:1988fe}. Leptonic decay modes are
accounted for in the narrow-width approximation and include all
spin-correlation effects. For predictions beyond the narrow-width
approximation, gauge invariance requires the inclusion of single
resonant diagrams in the calculations, which was first done in \cite{Campbell:1999ah}.
The next-to-next-to-leading order (NNLO) QCD calculation 
for the $ZZ$ production in $pp$ collisions became available recently \cite{Cascioli:2014yka}.

\begin{figure*}
\begin{center}
\begin{minipage}{0.38\textwidth}
\resizebox{1.0\textwidth}{!}{\includegraphics{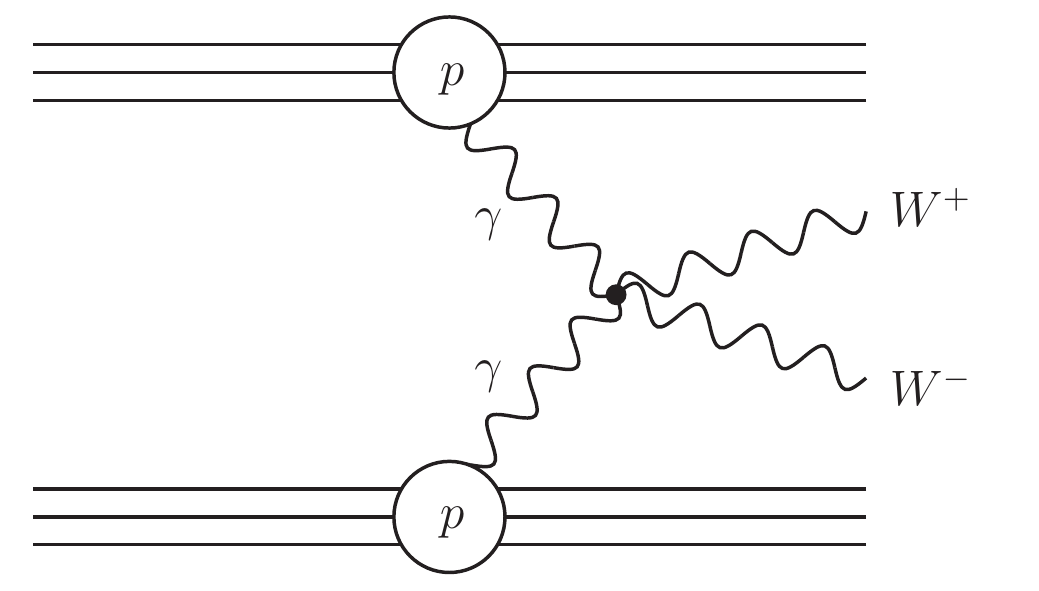}}
\caption{\label{Fig:PhotonFusion} $\gamma\gamma$ fusion contributing to the $W^+W^-$ production.}\vspace{0.7cm}
\end{minipage}
\hspace{0.3cm}
\begin{minipage}{0.48\textwidth}
\resizebox{1.45\textwidth}{!}{\includegraphics{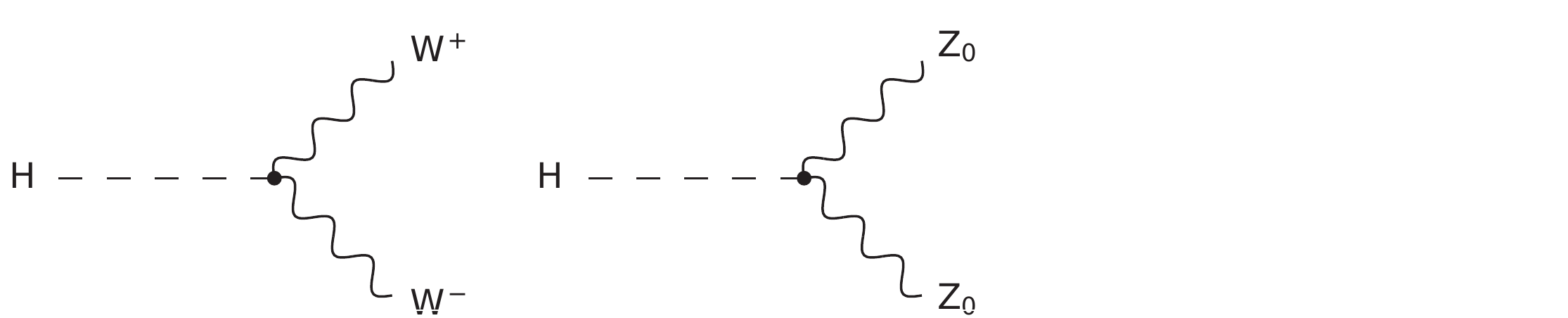}}
\caption{\label{Fig:HiggsDecay}Higgs decay to the $W^+W^-$ and $ZZ$ final states.}
\end{minipage}
\label{Labelname}
\end{center}
\end{figure*}

Electroweak corrections have been calculated so far only for the
$W^+W^-$ process \cite{Hollik:2004tm, Accomando:2004de, Accomando:2001fn, Accomando:2005xp, Billoni:2013pca}. Since the electroweak
coupling parameter ($\alpha_{\text{EW}}$) is small compared to the strong
coupling constant ($\alpha_s$), 
the effects of the EW corrections are expected to be minor. 
However, EW corrections increase with 
the square of $\log (\hat s)$ and can reach several percent at the energies 
While electroweak corrections modify the inclusive production cross section 
by less than $0.5\%$, they induce variations up to $10\%$ in the
rapidity distribution of the diboson pair. For large
diboson invariant masses ($M_{VV}>1\,\TeV$), EW corrections could modify the
inclusive production cross section by more than $15\%$~\cite{Billoni:2013pca}. 
EW corrections also open the $W^+W^-$ production
via photon-photon fusion (as shown in Fig.  \ref{Fig:PhotonFusion})
since the photon PDFs inside protons are non-vanishing. This channel
results in a predicted cross section enhancement of $\approx 1.5\%$
compared to the LO calculations at $\sqrt{s}=7\,\TeV$. 

Finally, the $W^+W^-$ and $ZZ$ boson pairs can also be produced via
the decay of the SM Higgs boson (as shown in
Fig. \ref{Fig:HiggsDecay}), which is described by the $\LC_{HVV}$ term
of the SM Lagrangian.  The Higgs boson production is known to NNLO in
$\alpha_s$ \cite{Harlander:2002wh} and to NLO in $\alpha_{\text{EW}}$
\cite{Actis:2008ug}.  Since diboson production from the decay of the 
SM Higgs bosons involves at least one off-shell vector
boson, the contribution of this decay channel to the full
inclusive diboson production is typically less than
$5\%$, depending on the experimental analysis
requirements. Further details are discussed in the experimental sections of this review.

In summary, the rich phenomenology of the electroweak sector can be
tested via studies of diboson production at hadron colliders. Four
terms on the Lagrangian contribute to the production, namely
$\LC_N$, $\LC_C$, $\LC_{HVV}$, $\LC_{WWV}$ and $\LC_{WWVV}$, and
therefore not only the $SU(2)$ gauge group structure but also 
higher-order calculations in QCD and partially the Higgs sector can be tested.

\subsection{\label{sec:eventgen}Event Generation and Available Computer Programs}

The central part of the prediction of the diboson production cross
sections at the LHC is the calculation of the matrix element of the
hard scattering process as introduced in Sect. \ref{sec:Theory}. The
integration over the full phase space in Eqn. \ref{EQN:FacTheorem} 
including spin- and color-effects is a complicated task, and can
be only achieved with Monte Carlo (MC) sampling methods that are
implemented in MC event generators.

The connection of hard scattering processes at high energy scales to
the partons within the proton at low scales is described by parton
shower models. The basic idea of these models is to relate the partons
from the hard interaction at $Q^2 \gg \Lambda^2 _{QCD}$ to partons near
the energy scale of $\Lambda_{QCD}$ via an all-order approximation in
the soft and collinear regions. The commonly used approach is the
leading-log approximation, where parton showers are modeled as a
sequence of the splitting of a mother parton to two daughter partons.
The implementation of parton showers is achieved with MC techniques. A
detailed discussion of these models can be found elsewhere 
\cite{Butterworth:2010ym}.  Phenomenological models have to be applied
at the scale $\Lambda _{QCD}$ to describe the process of
hadronization, i.e. a description for the formation of hadrons from
final state partons, such as the Lund string model 
\cite{Andersson:1983ia} and the cluster model \cite{Field:1982dg, Webber:1983if}.

Multiple purpose event generators include the following aspects of $pp$
collisions: the description of the proton via an interface to PDF
sets; initial state parton shower models; hard scattering processes and
the subsequent decay of unstable particles; and the simulation of the final
state parton shower and hadronization.

The commonly used event generators in the relevant analyses of this 
review article are {\sc pythia~} \cite{Sjostrand:2006za}, \Pythia8 \cite{Sjostrand:2007gs},
\Herwig \cite{Corcella:2000bw}, \HerwigPP \cite{Bahr:2008pv} and
\Sherpa \cite{Gleisberg:2008ta}. All generators contain an extensive
list of SM and BSM processes, calculated with LO matrix elements. 
Higher-order corrections are also available for some
important processes.  Several programs such as \MadGraph
\cite{Alwall:2011uj}, \MCFM \cite{Campbell:2003hd}, \Alpgen
\cite{Mangano:2002ea} and \Blackhat \cite{Berger:2008sj} calculate
matrix elements for LO and NLO processes, but do not provide a full
event generation with parton showers and hadronization
effects. The subsequent event generation starting from the final state
parton configuration is often performed by the \Herwig and \Sherpa
libraries. 

Since the matrix element calculations give not only a good description
of the hard emission of jets in the final states, but also handle
interferences of initial and final states correctly, it is desirable
to combine matrix element calculations and parton showers. This
combination is complicated by the fact that the phase space of
NLO matrix element calculation partially
overlaps with parton shower approaches. Several matching schemes
have been suggested to provide a combination methodology of LO 
matrix element calculations for different parton multiplicities
and parton shower models.  These schemes avoid double counting of
final state particles by reweighting techniques and veto-algorithms.
The Mangano (MLM) scheme \cite{Mangano:2001xp} and the
Catani-Krauss-Kuler-Webber (CKKW) matching scheme \cite{Catani:2001cc}
\cite{Krauss:2002up} are widely used for tree-level generators.

Matching schemes can only be applied for LO calculations, while the
matching between parton showers and NLO matrix element calculations is
more sophisticated and advanced methods have to be used.  The \MCAtNLO
approach \cite{Frixione:2002ik} was the first prescription to match
NLO matrix elements to the parton shower framework of the \Herwig
generator. The basic idea is to remove all terms of the NLO matrix
element expression which are generated by the subsequent parton
shower. The \Powheg procedure \cite{Frixione:2007vw} which is
implemented in the \PowhegBox framework \cite{Alioli:2010xd} was
the second approach developed for NLO matrix element matching. The \Powheg
procedure assumes that the highest energy emitted parton is generated
first and then feed into the shower generators of the subsequent
softer radiation. In contrast to the \MCAtNLO approach, only positive
weights appear and in addition the procedure can be interfaced to
other event generators besides {\sc herwig}.

The matching in the \Sherpa generator relies on the CKKW matching
scheme for LO ME and on the \Powheg scheme for NLO
calculation. \Pythia8 also  includes the possibility of matching NLO
matrix elements via the \Powheg approach.

A crucial ingredient for all MC event generators is the knowledge of
the proton PDF.  The determination of the PDFs has been performed by
several groups. The \CTEQ \cite{Nadolsky:2008zw}, \MRST
\cite{Martin:2009iq} and \NNPDF \cite{Ball:2011us} collaborations
include all available data for their fits but differ in the treatment
of the parametrization assumptions of $f_{q} (x, Q^2)$ in Eqn.
\ref{EQN:FacTheorem}. The \HeraPDF \cite{CooperSarkar:2011aa} group
bases their PDF fits on a subset of the available data, i.e. mainly on
the deep inelastic scattering measurements from the HERA collider.  The
results presented in this review rely mainly on the \CTEQ and \MRST PDFs.

\subsection{SM Predictions of Diboson Production Cross Sections}

Table \ref{tab:CrossSections} summarizes LO and NLO predictions of the
diboson production  in $pp$ collisions at
$\sqrt{s} = 7\,\TeV$ based on the MCFM generator. The uncertainties due to scales and PDF
variations are also shown.  The difference between LO and NLO
predictions varies up to 25\% depending on the final state. The
difference between different PDF sets is on the order of 3\%.

The most recent NNLO predictions for $ZZ$ suggests a further increase by 11\% on the inclusive 
cross section \cite{Cascioli:2014yka}. However, electroweak effects are not yet taken into account, which will lead to a reduction of the cross section.

\begin{table}[h]
\tbl{LO and NLO predictions of diboson production cross sections at 7
  TeV in $pp$ collisions based on MCFM for different PDF sets. The
  uncertainties include the PDF, renomalization, and factorization
  scale uncertainties. These uncertainties are evaluated based on the
  variation of renormalization and factorization scales up and down by a factor of
  two and from the eigenvector error sets of the CT10 PDF set. The $Z$
  boson is defined by the given mass range.}  {
\begin{tabular*}{\textwidth}{@{\extracolsep{\fill}}llclclclclc}
\hline \hline
Process 							&	$Z$ boson	& LO 			 	&	NLO 					&	NLO 				& Uncertainty			\\
								&	cut [GeV]	& MSTW2008			&	MSTW2008			& 	CT10			& CT10 + Scales		\\	[2pt]
\hline  
$\sigma(pp\rightarrow W^+W^-+X) $ [pb]	&	-		&	$29.1$			&	$46.3$				&	$45.3$			&	$^{+2.1}_{-1.9}$		\\ [2pt]
\hline
$\sigma(pp\rightarrow W^+Z+X) $ [pb]	&	$66-116$	&	$7.3$			&	$11.6$				&	$11.4$			&	$^{+0.7}_{-0.6}$		\\ [2pt]
$\sigma(pp\rightarrow W^-Z+X) $ [pb]	&	$66-116$	&	$4.1$			&	$6.5$				&	$6.2$			&	$^{+0.4}_{-0.4}$		\\ [2pt]
$\sigma(pp\rightarrow W^\pm Z+X) $ [pb]	&	$66-116$	&	$11.4$			&	$18.1$				&	$17.7$			&	$^{+1.1}_{-1.0}$		\\ [2pt]
$\sigma(pp\rightarrow W^\pm Z+X) $ [pb]	&	$60-120$	&	$11.4$			&	$18.1$				&	$17.7$			&	$^{+1.1}_{-1.0}$		\\ [2pt]
\hline
$\sigma(pp\rightarrow ZZ+X) $ [pb]		&	$66-116$	&	$8.8$			&	$12.3$				&	$12.1$			&	$^{+xx}_{-xx}$		\\ [2pt]
$\sigma(pp\rightarrow ZZ+X) $ [pb]		&	$60-120$	&	$9.0$			&	$12.5$				&	$12.3$			&	$^{+xx}_{-xx}$		\\ [2pt]
\hline \hline
\end{tabular*}
\label{tab:CrossSections} 
}
\end{table}

\subsection{\label{sec:theoryatgc}Description of SM Extensions and aTGCs}

Since this article focuses on the LHC results at $\sqrt{s} = 7 \TeV$,
we restrict ourselves to the discussion of the triple gauge boson
interaction vertices and their new physics modifications, commonly
called aTGCs.  There are only two triple gauge vertices allowed in the
SM, $WWV$ with $V=Z, \gamma$. The $WWV$ vertices are fully determined
by the $SU(2)_L \times U(1)_Y$ gauge group structure. A generic
anomalous contribution to the $WWV$ vertex can be parametrized in
terms of a purely phenomenological effective Lagrangian. The most
general Lagrangian that describes the trilinear interaction of
electroweak gauge bosons with the smallest number of degrees of
freedom \cite{Hagiwara:1986vm},
\cite{Zeppenfeld:1987ip}, \cite{Baur:1988qt}, reads as
\begin{eqnarray}
\label{eqn:EffLag}
i \LC_{WWV} / g_{WWV} 	&=&	 [1+ \Delta g_V^1] V^\mu (W^-_{\mu\nu} W^{+\nu} - W^+_{\mu\nu} W^{-\nu} ) 	\\
		& &
\nonumber
			+ [1+\Delta \kappa_V] W^+_\mu W^-_\nu V^{\mu\nu} \\
                     & &
\nonumber
+  \frac{\lambda_V}{m_W^2} V^{\mu\nu} W^{+\alpha}_\nu W^{-}_{\alpha\mu}  \\
		& &
\nonumber
			+ i g_V^4 W^-_\mu W^+_\nu (\partial ^\mu V^\nu + \partial^\nu V^\mu)  \\
		& &
\nonumber
			+ i g_V^5 \epsilon_{\mu\nu\alpha\beta} [(\partial^\alpha W^{-\mu})W^{+\nu} - W^{-\mu}(\partial ^\alpha W^{+\nu})] V^\beta 	\\
		& &
\nonumber
			- \frac{\tilde \kappa _V }{2} W^-_\mu W^+_\nu \epsilon^{\mu\nu\alpha\beta} V_{\alpha \beta} \\
		& &
\nonumber
			- \frac{\tilde \lambda _V }{2 m_W^2} W^-_{\rho\nu} W^{+\mu}_\nu \epsilon^{\nu\rho\alpha\beta} V_{\alpha\beta},
\end{eqnarray}
where $g_{WW\gamma} = -e$ and $g_{WWZ} = - e \cot \theta _W$ are the
two couplings, $m_W$ is the mass of the $W$ boson, $V^\mu$, $W^\mu$,
$W_{\mu\nu}=\delta _\mu W_\nu - \delta _\nu W_\mu$ and
$V_{\mu\nu}=\delta _\mu V_\nu - \delta _\nu V_\mu$ are the gauge boson
vector fields and their field tensors. The anomalous couplings
are described by seven parameters for each of the $WWV$ vertices,
$\Delta g_V^1$, $\Delta \kappa_V$, $\lambda_V$, $g_V^4$, $g_V^5$,
$\tilde \kappa_V$ and $\tilde \lambda_V$. All aTGCs are set to be zero
in the SM.

The first three terms in Eqn. \ref{eqn:EffLag} are $CP$-invariant,
while the remaining four terms violate the $C$- and/or $P$-
symmetry. Furthermore, electromagnetic gauge invariance requires that
$\Delta g_\gamma^1 = g_\gamma^4 = g_\gamma^5=0$, while the
corresponding $Z$ boson coupling parameters $\Delta g_Z^1, g_Z^4$ and
$g_Z^5$ can differ from their SM values. We are left with five
independent $C$- and $P$-conserving parameters $\Delta g_Z^1$, $\Delta
\kappa_\gamma$, $\Delta \kappa_Z$, $\lambda_\gamma$ and $\lambda _Z$,
and six $C$- and/or $P$-violating parameters $g_Z^4$, $g_Z^5$, $\tilde
\kappa_\gamma$, $\tilde \kappa_Z$, $\tilde \lambda_\gamma$ and $\tilde
\lambda_Z$. The studies presented in this paper assume gauge
invariance and conservation of $C$ and $P$ separately, resulting in
five independent aTGC parameters.

In order to further reduce the number of independent parameters and
therefore allow for a simpler experimental derivation of limits,
several additional assumptions can be made.  The `equal coupling'
scenario assumes that the anomalous triple gauge couplings are the
same for $\gamma$ and $Z$ bosons, i.e., $\kappa_Z = \kappa_\gamma$,
$\lambda_\gamma = \lambda _Z$ and $\Delta g_Z^1=0$, and hence leads to
two free parameters. The `LEP' scenario is motivated by the $SU(2)
\times U(1)$ gauge invariance \cite{Gounaris:1996rz} and assumes
$\Delta \kappa_\gamma = \cos^2\theta_W (\Delta g_1^Z - \Delta
\kappa_Z)/\sin^2\theta_W$ and $\lambda_Z = \lambda_\gamma$, leading to
three independent parameters. The `HISZ' scenario
\cite{Hagiwara:1993ck} assumes no cancellations between tree-level and
loop contributions, leading to the constraints, $\Delta g_1^Z = \Delta
\kappa _Z/(\cos^2\theta_W - \sin^2\theta)$, $\Delta \kappa_\gamma =
2\Delta \kappa _Z \cos^2\theta_W/(\cos^2\theta - \sin^2\theta_W)$, and
$\lambda_Z = \lambda_\gamma$, which leave two free parameters. A
review of various parametrization scenarios for aTGCs can be
found in \cite{aTGCReview}.

Non-zero aTGCs will lead to a change in the calculation of matrix
elements. As an example, we discuss here the change of the matrix
element $\Delta {\cal{M}}_{Z_H, W_H}$, describing the production of
the $q\bar q \rightarrow WZ$ process for large center-of-mass energies
$\sqrt{s} \gg M_W$ of the interacting partons
\cite{Zeppenfeld:1987ip}. The two subscripts $Z_H$ and $W_H$ denote
the helicity of the final state bosons (0, $+$, and $-$). The expected
changes read as
\begin{eqnarray}
\label{eqn:EffLagWZ}
\nonumber
\Delta {\cal{M}}_{\pm,0} 	&\sim&  \frac{\sqrt{\hat s}}{2m_W} [\Delta g_Z^1 + \Delta \kappa _Z + \lambda _Z] \frac{1}{2} (1\mp \cos \theta_Z^*), \\
\nonumber
\Delta {\cal{M}}_{\pm,\pm} 	&\sim&  \frac{\hat s}{2m^2_W} [\lambda _Z] \frac{1}{\sqrt{2}} (\sin \theta_Z^*), \\
\nonumber
\Delta {\cal{M}}_{0,\pm} 	&\sim&  \frac{\sqrt{\hat s}}{2m_W} [2\Delta g_Z^1 + \lambda _Z] \frac{1}{2} (1\pm \cos \theta_Z^*), \\
\nonumber
\Delta {\cal{M}}_{0,0} 		&\sim&  \frac{\hat s}{2m^2_W} [\Delta g^1 _Z] \frac{1}{\sqrt{2}} (\sin \theta_Z^*),
\end{eqnarray}
where $\theta^*_Z$ denotes the production angle of the $Z$ boson with
respect to the incoming quark direction. In general, the matrix
element $\cal{M}$ and thus the production cross section increases with
increasing center-of-mass energy of the interacting partons. The LHC allows for a larger
sensitivity to the $\lambda_Z$ and $\Delta g_Z^1$ parameters as their
contribution to the production cross section increases with the
squared center-of-mass energy $s$.
In addition, different aTGCs lead to different angular
distributions of the final state particles.  A multi-dimensional
differential cross section measurement can constrain these parameters
individually.

Even though the $ZZV$ vertex is forbidden in the SM, new physics
scenarios might allow for such interaction vertices. The corresponding
phenomenological effective Lagrangian reads as
\begin{equation}
\label{eqn:EffLagGamma}
i \LC_{ZZV}  	=	- \frac{e}{M_Z^2} [f_4^V ((\delta_\mu V^{\mu\beta}) Z _\alpha (\delta ^\alpha Z_\beta)) + f_5^V ((\delta^\sigma V_{\sigma\mu}) \tilde Z^{\mu\beta} Z_\beta)],
\end{equation}
where the anomalous on-shell $ZZ$ production is parametrized by two
$CP$-conserving ($f_5^V$) and two $CP$-violating ($f_4^V$) parameters.
Similarly, the vertex $ZV\gamma$ \cite{Baur:1992cd} can be described
by an effective Lagrangian, where the anomalous couplings are
described by $h_3^Z$ and $h_4^Z$ for the $ZZ\gamma$ vertex, and
$h_3^\gamma$ and $h_4^\gamma$ for the $Z\gamma\gamma$ vertex. It
should be noted that the parameters $h_i^V$ and $f_i^V$ are partly
correlated.  In contrast to the $WWV$ vertex, the effective parameters
for the $ZZV$ and $ZV\gamma$ vertices vanish at tree level in the SM
and only higher-order corrections allow for small $CP$-conserving
couplings in the order of $10^{-4}$.

An overview of the properties of all aTGC parameters is given in Table \ref{tab:aTGCParameter}.


\begin{table}[t]
\tbl{Summary of $CP$-conserving aTGC parameters and their relevant production processes.}  {
\begin{tabular*}{\textwidth}{@{\extracolsep{\fill}}llclclclclc}
\hline \hline
Parameter			& aTGC-Vertex			& 	Sensitive Process	\\
\hline
$\Delta g_Z^1$			&	$WWZ$			&	$q\bar q \rightarrow Z \rightarrow WW$, $q\bar q' \rightarrow W \rightarrow ZW$			\\
$\Delta \kappa_Z$		&	$WWZ$			&	$q\bar q \rightarrow Z \rightarrow WW$, $q\bar q' \rightarrow W \rightarrow ZW$			\\
$\Delta \kappa_\gamma$	&	$WW\gamma$		&	$q\bar q \rightarrow \gamma \rightarrow WW$, $q\bar q' \rightarrow W \rightarrow W\gamma$	\\
$\lambda_Z$			&	$WWZ$			&	$q\bar q \rightarrow Z \rightarrow WW$, $q\bar q' \rightarrow W \rightarrow WZ$			\\
$\lambda_\gamma$		&	$WW\gamma$		&	$q\bar q \rightarrow \gamma \rightarrow WW$, $q\bar q' \rightarrow W \rightarrow W\gamma$	\\
\hline
$f_5^\gamma$			&	$ZZ\gamma$		&	$q\bar q \rightarrow \gamma \rightarrow ZZ$			\\
$f_5^Z$				&	$ZZ\gamma$		&	$q\bar q \rightarrow \gamma \rightarrow ZZ$			\\
\hline
$h_4^\gamma$		&	$Z\gamma\gamma$	&	$q\bar q \rightarrow \gamma \rightarrow Z\gamma$		\\
$h_4^Z$				&	$Z\gamma\gamma$	&	$q\bar q \rightarrow \gamma \rightarrow Z\gamma$		\\
\hline \hline
\end{tabular*}
\label{tab:aTGCParameter} 
}
\end{table}

The unitarity of the SM electroweak Lagrangian is preserved due to 
gauge invariance. The introduction of aTGCs in the Lagrangian alters
its gauge structure and can lead to unitarity violations at relatively
low energies. This can be seen in Eqn. \ref{eqn:EffLagWZ}, where some
matrix elements are proportional to the center-of-mass energy.  To
avoid unitarity violations at high energies, the Lagrangian approach
in Eqn. \ref{eqn:EffLag} is replaced by a form factor as
\begin{equation}
\lambda_s = \frac{\lambda_0}{(1+s/\Lambda^2)^n}
\end{equation}
where $\lambda_0$ is the aTGC parameter at low energies and $\Lambda$
is the energy cut-off scale at which new physic effects become
dominant \cite{Ellison:1998uy}. By convention, $n=2$ is usually
chosen.  In some sense, the form factor can be interpreted by treating
the couplings in the Lagrangian as energy dependent\footnote{Strictly
  speaking, the Lagrangian couplings must remain constant and these
  are actually two different approaches.}.

The choice of the form factor parametrization and the cut-off scale
$\Lambda$ is arbitrary as long as it conserves unitarity for
reasonably small aTGC coupling parameters. It is not important at
$e^+e^-$ colliders as the fixed center-of-mass energy of the
interaction particles allows for a well-defined translation between
different choices of parametrization. The situation is different for
hadron colliders, where only the center-of-mass energy of the
interacting protons is known, but not the energy of the interacting
partons. The measured production cross sections are always
integrated over a certain energy range and their interpretation in
terms of aTGCs depends on the form factors chosen.

The cut-off scale $\Lambda$ is usually chosen such that the extracted
limits on aTGCs still preserve unitarity within a given analysis. In
order to give experimental limits that are free from the arbitrary
choice of the form factor, the cut-off scale is also set to $\Lambda =
\infty$, i.e. using a constant form factor and hence violating
unitarity at high energies.  However, the aTGC limits based on
$\Lambda = \infty$ are the more stringent and less conservative.

The introduction of the form factor is conceptually overly
restrictive, since the only physics constraint is that the theory
respects the unitarity bound in the region where there is data. A
newly-proposed approach for the study of aTGCs is based on effective
field theories \cite{Degrande:2012wf}.  An effective field theory of
the SM can be written as
\[ \LC_{eff} = \LC_{SM} + \sum _i \frac{c_i}{\Lambda ^2} O_i + \cdots \]
where $O_i$ are dimension-six operators, and $c_i$ are coupling
parameters that describe the interaction strength of new physics with
the SM fields. It can be shown \cite{Hagiwara:1993ck} that only three
dimension-six operators affect the electroweak gauge boson
self-interaction. The corresponding coefficients $c_i$ can be related
to the aTGCs that are discussed above.

There are several advantages of using effective field theories for the
description of aTGCs. By construction, an effective quantum field
theory is only useful up to energies of the order $\Lambda$. As long
as the effective theory describes data, it automatically respects the
unitarity bound. Hence, no further assumptions on the energy scale
$\Lambda$ have to be applied. Furthermore, the effective field theory
approach has fewer parameters and is renormalizable by
construction. While the results on aTGC presented in this review
article are based on the modified Lagrangian approach, future
measurements might use effective field theories for aTGC studies.


\section{\label{sec:ExperimentalAspects}Experimental Aspects of Diboson Measurements at the LHC}

\subsection{The LHC machine}

The LHC \cite{Evans:2008zzb} is currently the world's most powerful
particle accelerator. It consists of several stages that successively
increase the energy of the protons (and heavy ions).  Protons are
accelerated by the LINAC to 50 MeV, the Proton Synchrotron Booster to
1.4 \GeV, the Proton Synchrotron to 26 \GeV, and Super Proton
Synchrotron to 450 \GeV) before the final injection into the LHC
ring. With a circumference of 27 km and a magnetic field of 8.3 T, the
LHC can accelerate each proton beam up to 7 \TeV.  With a revolution
frequency of 11.25 kHz, and a maximum of 2808 bunches that can be
filled with up to 115 billion protons per bunch, the instantaneous
luminosity can reach $10^{34}$ cm$^{-2}$s$^{-1}$ with a beam emittance
of 16 $\mu$m, providing a bunch collision rate of 40 MHz. The two
proton beams are brought together and collide head-on in four points
around the LHC ring, where four large detectors - ALICE, ATLAS, CMS
and LHCb - are located.

First $pp$ collisions at the LHC were carried out in November 2009
with a proton-proton center-of-mass energy of 0.9 \TeV. The LHC
started operations at a center-of-mass energy of 7 \TeV\, in March
2010 and delivered an integrated luminosity of about 50 pb$^{-1}$ in
2010 and 5 fb$^{-1}$ in 2011 to both ATLAS and CMS experiments.  The
peak instantaneous luminosity delivered by the LHC at the start of a
fill increased from $2.0 \times 10^{32}$ cm$^{-2}$s$^{-1}$ in 2010 to
$3.6 \times 10^{33}$ cm$^{-2}$s$^{-1}$ by the end of 2011.  The
machine increased its center-of-mass energy to 8 \TeV\ in 2012 and
delivered 25 fb$^{-1}$ to both experiments. The peaks instantaneous
luminosity of $7.7 \times 10^{33}$ cm$^{-2}$s$^{-1}$ was reached in
November 2012. This is close to the design luminosity of $10^{34}$
cm$^{-2}$s$^{-1}$, albeit at twice the beam crossing time.  The data
analyses for the 2012 run are still ongoing, and this paper only
covers results from the 7 \TeV\ run in 2011.  During this period, the
maximum number of bunch pairs colliding was 1331, and the minimum
bunching spacing was 50 ns with a typical bunch population of $1.2
\times 10^{11}$ protons. The maximum number of inelastic interactions
per bunch crossing ('pile-up') was 20, and the average was 9.1.

\subsection{The ATLAS and CMS Detectors}

\begin{figure}[h]
\begin{center}
\resizebox{0.48\textwidth}{!}{\includegraphics{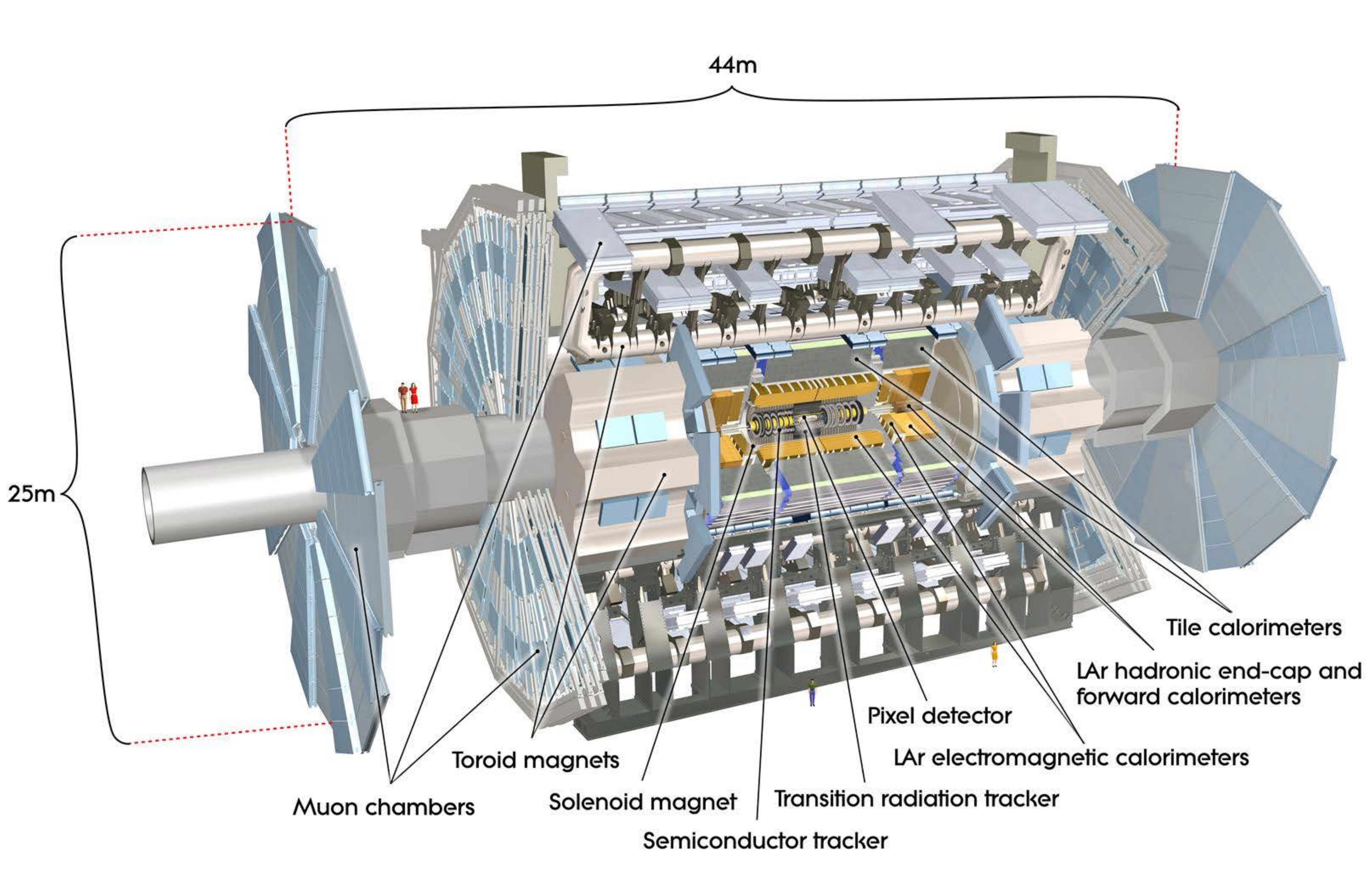}}
\resizebox{0.48\textwidth}{!}{\includegraphics{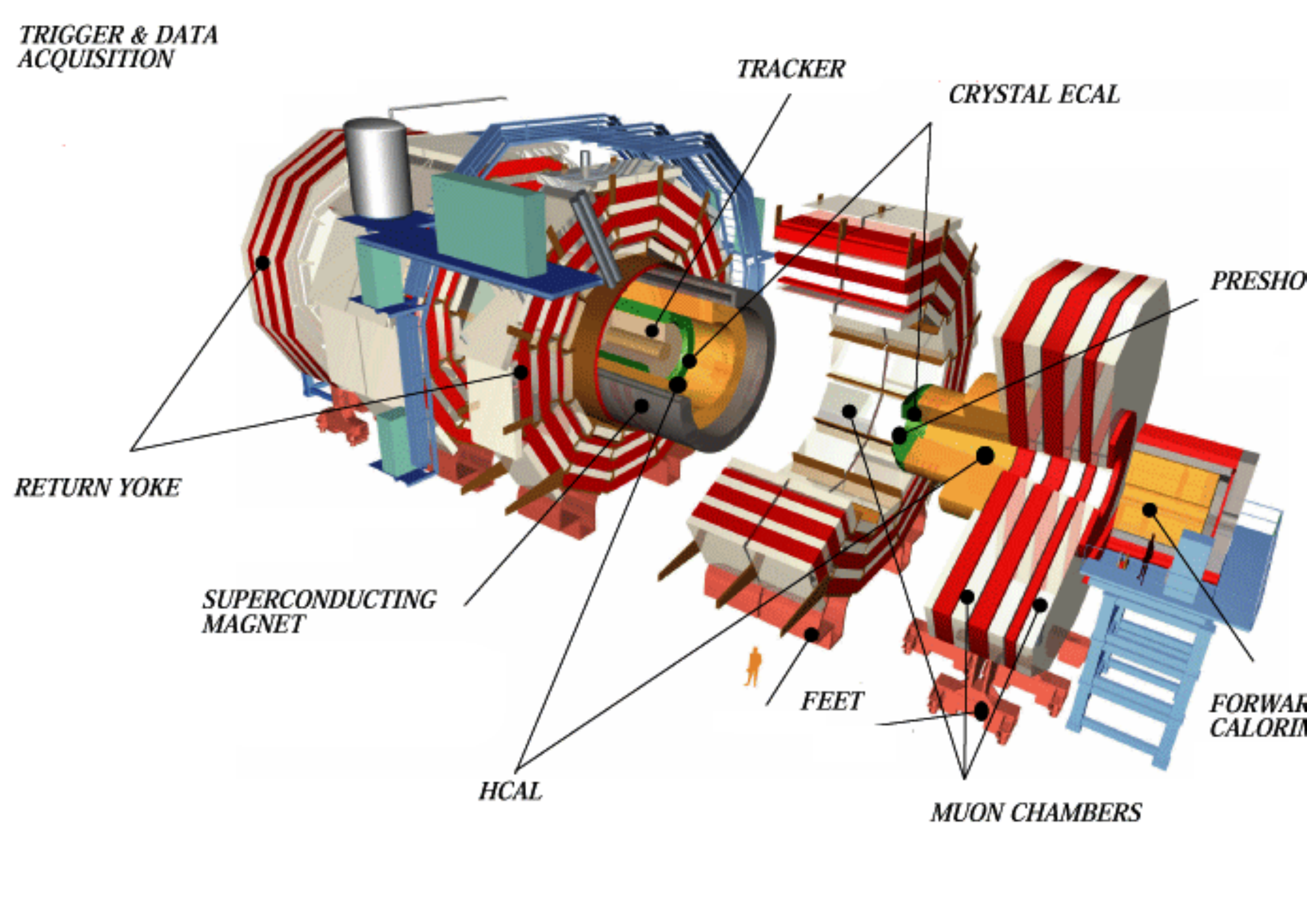}}
\caption{\label{fig:detectors} Layouts of the ATLAS (Left) and CMS (Right) detectors.}
\end{center}
\end{figure}

Since the topologies of new physics processes are unknown, detectors
should be designed to be sensitive to all detectable particles and
signatures ($e$, $\mu$, $\tau$, $\nu$, $\gamma$, jets, $b$-quarks)
produced in the interactions. Both ATLAS \cite{Aad:2008zzm} and CMS
\cite{Chatrchyan:2008aa} detectors are general-purpose detectors and
are composed of many sub-detectors, each of which has a specific task
in the reconstruction of the events. Although these two detectors are
differ in design and conception, the basic detection structure is
similar. Both detectors have fast, multi-level trigger systems to
select complex signatures online; fast data acquisition systems to
record all selected events; excellent inner tracking detectors
allowing efficient high-$p_T$ tracking and secondary vertex
reconstruction; fine-grained, high-resolution electromagnetic
calorimeters for electron and photon reconstruction; 
full coverage hadronic calorimetry for jet and missing transverse
energy measurements; and high-precision muon systems with standalone
tracking capability. The layouts of the ATLAS and CMS detectors are
shown in Fig.~\ref{fig:detectors}.  ATLAS emphasizes jet, missing
transverse energy, and standalone muon measurements, while CMS has
prioritized electron, photon, and inner tracking measurements.

CMS has chosen to have a single huge solenoid immersing the inner
tracker and electromagnetic and hadronic calorimeters inside a 4 T
axial magnetic field. To reduce the occupancy of the inner tracking
detector at the LHC design luminosity, the inner tracker consists
solely of silicon pixel and microstrip detectors, which provide high
granularity at all radii. CMS relies on the fine lateral granularity
of its lead tungstate scintillating crystal electromagnetic
calorimeter for electron and photon measurements.  The hadronic
sampling calorimeter system consists of brass absorbers and
scintillating tiles readout via wavelength-shifting optical fibers
guiding the light to photomultiplier tubes.  The strong constraints
imposed by the solenoid have resulted in a barrel hadronic calorimeter
with insufficient absorption before the coil, so a tail catcher has
been added around the coil to provide better protection against
punch-through to the muon system. Driven by the design of the magnet,
CMS relies on the high solenoidal field to bend muon tracks in the
transverse plane, requiring the extrapolation of the track into the
inner tracker.

ATLAS has chosen to have three different magnet systems: a thin
solenoid around the inner tracking system, one eight-fold barrel and
two eight-fold endcap air-core toroid magnets arranged radially around
the hadron calorimeters. The inner tracker consist of silicon pixel
and microstrip detectors at small radii and transition radiation
tracker (TRT) at large radii.  For electron and photon measurements,
ATLAS relies on the fine segmentation along both the lateral and
longitudinal directions of electromagnetic shower development using
a lead and liquid-argon sampling technique. The hadronic
calorimeter system uses a sampling technique similar to that used by
CMS except iron and copper are used as absorbers.  The structure of
barrel and endcap toroid magnets allows standalone muon tracking
inside the large area spanned by the toroids.

\subsection{\label{sec:RecoOb}Reconstruction Objects for Physics Analysis}

The $z$-axis of the coordinate system of both detectors is chosen to
be along the beam-direction, the $x$-axis points to the center of the
LHC, and the $y$-axis points upwards. The origin of the coordinate
system is placed at the nominal collision point, i.e. in the center of
the detectors. Two radial coordinates are used to describe event
kinematics: the azimuthal angle $\phi$ is defined in the $xy$-plane
and the polar angle $\theta$ is defined with respect to the
$z$-axis. The polar angle is commonly used to define the
pseudorapidity parameter $\eta = - \log \tan \frac{\theta}{2}$. 

Observations of heavy diboson pair production processes at the LHC
resulted from analyses of the fully leptonic (in which $W$ decays to
$\ell\nu$ and $Z$ bosons decay to $\ell\ell$, where $\ell=e, \mu$) and
semi-leptonic decay channels (in which one boson decays leptonically
while the other boson decays to hadrons or neutrinos).  The fully
leptonic decay channels produce clean experimental signatures of one
or more high-$p_T$ charged leptons and, in the case of $W \rightarrow
\ell\nu$, large missing transverse energy \ETMiss. The semi-leptonic
decay (or neutrino decay) channels result in the lepton$+$jets (or
lepton+\ETMiss) final states and are harder to detect experimentally
due to the large $V+$jets background despite higher production cross
sections.

The main analyses that are discussed in this review are based on the reconstruction of the electron and muon kinematics. 
Since the initial momenta of interacting partons in the plane transverse to
the beamline is zero, the projected momenta of reconstructed objects
in this plane, called transverse momenta $p_T$, are from special importance in event
reconstruction.  Similar concept is introduced for measured energies,
where the transverse energy is defined as $\ET = E \sin \theta$. 

The data analyzed were often selected online by a single lepton ($e$
or $\mu$) or dilepton trigger with a threshold on the transverse
energy in the electron case and on the transverse momentum in the muon
case. Different thresholds (normally around 20 \GeV\ for single lepton
triggers and around 12 \GeV\ for dilepton triggers) were applied
depending on the average instantaneous luminosity of running periods.

The reconstruction of electrons combines electromagnetic calorimeter
and inner tracker information \cite{Khachatryan:2010pw, Aad:2010bx} and makes use of standard electron
reconstruction algorithms at ATLAS and CMS.  Candidate electrons are
required to pass certain $p_T$ ($E_T$) threshold cuts and to be located
inside the detector fiducial regions.  Additional electron
identification requirements are imposed which rely on electromagnetic
shower shape observables, on associated track quality variables and on
track-cluster matching observables, so as to preserve the highest
possible efficiency while reducing the multijet background \cite{Aad:2011mk, CMS:ECAL, Chatrchyan:2013dga}. The
$\eta$-coverage of the electron and photon candidate reconstruction is
$|\eta| < 2.37$ and $|\eta|<2.5$ for ATLAS and CMS, respectively. The
regions $1.37<|\eta|<1.52$ in ATLAS and $1.44<|\eta|<1.57$ in CMS are
typically excluded for most analyses, as they contain a significant
amount of service infrastructure which reduces the
reconstruction quality.

The reconstruction of photon candidates is similar to the electron
case. However, specific cuts are applied on the shower-shape of the
reconstructed electromagnetic clusters and on tracking information. If
no track can be associated to the electromagnetic cluster, then the
photon is called `unconverted'. An association of two tracks to the
electromagnetic cluster imply a previous photon conversion into
$e^+e^-$ and therefore the corresponding photon candidates are
labeled `converted' photons.

Muon $p_T$ is reconstructed using hits collected in both the inner
tracker and the outer muon spectrometer and corrected for energy loss
measured by the calorimeter. Good quality reconstruction is ensured by
requiring a minimum numbes of hits associated with the track from
both inner and outer tracking systems. Due to limited pseudorapidity
coverage of the inner tracker and trigger detectors, muon candidates
are reconstructed within $|\eta|<2.4$ for CMS and $|\eta|<2.7$ for
ATLAS. However, the ATLAS inner detector only covers a region up to
$|\eta|<2.5$ and therefore most analyses using reconstructed muons
limit themselves also to $|\eta|<2.5$ in order to have combined muon
candidates, i.e. tracks which are reconstructed in the inner detector
and the muon spectrometer \cite{ATLAS:Muon, Chatrchyan:2012xi}.

To ensure candidate electrons and muons originate from the primary
interaction vertex, some analyses required these lepton candidates to
have small longitudinal and transverse impact parameters. These
requirements reduce contamination from heavy flavor quark decays and
cosmic rays.  Leptons from heavy boson decays tend to be isolated from
other particles in the event, while fake leptons or leptons from heavy
quark decays will usually be close to a jet.  To suppress the
contribution from hadronic jets which are misidentified as leptons,
electron and muon candidates are often required to be isolated in the
inner tracker and (or) the electromagnetic calorimeter. Certain cuts
are made on the sum of transverse energies of all clusters around the
lepton or the sum of the $p_T$ of all tracks that originate from the
primary vertex and are within a certain cone around the lepton
candidate. A typical cone-size $\Delta R=\sqrt{(\Delta \eta)^2 + (\Delta \phi)^2}$ is 0.3 in the
$(\eta,\phi)$-plane. For the CMS case, a relative isolation variable
combining the tracker and calorimeter isolation information is used.

Jets are reconstructed from topological clusters of energy in the
calorimeter using the anti-$k_T$ algorithm with a certain radius
parameter \cite{Cacciari:2008gp}.  Jet energies are often calibrated using $p_T$- and
$\eta$-dependent correction factors derived from studies based on the
{\sc Geant4} simulation \cite{Agostinelli:2002hh}, dijet,
$\gamma+$jets, and $Z+$jets collision data. Jets are classified as
originating from $b$-quarks by using algorithms that combine
information about the impact parameter significance of tracks in a jet
which has a topology of semileptonic $b$- or $c$-hadron decays. ATLAS
and CMS reconstruct particle jets within regions of $|\eta|<4.9$ and
$|\eta|<5.0$, respectively \cite{Aad:2011he, CMS:2009nxa}.

A summary of the identification and reconstruction features for
electrons, muons, and jets of the ATLAS and CMS experiment is given in
Tab. \ref{Tab:KinID:Summary}. The given kinematic constraints are
also the basis for event selections that are discussed in the
following chapters.

\begin{table}[h]
\tbl{Summary of electron, muon and jet reconstruction in the ATLAS and CMS detectors.}
{
  \begin{tabular}{l l l} 
\hline
\hline
							&	ATLAS  												& CMS 										\\
\hline
Electron						&	 \multicolumn{2}{c}{Combination of inner tracker and calorimeter}  	\\
             						&	 \multicolumn{2}{c}{Shower shape cuts on calorimeter cluster}  	\\
							&	`loose', `medium', `tight' identification qualities					&	`loose', `tight' identification qualities\\
							&	$0<|\eta|<1.37$ and $1.52<|\eta|<2.37$						&	$0<|\eta|<1.44$ and $1.57<|\eta|<2.5$\\
							&	forward electrons: $2.37<|\eta|<4.9$	(w/o tracker)				&	\\
\hline
Muon						&	 \multicolumn{2}{c}{Combination of inner tracker and muon system}  	\\
							&	combined muons up to $|\eta|<2.5$							&	combined muons up to $|\eta|<2.4$	\\
							&	muon system coverage up to $|\eta|<2.7$						&	muon system coverage up to $|\eta|<2.4$	\\
							&	muon trigger available up to $|\eta|<2.4$						&		\\
\hline
Jets							&	Anti-$k_T$ algorithm with $\Delta R =0.4$ 					&	Anti-$k_T$ algorithm with $\Delta R =0.5$\\
							&	based on calorimeter information							&	based on particle flow algorithm\\
							&	coverage up to $|\eta|<4.9$								&	coverage up to $|\eta|<5.0$\\
\hline
\hline
   \end{tabular}
   \label{Tab:KinID:Summary}
}
\end{table}

Weakly interacting particles such as neutrinos leave the detector
unseen and can be only reconstructed indirectly.  The concept of this
indirect measurement is based on the fact that the momentum in the
transverse plane before the $pp$ collision is zero.
Undetected energy and momentum carried out of the detector will
therefore lead to a missing transverse energy, $\ETMiss = |\vec{E}_T^{\text{miss}}| $, in an
event. The two-dimensional vector of \ETMiss\ is based on the
calorimeter information and is calculated as the negative vector sum
of the transverse energies deposited in the calorimeter towers. The
latter is corrected for the under-measurement of the hadronic energy
in the calorimeters and muon tracks reconstructed by the inner tracker
and the muon spectrometer, leading to the following schematic
definition:
\begin{equation}
\vec{E}_T^{\text{miss}} = - \sum_{i} \vec E^{\text{calo}}_{T,i} - \sum_{i} \vec p_{T,i}^{~\mu}.
\end{equation}

For the traditional calorimeter-based algorithm, the correction for
the under-measurement of the hadronic energy in the calorimeter is
performed by replacing energies deposited by reconstructed jets with
those of the jet energy scale corrected jets.  A different track-based
algorithm to correct for the under-measurement of the hadronic energy
in the calorimeter was also developed by both experiments. In this
algorithm, the transverse momentum of each reconstructed charged
particle track is added to the total missing transverse momentum, from
which the corresponding transverse energy expected to be deposited in
the calorimeters is subtracted.

At both ATLAS and CMS, tau reconstruction and identification
concentrates on the tau hadronic decay modes which are characterized
by either one or three charged pions accompanied by neutral
pions \cite{ATLAS:2011oka, Chatrchyan:2012zz}. They are classified according to the number of reconstructed
charged decay particles (prongs).  Several sophisticated tau
identification algorithms have been developed by both collaborations
using different sets of identification variables such as tracking and
calorimeter information to find the optimal set of cuts in a
multi-dimensional phase space.

Detailed simulations of the ATLAS and CMS detector response have been
developed over the recent years. Both simulations are based on the
{\sc Geant4} package, which describes the interactions of all final
state particles with the detectors at a microscopic level. In a second
step, the digitization of the simulated detector interactions is
performed and the nominal data reconstruction algorithms are applied.

Despite of the great detail of the simulation software, several
differences between data and MC predictions remain. To improve the
agreement between data and MC simulations, several quantities
such as reconstruction efficiencies or energy scales are
measured independently in data and simulation. Correction factors are
then determined and applied to the simulations to account for the
observed differences. 

\subsection{\label{sec:Methodology}Methodology of Cross-Section Measurements at the LHC}

For the measurement of the diboson production at the LHC, it is
generally assumed that both bosons are on-shell. Three different modes
can be considered in the decay of heavy diboson pairs: the full
hadronic decay channel where both bosons decay into quarks; the
semi-leptonic decay channel in which one boson decays into quarks and
the other to leptons; and the leptonic decay channel where the final
state contains four leptons.  The hadronic decay modes of the vector
bosons are hard to be distinguished in hadron colliders due to the
overwhelmingly large cross-section of jet-induced background
processes. The CMS collaboration has published a combined cross
section measurement based on semi-leptonic decays in $WW$ and $WZ$
pairs where the $W$ boson decays leptonically while the second boson
decays hadronically \cite{Chatrchyan:2012bd}. However, the systematic
and statistical uncertainties of this measurement are significantly
larger than the results based on studies of the fully leptonic decay
channel, which allows for a rather clean signal selection.  Due to
this precise signal selection and the fact that the branching ratios
of vector bosons are well known, the fully leptonic decay channel is
the best channel in which precision measurements of the production
cross section of diboson pairs at the LHC can be performed.

The theoretical prediction and calculation of diboson production cross
sections has been discussed in Sect. \ref{sec:DiBosonProd}. On the
experimental side, the inclusive production cross section can be
calculated via the following equation:
\begin{equation}
\label{EQN:CrossSectionExp}
\sigma_{VV}^{incl} = \frac{N_{s}}{\epsilon \cdot {\text{BR}} \cdot \IntLumi}.
\end{equation}
The number of signal events is determined as $N_{s} = N_{d} - N_{b}$,
where $N_{d}$ is the number of selected events in data, and $N_{b}$ is the
number of background events surviving the signal selection.  The
factor $\epsilon$ gives the fraction of signal events which pass the
signal selection criteria. In order to correct the inclusive cross
section for the choice of a specific decay channel, the total value
has to be corrected by the appropriate branching ratio BR. These
ratios are known to high accuracy from the LEP experiments
 \cite{Beringer:1900zz}.
The last term in the denominator of Eqn. \ref{EQN:CrossSectionExp} is
the integrated luminosity, i.e. a measure of the size of the data sample used.

The efficiency correction factor $\epsilon$ is usually estimated from
the fraction of signal MC events passing all cuts through the full
detector simulation.  It should be noted that certain requirements on
the final states can be applied directly at the generator level, for
example, final state leptons may have to pass some minimal $p_T$ cut.
The factor $\epsilon$ is thus defined as the ratio of all events which
pass the signal selection at the reconstruction level over the number
of all generated events.
However, the simulation usually exhibits some differences compared to
the real detector. These differences are corrected in the estimation
of $\epsilon$, following the methods described in
Sect. \ref{sec:RecoOb}.

The efficiency correction $\epsilon$ can be decomposed as the product
of a fiducial acceptance ($A$) and a detector-induced correction
factor ($C$): $\epsilon = A \cdot C$.  The fiducial acceptance, which
is the ratio of the number of events which pass the geometric and
kinematic cuts in the analysis at the generator level
($N^{\textrm{selected}}_{\textrm{gen.}}$) over the total number of
generated events in a simulated sample of signal process
($N^{\textrm{all}}_{\textrm{gen.}}$).  These selection cuts at the
generator level usually require geometric and kinematic constraints
close to the cuts applied on the reconstructed objects.
The dominant uncertainties on the fiducial acceptance are the scale
and PDF uncertainties.

The detector correction factor, $C$, is defined as the number of
selected events in the simulated sample
($N^{\textrm{selected}}_{\textrm{reco.}}$) over the number of events
in the fiducial phase space at the generator level
($N^{\textrm{selected}}_{\textrm{gen.}}$). Hence $\epsilon$ can be
written as
\begin{equation}
\epsilon = C\cdot A =  
\frac{N^{\textrm{selected}}_{\textrm{reco.}} }{N^{\textrm{selected}}_{\textrm{gen.}} } 
\cdot 
\frac{N^{\textrm{selected}}_{\textrm{gen.}}}{N^{\textrm{all}}_{\textrm{gen.}}} 
= \frac{N^{\textrm{selected}}_{\textrm{reco.}} }{N^{\textrm{all}}_{\textrm{gen.}} }
\end{equation}

The separation of $\epsilon$ into $A$ and $C$ allows a separation
of theoretical and experimental uncertainties, assuming the definition
of the fiducial volume at the generator level resembles to a good
extent the signal selection cuts at the reconstruction level. 
The fiducial cross section is defined as 
\begin{equation}
\label{EQN:CrossSectionFid}
\sigma_{VV}^{fid.} = \frac{N_{d} - N_{b}}{C \cdot BR \cdot
\IntLumi} = \sigma_{VV}^{incl.} \cdot A,
\end{equation}
which allows a measurement only effected to a small extent by
theoretical uncertainties. It can therefore be used to compare
measurements to theoretical predictions which might become available
at a later date.

The uncertainties associated with the detector correction parameter
$C$ are dominated by experimental sources, such as limited knowledge
of reconstruction or cut efficiencies and the accuracy of the
energy/momentum measurements. In principle, this parameter can be
larger than unity due to events outside of the fiducial region at the
generator level which may migrate to the fiducial region defined at the
reconstruction level. However, in practice this is usually not the
case, as detector inefficiency and quality criteria on reconstructed
objects have to be considered.

Equation \ref{EQN:CrossSectionFid} can be interpreted as the removal
of all experimental effects due to detector acceptance, efficiencies,
and resolutions from an experimental quantity to make it comparable to
the theoretical prediction.  Within this fiducial region, the
distributions of variables can be unfolded, e.g. the transverse
momenta of leptons in the final state. These distributions provide a
differential cross section measurement and allow a full shape
comparison with theoretical predictions.

The so-called `bin-by-bin' unfolding method can be used if the purity
of the underlying distribution is high, typically above $90\%$.  This
method is equivalent to calculating the cross section for each bin
using Eqn. \ref{EQN:CrossSectionFid}.  The purity of one bin is
defined as the ratio of events which have been reconstructed in the
same bin as they have been generated in to the number of events
generated in the chosen bin. For lower purities, advanced unfolding
methods have to be used. One commonly used approach in diboson studies
at the LHC is the Bayesian unfolding \cite{D'Agostini:1994zf} which
takes into account bin migration effects and reduces the impact of the
underlying theoretical distribution which is used as input information.


\section{\label{sec:wwvvertex}Studies of the $WW$, $WZ$ and $W\gamma$ final states}

The production of $W^+W^-$, $W^\pm Z$ and $W^\pm \gamma$ final states
can occur via $s$-channel processes in the SM and therefore provide a
test of the $WWV$ vertex (with $V=Z, \gamma$). In the following
sections we describe in detail the event selection, the background
estimation methods, and results of both experiments for the production
cross section measurements of the $W^+W^-$, $W^\pm Z$ and $W^\pm
\gamma$ final states. In particular, we will highlight the differences
in the published analyses and derive combinations of the measured
cross sections.

\subsection{\label{sec:ww}WW Analysis}

ATLAS and CMS have published analyses of the $WW$ boson production
based on the full available dataset at $\sqrt{s}=7\,\TeV$
\cite{ATLAS:2012mec, Chatrchyan:2013yaa}.  Both analysis are
based on final states in which both \Wboson bosons decay leptonically
$W^\pm\rightarrow \ell^\pm\nu$ with $\ell=e,\mu$. We discuss and
compare these two measurements in the following sections.

\subsubsection{Event Selection}

The experimental signature of the $WW$ production in the leptonic
decay channel is two high energetic leptons with no distinct invariant
mass peak, together with a significant missing transverse momentum due
to the presence of two neutrinos in the final state. Since two
different-flavor leptons appear in the final state, three decay
channels are studied, namely $ee$, $e\mu$ and $\mu\mu$. The
$\tau$-decays of the $W$ bosons are not directly taken into account 
due to the relatively low $\tau$-reconstruction efficiency and
significant fake rate. However, the cascade decay of the $W$ boson via
$W \rightarrow \tau\nu \rightarrow \ell\nu\nu\nu$ is considered as signal.

\begin{figure}[t]
    \begin{center}
        \includegraphics[width=0.19\textwidth]{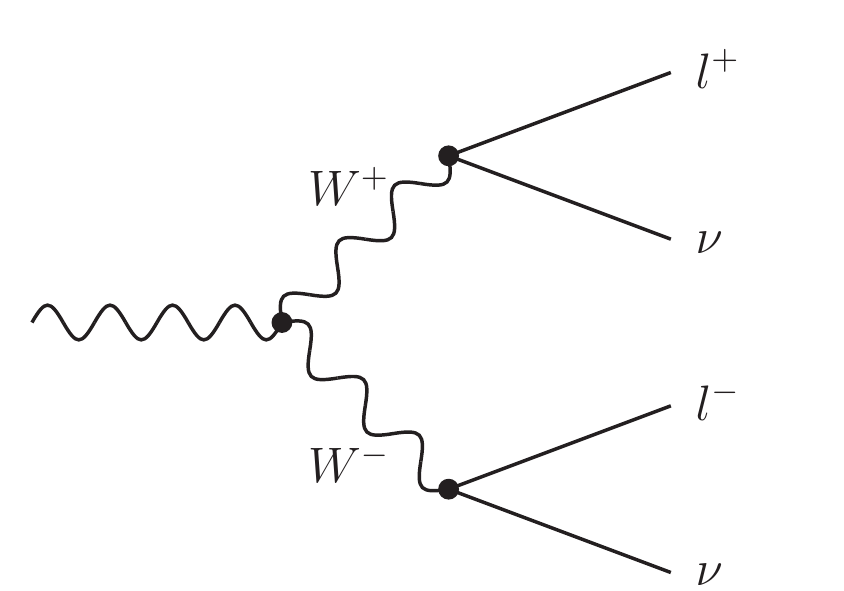}
        \includegraphics[width=0.19\textwidth]{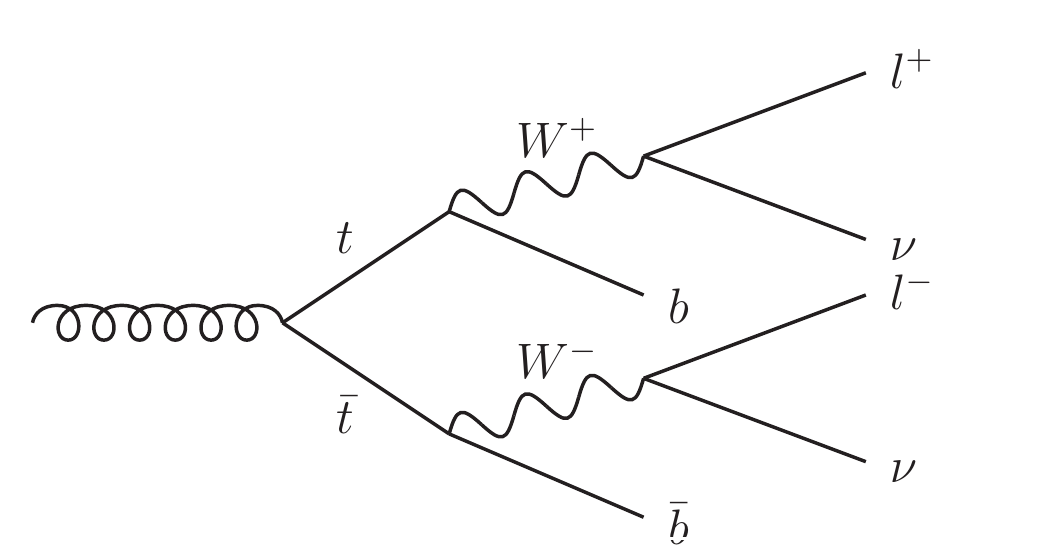}
        \includegraphics[width=0.19\textwidth]{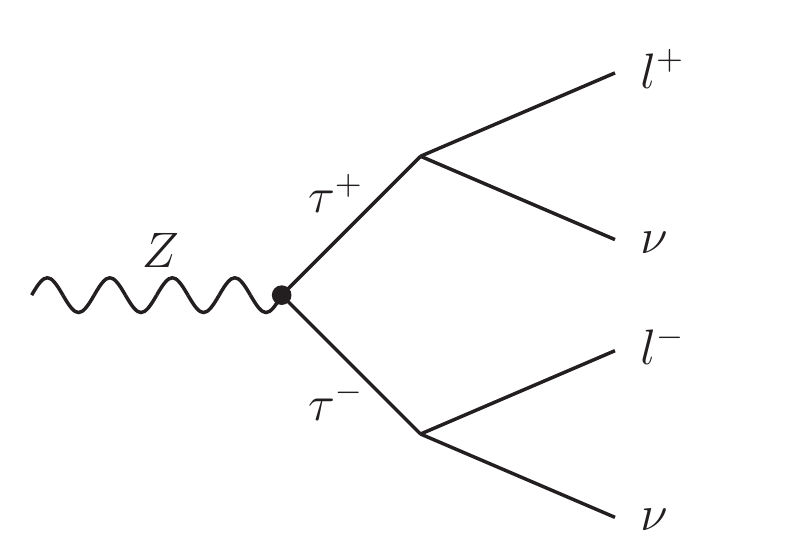}
        \includegraphics[width=0.19\textwidth]{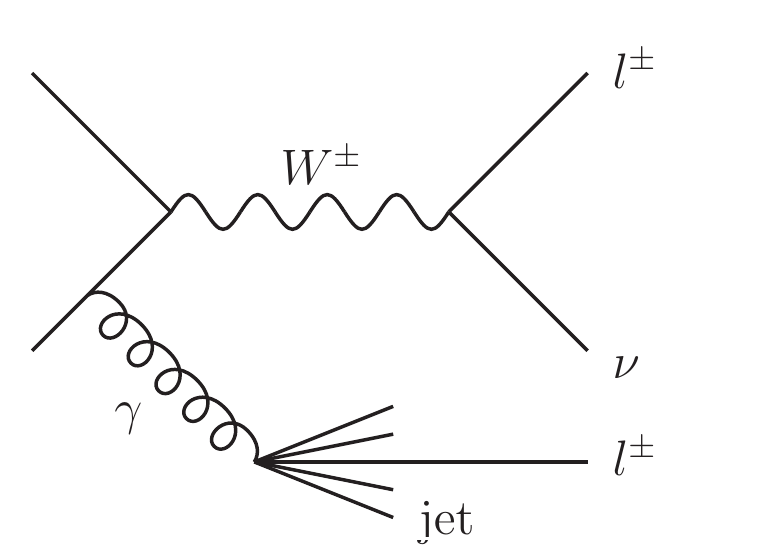}
        \includegraphics[width=0.19\textwidth]{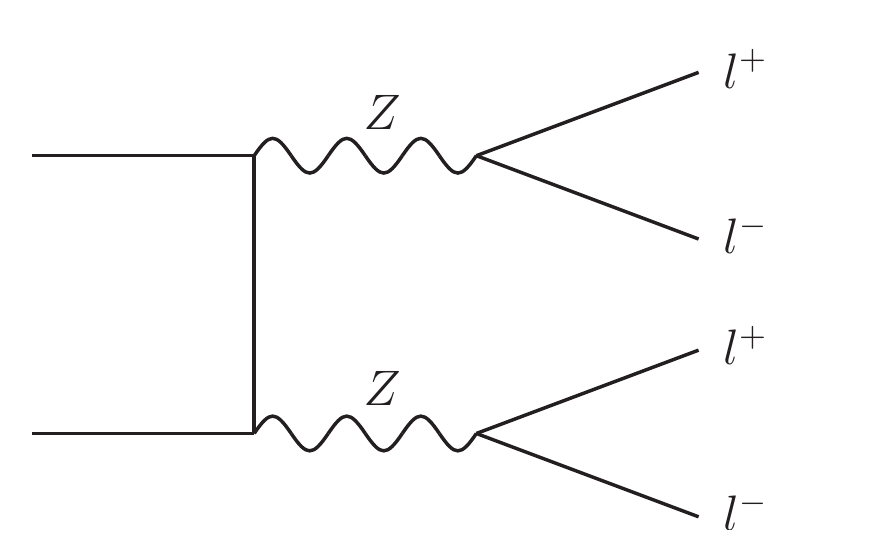}
        \caption{$WW$ production in the $s$-channel in its leptonic decay mode and background processes 
($t \bar t$, $Z\rightarrow \tau\tau$, $W+jets$ and diboson $ZZ$ production).}
        \label{Fig:WW:Processes}
    \end{center}
\end{figure}

The dominant SM background processes are the Drell-Yan process ($Z/\gamma^*\rightarrow \ell^+ \ell^-$), the
top-pair production, the production of $W$ bosons in association with
jets where one jet is incorrectly identified as a signal lepton, and
other diboson processes. The relevant Feynman diagrams are shown in
Fig. \ref{Fig:WW:Processes}.

The Drell-Yan process $Z/\gamma^*\rightarrow \ell^+ \ell^-$ has
two same-flavor opposite-charge leptons with an invariant mass
peaks at the \Zboson boson mass. A mis-measurement of the missing
transverse energies of the two leptons can lead to a signal similar to
the signal process. It should be noted that the $e\mu$ signal region
is affected by the Drell-Yan process via the $\tau$-decay channel
which could also lead to an $e\mu$ final state. The top-pair production
and its subsequent leptonic decay $t\bar t \rightarrow W^+bW^-b
\rightarrow \ell^+ b \nu \ell^- b \nu$ also leads to a signal-like
signature. These events are always produced with at least two
additional $b$-jets. $W$ production in association with jets, if the $W$
boson decays leptonically and one of the jets is falsely identified as
a signal lepton, will also fake the signal process. The leptonic
decays of other diboson processes such as $WZ$ and $ZZ$ are also
considered as background processes, in the case that one or two
leptons are not reconstructed and cause large missing transverse
energy in the event.

The full signal selections of the ATLAS and CMS measurements are
rather complex. Since the final results do not explicitly depend on
the details of the signal selection, we discuss here only the basic
concept of the signal selection and the reasoning behind it. A
detailed discussion can be found in \cite{ATLAS:2012mec} and
\cite{Chatrchyan:2013yaa}.

A schematic $WW$ signal selection requires exactly two high
energetic, oppositely-charged and isolated leptons (e.g.  $p_T>20$
GeV) are reconstructed and a minimal missing transverse momentum in
each event. Since the latter cut is intended to suppress the Drell-Yan
background, the $\METREL$ variable is used since it reduces the impact
of mis-measured leptons or jets compared to the standard
\MET~definition. The background events due to the top-pair production
can be effectively rejected by vetoing events with additional
jets. The ATLAS analysis vetoes events that contain a jet with $p_T>25\,\GeV$, while
CMS vetoes events with jets above 30 GeV. CMS imposes additional
vetoes from two top-quark tagging techniques. It should be noted that
the jet-veto criteria also rejects a significant fraction of $WW$ events which
are predicted by NLO QCD corrections. A summary of the detailed
selection cuts is given in Tab. \ref{TAB:WW:Selection}.

%

\begin{table}[h]
\tbl{Summary of $WW$ selection requirements for the ATLAS and CMS
  experiments on the lepton transverse momentum $p_T$, the invariant
  mass ($m_{\ell\ell}$) and transverse momentum ($p_T(\ell\ell)$) of
  the dilepton system, and the relative transverse energy $\METREL$
  ($N_{vtx}$ denotes the number of reconstructed vertices in the
  event). For ATLAS, these are also the basis for the definition of
  the fiducial phase-space region.}  {
  \begin{tabular}{l l l} 
\hline
\hline
							&	ATLAS  									& CMS 										\\
\hline
$WW\rightarrow \ell\nu \ell'\nu$		&	2 leptons with $p_T>20\,\GeV$					&	2 leptons with $p_T>20\,\GeV$					\\
$(\ell,\ell'=e, \mu)$					&	$|\eta^\mu|<2.5$, $|\eta^e|<2.47$			&	$|\eta^\mu|<2.5$, $|\eta^e|<2.47$				\\
							&	Dilepton $p_T(\ell\ell)>30$ GeV		                   &	Dilepton $p_T(\ell\ell)>45$ GeV	\\
							&	No jet with $p_T>25$ GeV and 				& 	No jet with $p_T>30$ GeV and			\\
							&	rapidity $Y<4.5$							& 	rapidity $Y<4.5$			\\
\hline
$WW\rightarrow \ell\nu\ell\nu$			&	dilepton mass $m_{\ell\ell}>15 \,\GeV$	 			&	di-electron mass $m_{ee}>15 \,\GeV$			\\
							&											&	di-muon mass $m_{\mu\mu}>12 \,\GeV$			\\
							&	$Z$ boson veto $|m_{\ell\ell}-m_Z|>15\,\GeV$			&	$Z$ boson veto $|m_{\ell\ell}-m_Z|>15\,\GeV$			\\
							&	$\METREL>45\,\GeV$						&	$\ETMiss>(37+0.5\cdot N_{vtx})\,\GeV$					\\
\hline
$WW\rightarrow e\nu \mu\nu$		&	dilepton invariant mass $m_{e\mu}>10 \,\GeV$ 		&	dilepton invariant mass $m_{e\mu}>12 \,\GeV$		\\
							&	$\METREL>25\,\GeV$						&	$\ETMiss > 20\,\GeV$						\\
\hline
\hline
   \end{tabular}
   \label{TAB:WW:Selection}
}
\end{table}


The resulting signal and background predictions compared to the yield
in data, where the SM prediction of the $W^+W^-$ cross-section is
used, are shown in Tab. \ref{TAB:WW:SigBack}. ATLAS expects $12\%$,
$23\%$ and $65\%$ of signal events in the $ee$, $\mu\mu$ and $e\mu$
channels, respectively. CMS has not published number for each
individual channel in \cite{Chatrchyan:2013yaa}. However, previous
studies based on a smaller integrated luminosity \cite{CMS:2011dqa},
suggest very similar numbers. The final results are dominated by the
contribution from the $e\mu$ channel, as the corresponding selection
cuts are  relaxed due to the reduced background from the Drell-Yan
process.

\begin{table}[h]
\tbl{Event yields for ATLAS and CMS for different decay
  channels. Signal and background estimates are also given.}  {
  \begin{tabular}{l |cccc|c} 
\hline
\hline
Experiment		& \multicolumn{4}{c|}{ATLAS}  & CMS \\
  \hline
  Channel			&	$ee$				&	$\mu\mu$				&		$e\mu$		&	Combined		& Combined	\\
  \hline
  Data			&	174				&	330					&		821			&	1325				& 1134		\\
\hline
  $WW$ (MC)		&	$100\pm9$	&	$186\pm15$		&	$538\pm45$	&	$824\pm4\pm69$	& $751\pm4\pm55$\\
  \hline
  Top 			&	$22\pm12$	&	$32\pm15$		&	$87\pm26$	&	$141\pm30\pm22$	& $129\pm13\pm20$\\
  $W+$jets 			&	$21\pm11$	&	$7\pm3$			&	$70\pm31$	&	$98\pm2\pm43$	& $60\pm4\pm21$\\
  Drell-Yan 		&	$12\pm4$		&	$34\pm12$		&	$5\pm2$		&	$51\pm7\pm12$	& $11\pm5\pm3$\\
  Other DiBosons 	&	$13\pm2$		&	$21\pm2$			&	$44\pm6$		&	$78\pm2\pm10$	& $48\pm3\pm5$\\
  \hline
  Total Background 	&	$68\pm18$	&	$94\pm13$		&	$206\pm42$	&	$369\pm31\pm53$	& $247\pm15\pm30$\\
  \hline
  Total Expected 	&	$169\pm20$	&	$280\pm25$		&	$744\pm61$	&	$1192\pm31\pm87$	& $1044\pm15\pm62$\\
\hline
\hline
   \end{tabular}
   \label{TAB:WW:SigBack}
}
\end{table}

\subsubsection{Background Estimation}

The expected background contributions are summarized in
Tab. \ref{TAB:WW:SigBack}. The dominant background contributions come
from top-pair and $W+$jets events, which will be discussed in more
detail in this section.

The background contribution due to top-quark pairs is estimated in
both analysis with a data-driven method. ATLAS defines a data sample,
named extended signal region (ESR), including all events which pass
the signal selection cuts but without the jet-veto requirement.  This
sample is dominated by events from $t\bar{t}$ and single top processes
with more than one jet in each event.  A control region is defined by
applying the full signal selection criteria except requiring at least
one $b$-tagged jet with $p_T>20\,\GeV$.
The expected jet multiplicity distribution in the ESR is estimated
from the measured control region distribution, and extrapolated to the
ESR using MC predictions. The expected jet-multiplicity distribution
in the ESR is fitted to the measured ESR in the higher
jet-multiplicity region and the nominal value in the 0-jets bin taken
as background estimate. The dominating uncertainty of this approach is
due to the limited statistics in the control region.

CMS also defines a control region dominated by top-quark background
events by requiring that a positive top-quark identification algorithm
tags the given event.  The normalization of the top-quark background
is estimated via $N^{top} (\mbox{not-tagged}) = N^{top}
(\mbox{tagged}) \times (1-\epsilon_{tt})/{\epsilon_{tt}}$, where
$\epsilon_{tt}$ is the efficiency to tag a $t\bar{t}$ event. This
efficiency is estimated in a data sample selected by the nominal
signal selection criteria but requiring one jet with
$p_T>30\,\GeV$. The dominant uncertainty in the estimation of
$N^{top} (\mbox{not-tagged})$ is due to statistical and
systematic uncertainties on $\epsilon_{tt}$.

Since the probability for a jet to be misidentified as an isolated
lepton might not be modeled correctly in the MC simulations, a similar
data-driven method has been used in both experiments for the $W+$jets
background estimation.  A $W+$jets enriched sample is selected by
loosening the isolation or identification requirements on one
lepton. The number of $W+$jet events in the signal region is then
estimated via fake-factor $f$, which is defined as the ratio of the probability of
a jet passing the nominal lepton selection over the probability of
passing the loosened selection.  The factor $f$ is determined in data
for muons and electrons, separately, using QCD multijet events.

The Drell-Yan background in both the $ee$ and $\mu\mu$ channels
is estimated by inverting the $Z$ boson veto cuts and then
extrapolating from this control region into the signal region. The
remaining background contributions are estimated with MC simulations,
where all theoretical and experimental uncertainties have been taken
into account.

\begin{figure*}
\resizebox{0.49\textwidth}{!}{\includegraphics{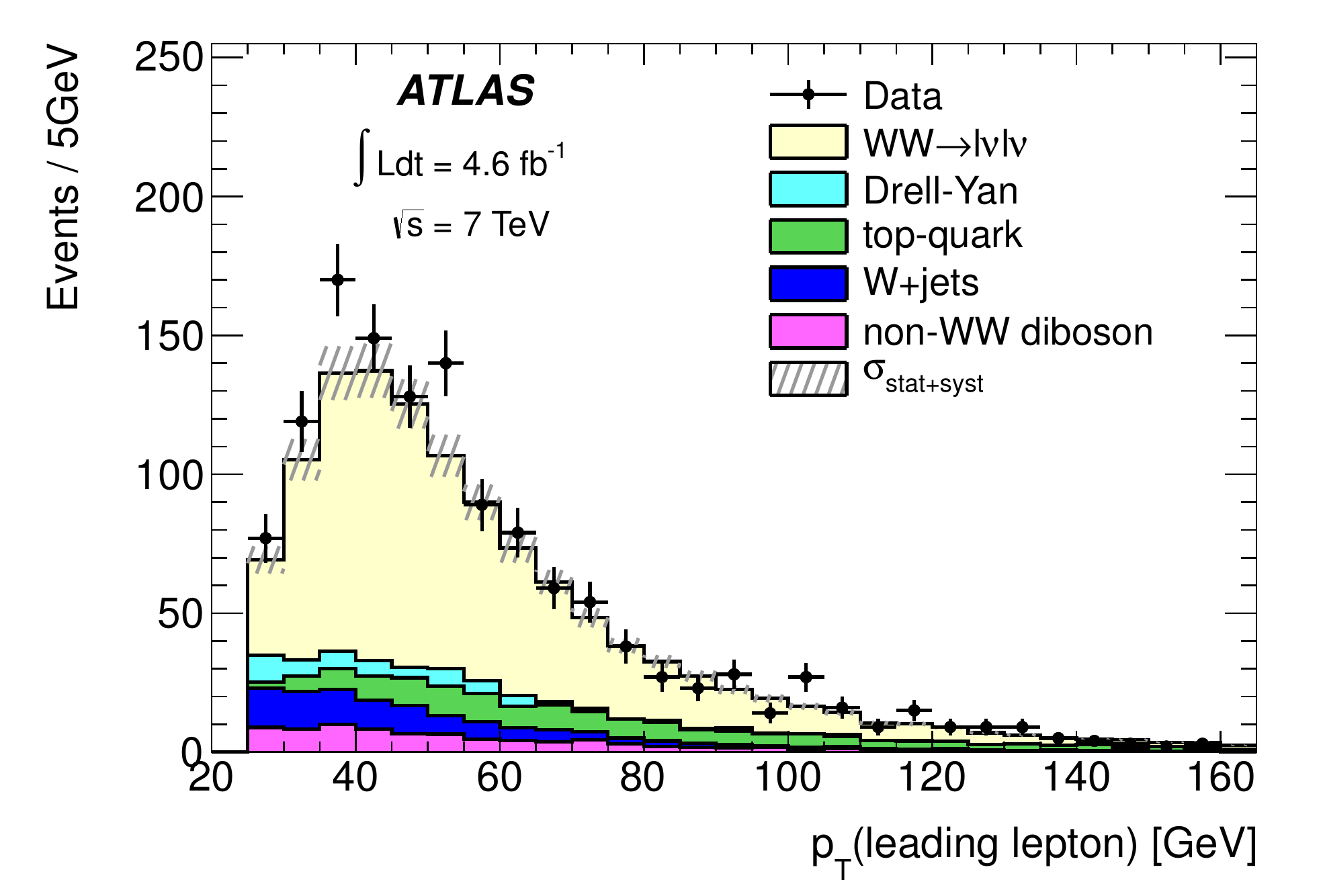}}
\resizebox{0.49\textwidth}{!}{\includegraphics{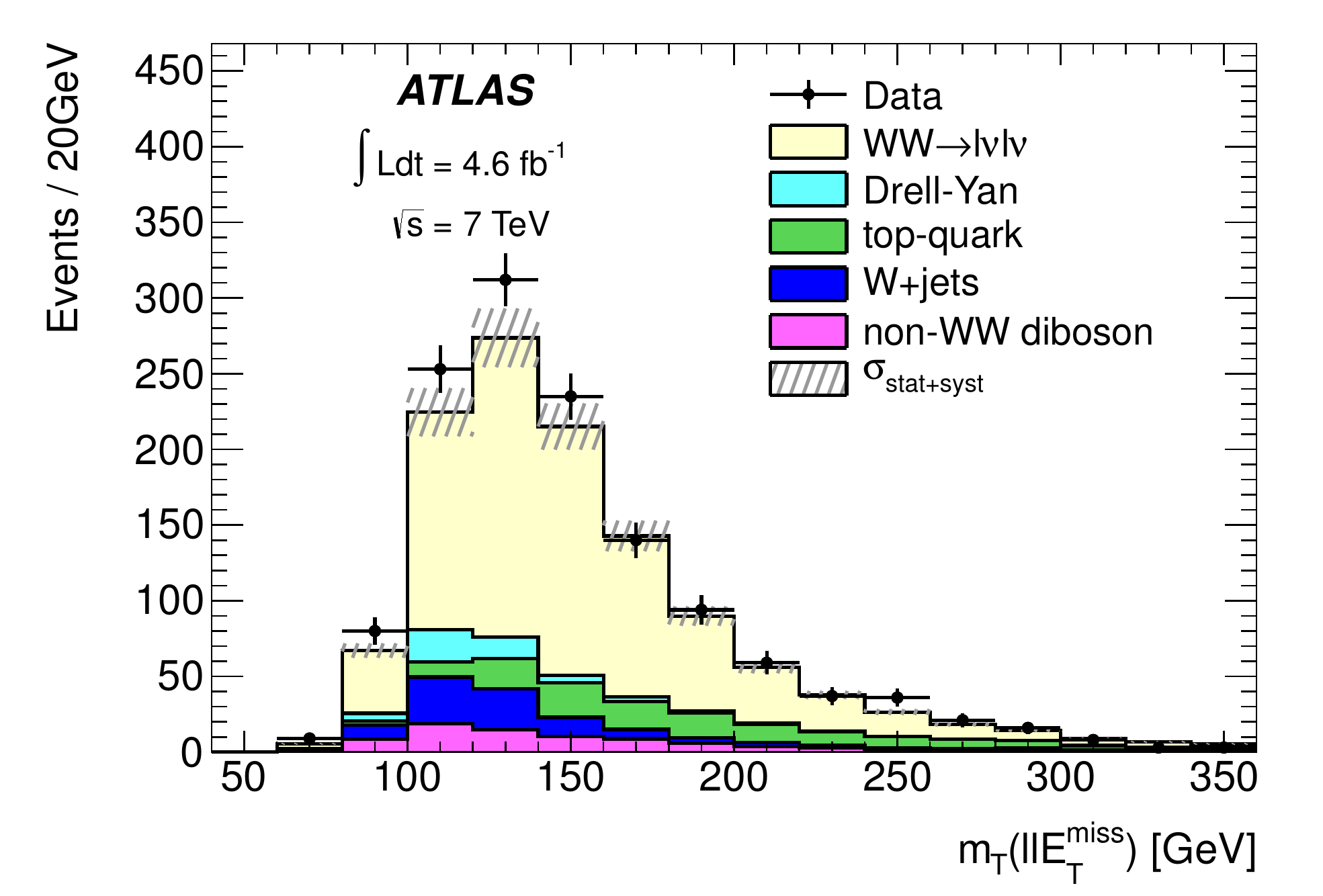}}
\caption{ATLAS: Leading lepton $p_T$ (left) and
  $m_T$ (right) distributions of the dilepton \ETMiss\ system for $WW$ candidates with all selection
  criteria applied and combining $ee$, $\mu\mu$ and $e\mu$ channels. The points
  represent data. The statistical and systematic uncertainties are
  shown as grey bands. The stacked histograms are from MC predictions
  except the background contributions from the Drell-Yan, top-quark
  and $W+$jets processes, which are obtained from data-driven
  methods. The prediction of the SM $WW$ contribution is normalized to
  the inclusive theoretical cross section of 44.7 pb. }
\label{fig:ALTASWWDis}
\end{figure*}

\begin{figure*}
\resizebox{0.49\textwidth}{!}{\includegraphics{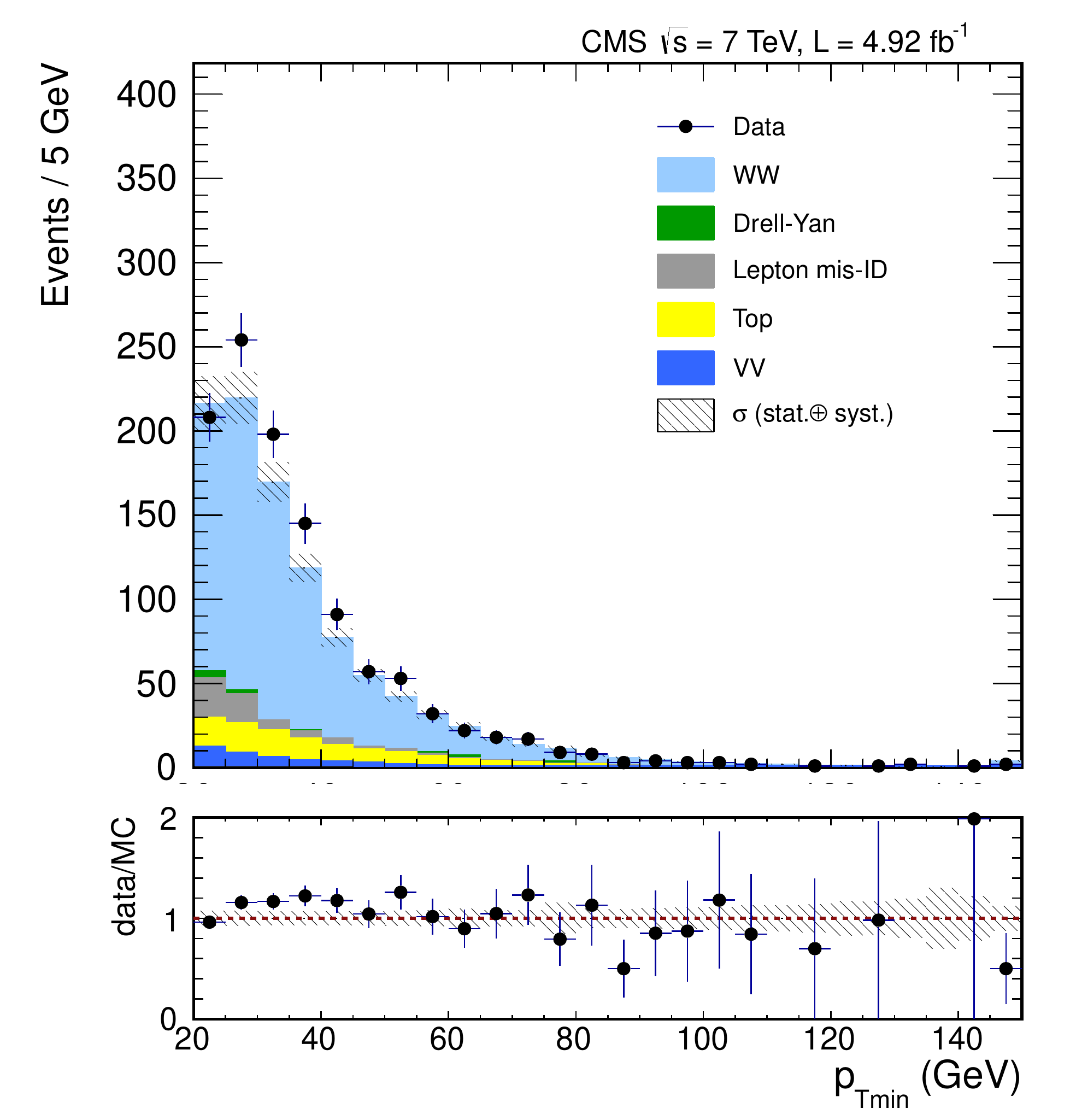}}
\resizebox{0.49\textwidth}{!}{\includegraphics{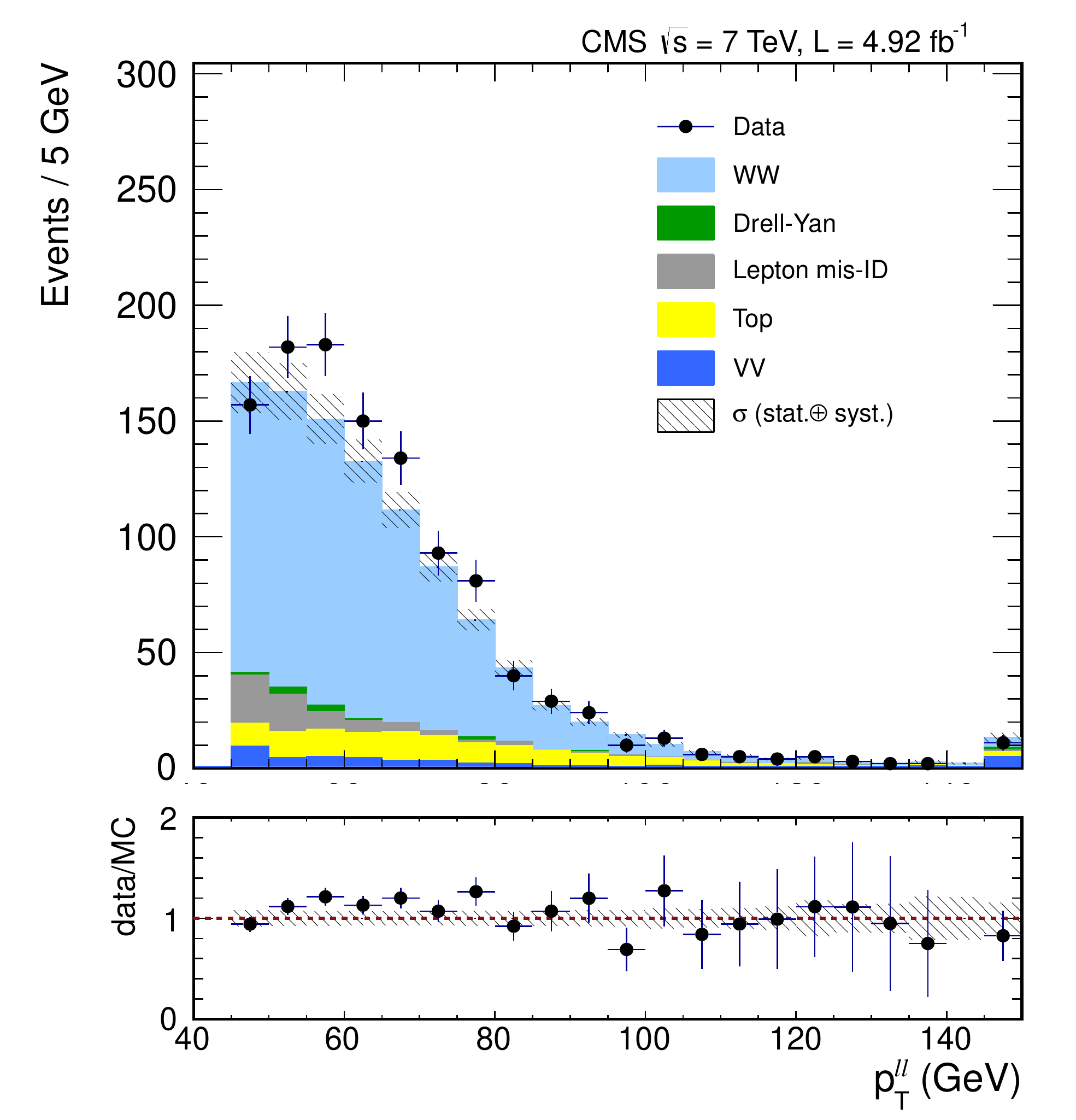}}
\caption{CMS: $p_T$ of the leading lepton (left) and the
  invariant mass distribution of the dilepton system (right) at the $WW$
  selection level, reweighted to the data-driven estimates. All three
  channels ($ee$, $\mu\mu$ and $e\mu$) are combined, and the
  uncertainty band corresponds to the statistical and systematic
  uncertainties on the predicted yield.}
\label{fig:CMSWWDis}
\end{figure*}

The resulting distributions for the selected $WW$ candidate events are
shown in Figs. \ref{fig:ALTASWWDis} and \ref{fig:CMSWWDis} for ATLAS
and CMS, respectively. The expected $WW$ signal contribution is based
on MC expectations with \MCAtNLO\ used by ATLAS and \MadGraph\ used by
CMS.

\subsubsection{Cross Section Measurement}

The ATLAS cross section measurement is performed in each of the three
decay channels and then combined, while CMS does not distinguish
between the final states and directly derives a combined cross
section.  Furthermore, ATLAS defines a fiducial volume and separates
the signal selection efficiency in acceptance and detector effects. The
corresponding parameters $\epsilon_{WW}$, $A_{WW}$ and $C_{WW}$ are
shown in Tab. \ref{TAB:WWSelectionEff}.

\begin{table}[h]
\tbl{Selection efficiencies and acceptance factors for ATLAS and CMS
  including their respective uncertainties.}  {
  \begin{tabular}{l | c c c | c} 
\hline
\hline
Experiment				& \multicolumn{3}{c|}{ATLAS}  & CMS \\
\hline
Channel					& $ee$     					& $\mu\mu$    			& $e\mu$    				& combined \\
\hline
$A_{WW}$    				& $(7.5\pm0.4)\%$			& $(8.1\pm0.5)\%$ 		& $(15.9\pm0.9)\%$			& - 		\\
$C_{WW}$    				& $(40.3\pm1.8)\%$		& $(68.7\pm2.1)\%$ 		& $(50.5\pm1.6)\%$			& - 		\\
\hline
$A_{WW}\times C_{WW}$    	          & $(3.0\pm0.1)\%$			& $(5.6\pm0.2)\%$		& $(8.0\pm0.3)\%$ 			& $(3.28\pm0.26)\%$ \\
\hline
Stat. Unc.    				& 1.2\%     				& 0.7\%    				& 0.4\% 					& 0.6\%        \\
Detector.Unc.    			& 3.1\%     				& 1.3\%    				& 1.6\% 					& 4.9\%        \\
Theo.Unc.    				& 1.0\%     				& 1.0\%    				& 1.0\% 					& 2.7\%        \\
Jet.Veto Unc.    				& 3.7\%     				& 3.7\%    				& 3.7\% 					& 4.7\%        \\
\hline
Combined Unc.    			& 4.9\%     				& 4.0\%    				& 4.1\% 					& 7.3\%        \\
\hline
\hline
\end{tabular}
\label{TAB:WWSelectionEff} 
}
\end{table}

The efficiency factor $\epsilon$ of the CMS analysis averages over all
lepton flavors and is defined with respect to a phase space that
includes all possible leptonic decay modes.
The correction factor $\mbox{BR}$ due to the branching ratio in
Eqn. \ref{EQN:CrossSectionExp} is therefore given by BR$=[3\cdot
  \mbox{BR}(W\rightarrow \ell\nu)]^2$. The efficiency correction
factor $\epsilon$ in the ATLAS analysis is defined for each decay
channel with respect to a phase space that includes only the
respective final states. The contributions from the cascade decay
$WW\rightarrow \tau \nu \ell\nu \rightarrow \ell' \nu \nu \nu \ell
\nu$ and $WW\rightarrow \tau \nu \tau \nu \rightarrow \ell \nu \nu \nu
\ell' \nu \nu \nu$ are also included.

Experimental uncertainties in Tab. \ref{TAB:WWSelectionEff} are
dominated by lepton reconstruction efficiencies and energy/momentum
scale uncertainties.  Theoretical uncertainties are significantly
different in both analyses even though the signal selection
requirements are similar. Theoretical uncertainties contain
contributions from the uncertainties on strong coupling constant ($\alpha_s$), renormalization
($\mu_r$) and factorization ($\mu_f$) scales, and PDFs.  The two
scales are varied in the range of $\mu_0/2$ and $2\mu_0$ ($\mu_0=\mu_r=\mu_f$) to estimate
the uncertainty. ATLAS chose to use $\mu_0=m_{WW}$ while CMS chose to
use $\mu_0=m_W$.  A second and more significant difference comes from
the fact that CMS calculates the above theoretical uncertainties with
the jet-veto scale factor applied, while ATLAS estimates the
theoretical uncertainties before applying the jet-veto requirement.

Hence the estimation of the jet-veto scale factor needs to be
discussed in more detail.  Both analyses use a data-driven approach to
estimate the probability ($P^{data}_{WW}$) for a $WW$ signal event
failing the jet-veto requirement in data. This probability is
calculated as
\begin{equation}
\label{EQN:JetVeto}
P^{data}_{WW} = \frac{P^{MC}_{WW}}{P^{MC}_{Z/\gamma^*}} \times P^{Data}_{Z/\gamma^*}
\end{equation}
where $P_{Z/\gamma^*}$ denotes the probability of $Z/\gamma^*$ boson
events to pass a jet-veto requirement. Events containing a $Z/\gamma^*$
boson can be selected with a high purity in data and the kinematic
distributions of jets are expected to be similar to those in $WW$
events.  Most uncertainties on the jet-veto requirement cancel in the
ratio ($P^{MC}_{WW}/P^{MC}_{Z/\gamma^*}$) and therefore a reduction on
the uncertainty of $P^{data}_{WW}$ is achieved.  The cancellation in
the ratio is also the reason why ATLAS chose to estimate the PDF and
scale uncertainties on $\epsilon$ before applying the jet-veto requirement.
The overall uncertainty on the jet-veto requirement is significantly
lower in ATLAS compared to that in the CMS analysis. Another possible
contribution is that ATLAS estimates the effect of higher order
corrections using \MCAtNLO, while CMS uses MCFM. Since \MCAtNLO
includes parton shower effects in contrast to MCFM, a better
description of the jets is expected which could lead to a smaller
effects from higher order corrections.

It is worthwhile noting that the uncertainties of the jet-veto
probability on $A_{WW}$ are approximately $5.6\%$ and therefore
significantly larger than its impact on $\epsilon_{WW}$.  This is
mainly due to the fact that the method of Eqn. \ref{EQN:JetVeto}
cannot be applied directly at the generator level. A naive estimate of
the scale and PDF uncertainties on the jet-veto requirement leads to
an underestimate of the corresponding uncertainty and therefore more
sophisticated methods have to be applied \cite{Stewart:2011cf}.

\begin{table}[h]
\tbl{Summary of measured and predicted inclusive cross sections $\sigma WW$ for
  the $WW$ process from the ATLAS and CMS collaborations.}  {
\begin{tabular}{lll}
\hline
\hline
Channel 									& Measured [pb]				& Predicted [pb] 						\\
\hline
ATLAS~~~~~~~~~~~~~~~~~~~~~~~~~~~~~~~~~~	&  							& 									\\
$WW\rightarrow e\nu e\nu$				& $46.9~\pm~5.7 \,(stat.)~\pm ~8.2\,(sys.)~\pm ~1.8 \,(lumi.)$	&	$45.3 \pm 2.0$	\\
$WW\rightarrow \mu\nu \mu\nu$			& $56.7~\pm~4.5 \,(stat.)~\pm ~5.5\,(sys.)~\pm ~2.2 \,(lumi.)$	&	$45.3 \pm 2.0$	\\
$WW\rightarrow e\nu \mu \nu$			& $51.1~\pm~2.4 \,(stat.)~\pm ~4.2\,(sys.)~\pm ~2.0 \,(lumi.)$		&	$45.3 \pm 2.0$	\\
$WW\rightarrow \ell\nu \ell'\nu$			& $51.9~\pm~2.0 \,(stat.)~\pm ~3.9\,(sys.)~\pm ~2.0 \,(lumi.)$		&	$45.3 \pm 2.0$	\\
\hline
CMS 									& 							& 									\\
$WW\rightarrow \ell\nu \ell'\nu$			& $52.4~\pm~2.0 \,(stat.)~\pm ~4.5\,(sys.)~\pm ~1.2 \,(lumi.)$		&	$45.3 \pm 2.0$	\\
\hline
\hline
\end{tabular}
\label{TAB:WWResults}
}  
\end{table}

The cross section results from both experiments are summarized in
Tab. \ref{TAB:WWResults}. The cross section results from three
individual channels are combined by minimizing a negative
log-likelihood function, which depends on the expected number of
signal events in each decay channel.

The measured inclusive cross sections agree within their uncertainties
for both experiments.  The overall uncertainty is dominated by the
systematic uncertainty, which is $7.5\%$ for ATLAS and $8.6\%$ for
CMS.  The dominant uncertainty sources come from the top-quark
background estimation ($3.6\%$ at ATLAS and $2.9\%$ at CMS) and the
jet-veto selection efficiency ($3.6\%$ at ATLAS and $4.7\%$ at CMS).

CMS also published a measurement of the cross section ratio of
$\sigma_{WW}$ to $\sigma_Z$, where $\sigma_Z$ is the $Z$ boson
inclusive production cross section measured in the $ee$ and $\mu\mu$
channels.  The ratio measurement helps to reduce or cancel
experimental uncertainties on the lepton identification efficiency and
the integrated luminosity, as well as theoretical uncertainties on
scales and PDFs. The latter is mainly due to the fact that the
dominant production mechanism for both $Z$-boson and $WW$-pair is
$q\bar q$ annihilation. The measured cross section ratio is found to
be
\[ \sigma_{WW} / \sigma_Z = (1.79 \pm 0.16) \times 10^{-3}. \]
ATLAS published in addition a fiducial cross section in each decay
channel.  The fiducial volume of each decay channel reassembles the
corresponding selection cuts used and hence leads to a rather complex
definition and is also different for each decay channel. Hence, only
individual results are available and no combination has been
performed.

\begin{figure*}
\begin{minipage}{0.49\textwidth}
\resizebox{1.0\textwidth}{!}{\includegraphics{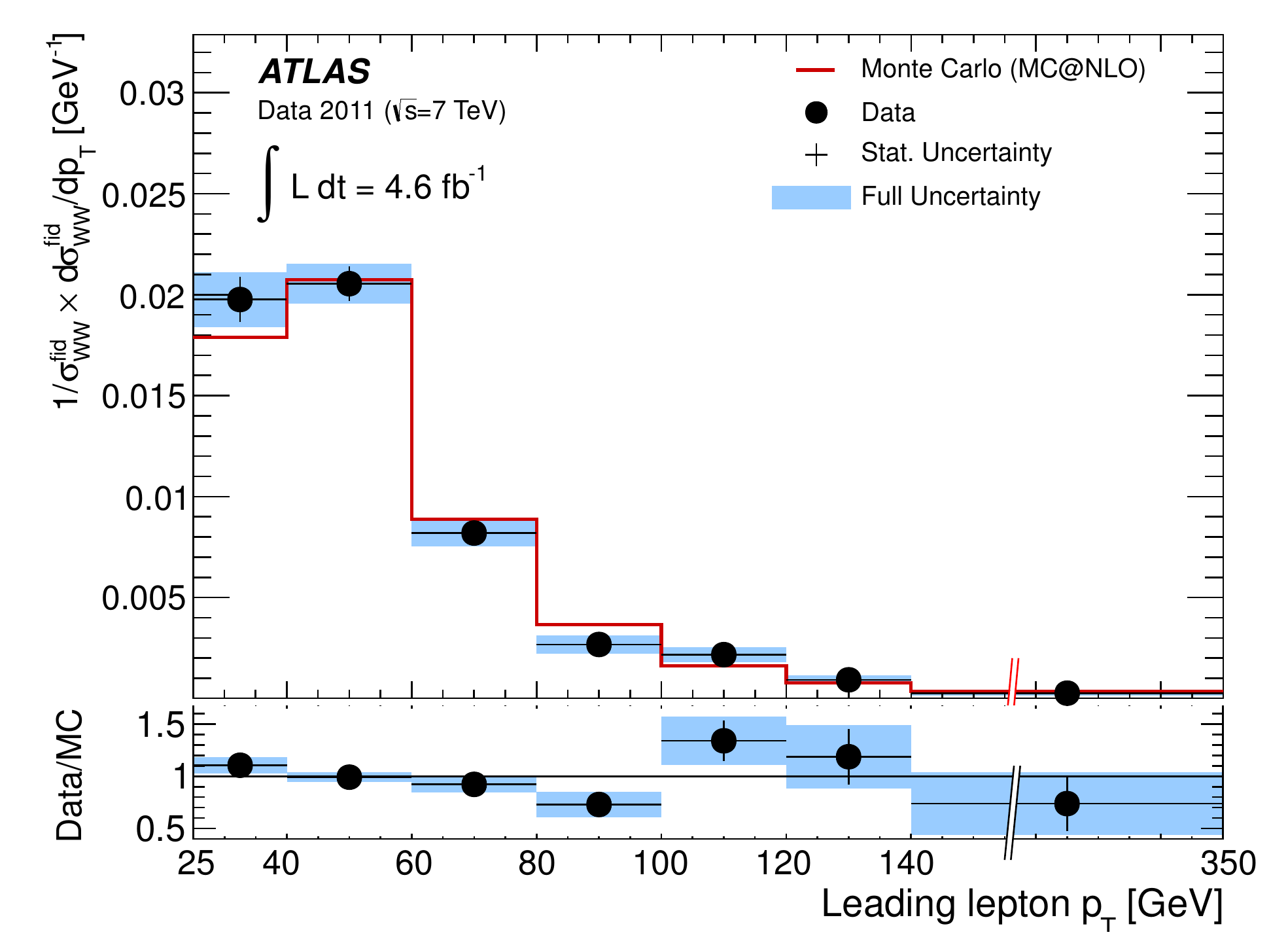}}
\caption{\label{Fig:ATLASWWDiff}ATLAS: The normalized differential
  $WW$ fiducial cross section as a function of the leading lepton pT
  compared to the SM prediction.}
\end{minipage}
\begin{minipage}{0.49\textwidth}
\resizebox{1.0\textwidth}{!}{\includegraphics{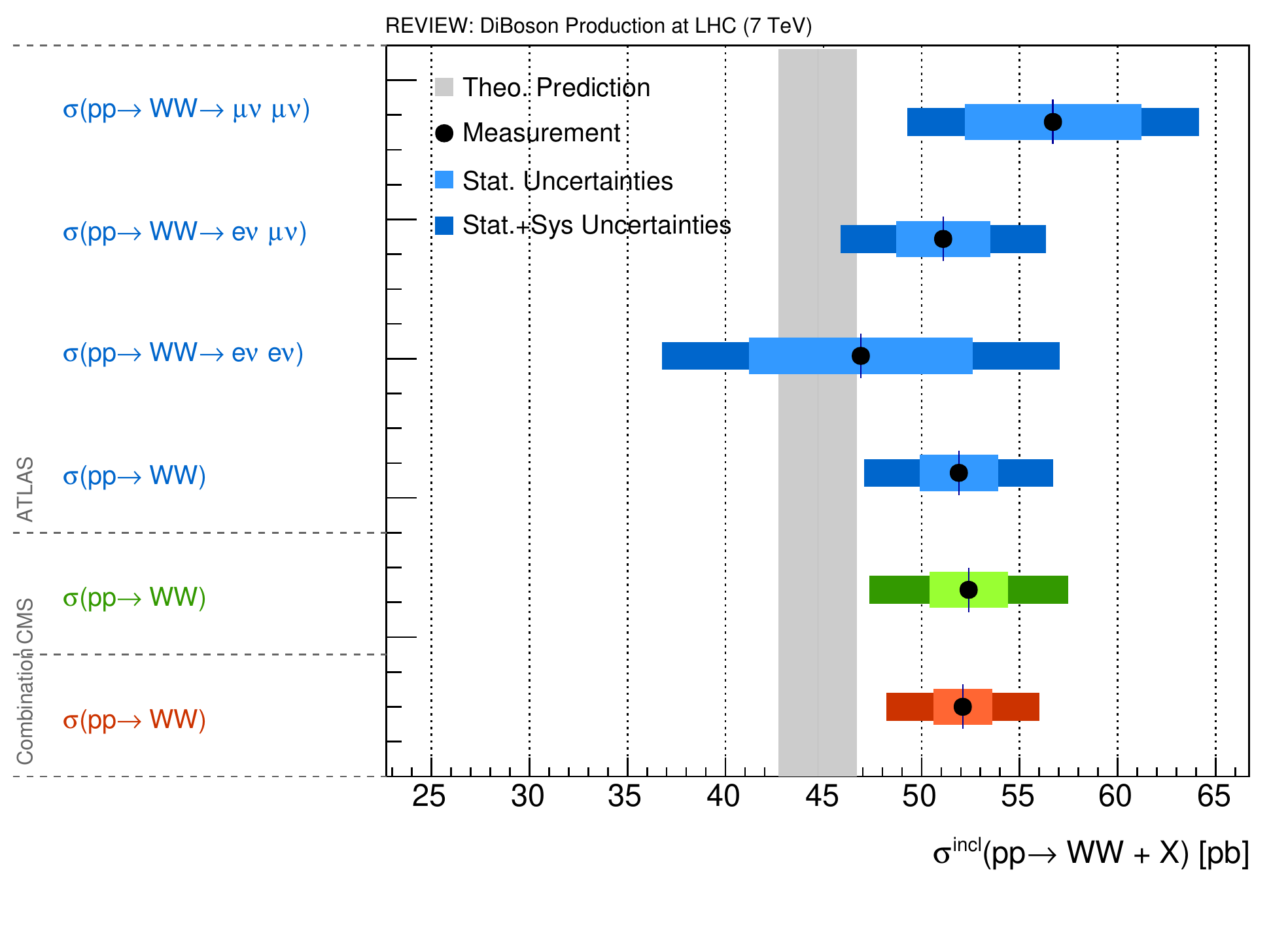}}
\caption{\label{Fig:SummaryWW}Summary of measured inclusive $WW$
  production cross-sections at $\sqrt{s}=7\,TeV$ at the ATLAS and CMS
  Experiments.}\vspace{0.3cm}
\end{minipage}
\end{figure*}

Within the defined fiducial phase space, ATLAS also provides a
normalized unfolded distribution of the leading lepton $p_T$ in the
final state with all three channels combined. The unfolding procedure
follows the iterative Bayesian method introduced in
Sect. \ref{sec:Methodology}.  The resulting distribution is shown in
Fig. \ref{Fig:ATLASWWDiff} and is compared to the \MCAtNLO prediction.

\subsubsection{\label{sec:WWDerived}Derived Results and Discussion}

We first combine the measurements of the inclusive $WW$ cross section
for ATLAS and CMS at 7 \TeV. For this, we assume that the
uncertainties on the integrated luminosity are fully correlated among
two experiments.  Theoretical uncertainties on the fiducial acceptance
are also assumed to be fully correlated. Even though there are
differences in the signal selection criteria, the main sources of
theoretical uncertainties are similar and the methods for estimating
their uncertainties are the same. In particular this holds to a large
extent also for uncertainties on the jet-veto requirement, which is a
large source of uncertainty. The remaining experimental uncertainties
on the selection efficiency, signal acceptance, and background
estimations are treated fully uncorrelated.

The combination procedure follows the BLUE method \cite{Lyons:1988rp},
leading to a combined measurement of the $WW$ production cross section
\[ \sigma_{WW} = 52.1 \pm 1.5 (\mbox{stat.}) \pm 3.6 (\mbox{syst.})\,\pb.\]
When no correlations are assumed between theoretical and jet-veto
uncertainties, the mean value stays unchanged, while the systematic
uncertainty reduces to 3.4 pb.  The NLO QCD calculation of the cross
section is $45.3\pm2.0$ pb (as shown in Tab. \ref{tab:CrossSections})
and compatible at the $1.6\sigma$-level with the measured cross
section.  This deviation cannot be caused by a SM Higgs boson with a
mass around 125 GeV decaying into $W^+W^-$, as already discussed.  A
summary of the measured inclusive cross sections and their combination
is shown in Fig. \ref{Fig:SummaryWW}.

The definition of the fiducial phase space in the ATLAS analysis is
highly complex. 
Hence we provide here an extrapolation to a fiducial phase space
region, which is significantly simpler but keeps the
theoretical uncertainties due to the required extrapolations to a
minimum. Our simplified fiducial phase space is defined only in the
prompt $e$ and $\mu$ channels (events with $e$ or $\mu$ from the
$\tau$ decays are removed) via the following criteria: two leptons
with $p_T>20$ GeV and $|\eta|<2.5$, $p_T(\nu\nu)>25$ GeV,
$p_T(\ell\ell)>30$ GeV and $m_{\ell\ell}>20$ GeV. The extrapolation
factor from the ATLAS fiducial phase space to the simplified phase
space is evaluated using \MCAtNLO and is found to be $2.06 \pm
0.18$. Uncertainties due to scales and PDFs are estimated along with the
above prescription. The largest uncertainty is due to the unused
jet-veto criteria. This leads to a simplified fiducial cross section
of $\sigma ^{fid,simp}_{WW} = 0.81 \pm 0.03 (\mbox{stat.}) \pm 0.09 (\mbox{syst.})$ pb.

Within this simplified fiducial phase space, one can also calculate
the production cross section for $WW$ events with the leading lepton
$p_T$ between $140\,\GeV$ and $180\,\GeV$ based on the unfolded
results of the ATLAS experiment. The cross section we obtained is
$\sigma ^{fid,sim}_{WW}(140~\mbox{GeV}<p_T^{leading\,lepton}<350~\mbox{GeV}) = 24\pm10$ fb.  This fiducial cross
section is sensitive to the $s$-channel production of the $WW$ process
and can therefore be used to make constraints on aTGC parameters
without any further experimental information.


\subsection{\label{sec:wz}WZ Analysis}

ATLAS has published results on the $WZ$ production using the full
$\sqrt{s}=7\,\TeV$ dataset \cite{Aad:2012twa}, while only preliminary
CMS results are available at this point \cite{CMS:2011wz}. Similar to
the $WW$ analysis, only final states involving muons and/or electrons
are considered.

\subsubsection{Event Selection}

The process $W^\pm Z\rightarrow\ell^\pm \nu \ell'^+\ell'^-$ with
$\ell=e,\mu$ is characterized by three highly energetic and isolated
leptons of which at least two have the same flavor and opposite charge
with an invariant mass close to the \Zboson boson
pole mass. In addition a significant amount of $\MET$ is expected due
to the escaping neutrino from the \Wboson boson decay. The possible
final states are therefore $e^+e^-e^\pm\nu$, $e^+e^-\mu^\pm\nu$,
$\mu^+\mu^-e^\pm\nu$ and $\mu^+\mu^-\mu^\pm\nu$.  The contribution of
secondary $\tau\rightarrow \ell \nu \nu$ decays of one or both bosons
in $WZ$ events is also regarded as signal.

\begin{figure}[bt]
    \begin{center}
        \includegraphics[width=1.0  \textwidth]{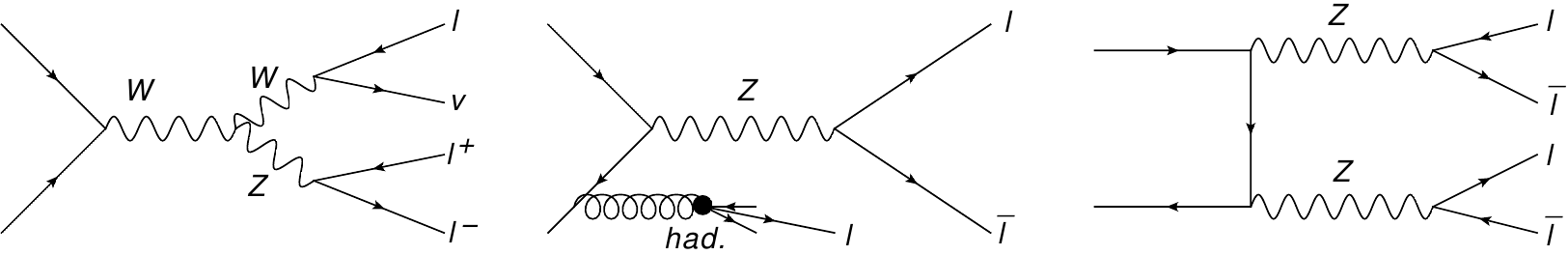}
        \caption{Feynman diagrams for $s$-channel signal processes
          (left) and primary background processes (middle: $Z$+jets,
          right: $ZZ$) relevant for the $WZ$ diboson analysis}
        \label{Fig:WZLOFeynman}
    \end{center}
\end{figure}

SM background processes which lead to similar signatures in the
detector are $Z+$jets, $ZZ$, $Z\gamma$, $t\bar t$ and $t\bar t + W/Z$.
The dominate background comes from $Z$ boson production in association
with at least one jet which is incorrectly identified as a signal
lepton.
For $t\bar t$ events, two real leptons are expected in the fully
leptonic decay channel and an additional third reconstructed lepton
can come from a mis-identified jet.  It should be noted that the
probability for a jet to be identified as a signal lepton in $t\bar t$
events is significantly higher than the probability in $Z+$jets
events, since at least two $b$-jets are in the final state that can
decay into real muons. The $t\bar t$ production in association with a
vector boson, i.e.  $t\bar t + W/Z$, can lead to three real leptons in
the final state if the associated vector boson also decays
leptonically.  The full leptonic decay of the $ZZ$ pair can mimic the
signal if one lepton is not reconstructed or is outside the detector
acceptance, resulting in missing transverse energy. The $Z\gamma$
process can fake a $WZ$ signature only in the $eee$ and $\mu\mu e$
channels as the photon can be falsely reconstructed as an isolated
electron. LO Feynman diagrams for the signal and background
processes are shown in Fig. \ref{Fig:WZLOFeynman}.

The generic signal selection of ATLAS and CMS requires two isolated
leptons whose invariant mass is close to the \Zboson boson mass. This
requirement strongly suppresses the background from top-quark pair
events. A third isolated lepton is assumed to come from the \Wboson
boson decay and is required to pass more stringent requirements on the
identification quality and/or transverse momentum. The neutrino
carries significant \MET~in the signal events and hence a minimal
\MET~requirement is imposed. In addition, a cut on the transverse mass
($M_T^W$) of the third lepton and the \MET~can be used to further
reduce the background from $Z+$jets, $Z\gamma$ and $ZZ$ production. A
summary of the detailed selections criteria used is given in
Tab. \ref{TAB:WZ:Selection}.

\begin{table}[h]
\tbl{Summary of $WZ$ selection requirements for the ATLAS and CMS
  experiment on the lepton transverse momenta $p_T$ due to the
  $W$ and $Z$ bosons, the missing transverse energy $\MET$, the
  invariant mass $m_{\ell\ell}$ and
  $p_T(\ell\ell)$ of $Z$ boson candidate and the transverse mass of the $W$ boson system.
  ATLAS requires also a minimal $\Delta R$ distance between all
  leptons.}  {
  \begin{tabular}{l l l} 
\hline
\hline
$WZ\rightarrow \ell^+\ell^- \ell'\nu$		&	$Z$ decay leptons $p_T>15\,\GeV$			&	$Z$ decay leptons $p^{1,2}_T>10,20\,\GeV$						\\
$(l,l'=e, \mu)$					&	$W$ decay lepton $p_T>20\,\GeV$			&	$W$ decay lepton 	$p_T>20\,\GeV$			\\
							&	$|\eta^{\ell}|<2.5$					&	 $|\eta^{\ell}|<2.5$							\\
							&	$|m_Z-m_{\ell^+\ell^-}|<10\,\GeV$						& 	$|m_Z-m_{\ell^+\ell^-}|<20\,\GeV$	\\
							&	$\MET>25\,\GeV$								&	$\MET>30\,\GeV$	\\
							&	Minimal distance of leptons $\Delta R>0.3$			&	\\
							&	$m_T>20\,\GeV$ of $W$ system		&	\\
\hline
\hline
   \end{tabular}
   \label{TAB:WZ:Selection}
}
\end{table}


The expected signal and backgrounds together with the observed data
are shown in Tab. \ref{Tab:WZ:Yields}. ATLAS expects $17\%$ of events
in the $eee$ channel, $23\%$ in the $ee\mu$ channel, $24\%$ in the
$\mu\mu e$ channel, and $35\%$ in the $\mu\mu\mu$ channel. The
increase in signal with decreasing number of electrons in the final
state channels is due to the lower electron identification efficiency.
A similar relationship is observed in the CMS analysis. It should be
noted that the background expectation in the CMS analysis is
significantly smaller compared to the ATLAS approach. This cannot be
explained by kinematic cuts, but can be attributed to an enhanced
fake-lepton rejection methodology of the CMS approach.

\begin{table}[h]
\tbl{Event yields in data and estimations of signal and background of
  the $WZ$ production in the ATLAS and CMS analyses.}  {
  \begin{tabular}{l |cccc|c} \hline
Experiment			& \multicolumn{5}{c}{ATLAS}   \\
  \hline
  				&	$eee$			&	$ee\mu$				&	$\mu\mu e$	&	$\mu\mu\mu$		& combined\\
  \hline
  Data			&	$56$				&	$75$					&	$78$			&	$108	$			&	$317$		\\
\hline
  $WZ$	(MC)			&	$38.9\pm3.1$		&	$54.0\pm2.2$			&	$56.6\pm1.7$	&	$81.7\pm2.1$		&	$231.2\pm4.7$		\\
  \hline
  $Z+$jets 			&	$8.8\pm2.8$		&	$3.7\pm2.3$			&	$10.2\pm3.3$	&	$9.1\pm5.5$		&	$31.8\pm7.4$		\\
  $ZZ$ 			&	$3.2\pm0.2$		&	$4.9\pm0.2$			&	$5.0\pm0.1$	&	$7.9\pm0.2$		&	$21.0\pm0.4$		\\
  $Z\gamma$ 		&	$1.4\pm0.7$		&	$0$					&	$2.3\pm0.9$	&	$0$				&	$3.7 \pm1.1$		\\
  top 			&	$1.1\pm0.4$		&	$2.9\pm0.9$			&	$3.6\pm1.1$	&	$4.0\pm1.2$		&	$11.6\pm1.9$	\\
  \hline
  Total Background 	&	$14.5\pm2.9$		&	$11.5\pm2.5$			&	$21.0\pm3.5$	&	$21.0\pm5.6$		&	$68.0\pm7.6$		\\
  \hline
  Total Expected 	&	$53.4\pm4.2$		&	$65.5\pm3.3$			&	$77.6\pm3.9$	&	$102.7\pm6.0$		&	$300.2\pm8.9$		\\
  \hline
  \hline
Experiment			& \multicolumn{5}{c}{CMS}   \\
  \hline
  				&	$eee$			&	$ee\mu$				&	$\mu\mu e$	&	$\mu\mu\mu$		& combined\\
  \hline
  Data			&	$64$				&	$62$					&	$70$			&	$97$				&	$293	$	\\
\hline
  $WZ$	(MC)			&	$44.7\pm0.5$		&	$55.0\pm1.0$			&	$56.0\pm0.5$	&	$73.8.7\pm0.6$	&	$229.5\pm1.4$		\\
  \hline
  $Z+$jets 			&	$1.2\pm0.8$			&	$1.2\pm0.9$		&	$0.8\pm0.6$	&	$0.6\pm0.6$		&	$3.8\pm1.5$		\\
  $ZZ$ 			&	$2.0\pm0.1$			&	$3.5\pm0.1$		&	$2.7\pm0.1$	&	$5.1\pm0.1$		&	$13.3\pm0.2$		\\
  $Z\gamma$ 		&	$0		$		&	$0$					&	$0.5\pm0.5$	&	$0$				&	$0.5\pm0.5$		\\
  top 			&	$0.3\pm0.1$			&	$0.6\pm0.1$		&	$0.6\pm0.1$	&	$1.0\pm0.1$		&	$2.5\pm0.2$		\\
  \hline
  Total Background 	&	$3.5\pm0.8$			&	$5.3\pm1.0$		&	$4.6\pm0.8$	&	$6.7\pm0.6$		&	$20.1\pm2.6$		\\
  \hline
  Total Expected (MC)&	$48.2\pm1.0$		&	$55.0\pm1.0$			&	$60.5\pm1.1$	&	$80.5\pm0.9$		&	$244.2\pm2.0$		\\

  \hline
   \end{tabular}
   \label{Tab:WZ:Yields}
}
 \end{table}


\subsubsection{Background Estimation}

Because jets faking signal leptons may not be modeled correctly in MC
the background contributions from $Z+$jets and $t\bar t$ are estimated
in a data-driven method in the ATLAS analysis.  The basic idea of the
$Z+$jets background estimate is a data-driven estimate of the
probability, $p_F$, for lepton-like jets, i.e.  jets that pass loose
lepton identification requirements, to pass the full lepton
identification. This probability $p_F$ is determined in a control
region of events which pass \Zboson boson selection cuts, contain at
least one lepton-like jet, and have no significant \MET. The measured
fake probability is then applied in the signal region, where only a
loose lepton identification is required.  The background due to $t\bar
t$ is the only background whose leptonic invariant mass does not peak
at the \Zboson boson mass.  Hence, by requiring two same-sign leptons
to pass the \Zboson boson mass constraint leads to a top-quark
enriched control region which is used to estimate the $t\bar t$ contribution in
the signal region.  All other backgrounds are theoretically well
understood and are estimated with MC simulations. In particular the
$ZZ$ background has a sizable contribution, but since it is mainly due
to limited detector acceptance it can be precisely estimated using simulated MC events.

The CMS analysis uses a common data-driven technique to estimate the
background from $t\bar t$ and $Z+$jets, which are due to fake leptons.
The basic idea is to define some looser lepton identification
requirements to enhance the background contribution.  The probability
for leptons that have passed the `loose' identification criteria to also pass the
nominal identification is estimated from data using independent
samples. Knowing these probabilities, it is possible to construct a
set of linear equations which relate the number of selected events in
different event categories. Each event category is defined by the
number of prompt and fake leptons. This set of equations leads to a
unique solution for the number of background events in the signal
region.  The backgrounds from $ZZ$ and $Z\gamma$ are also estimated
using simulated MC events.

\begin{figure*}
\resizebox{0.475\textwidth}{!}{\includegraphics{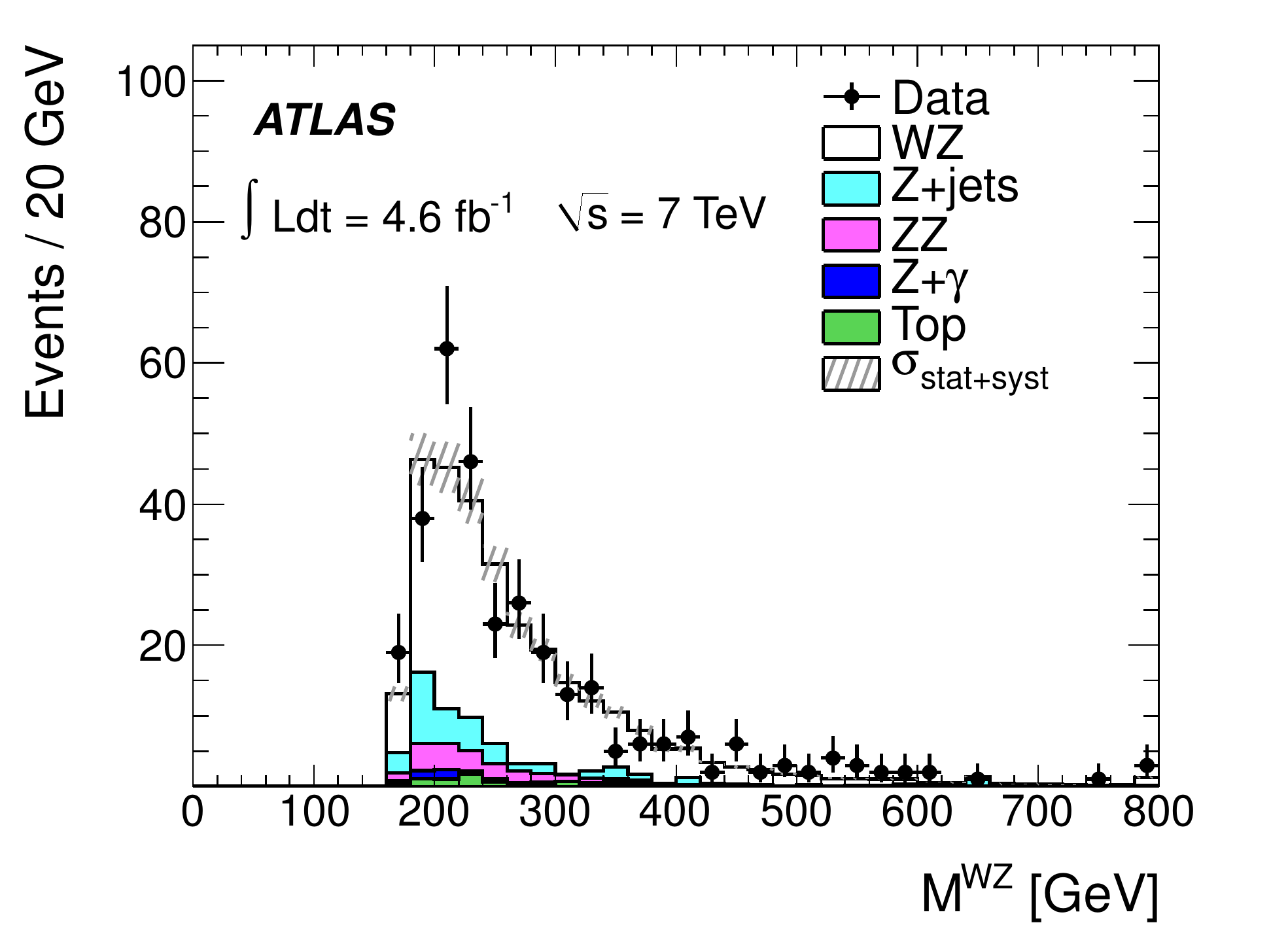}}
\resizebox{0.505\textwidth}{!}{\includegraphics{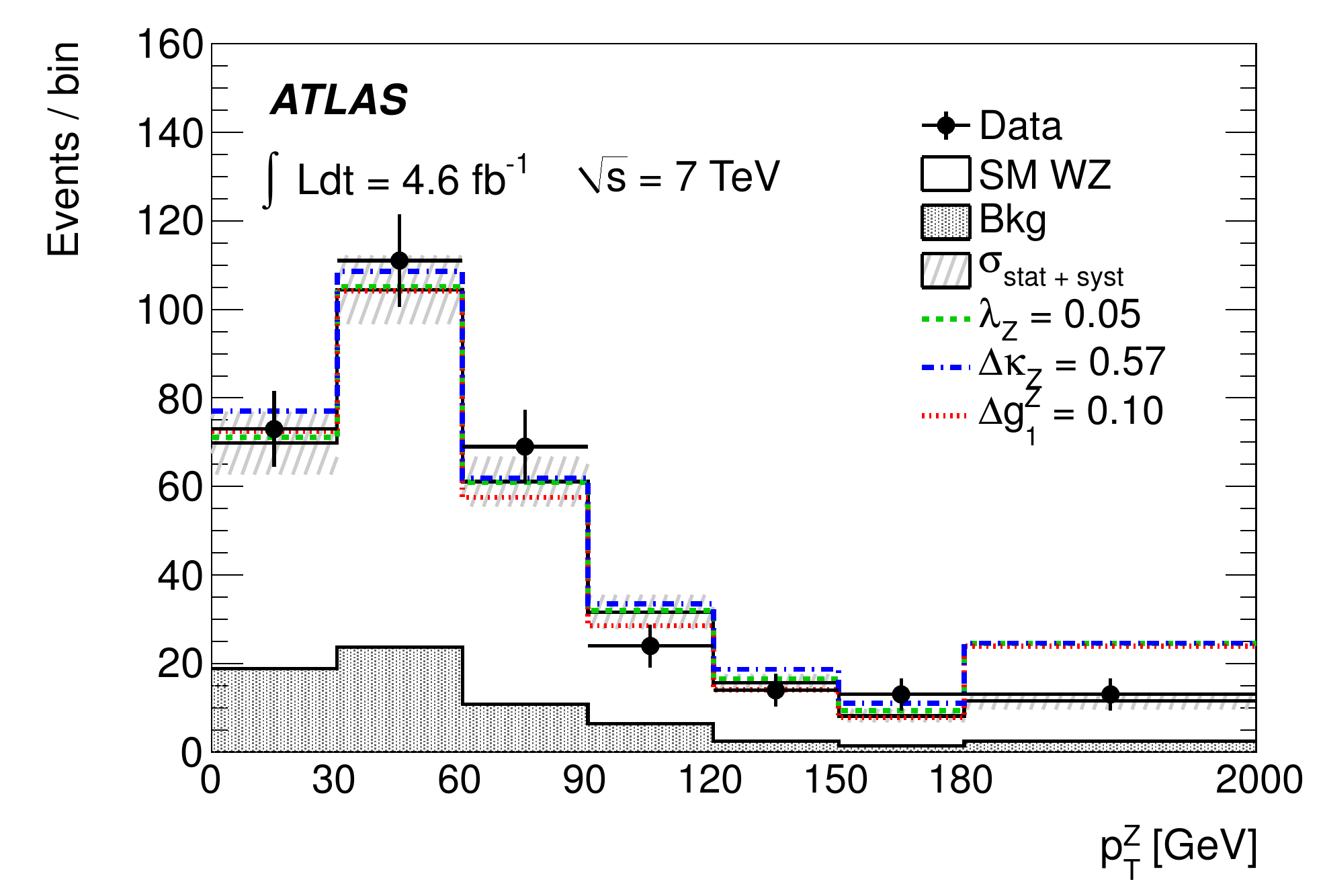}}
\caption{ATLAS: Invariant mass $m_{WZ}$-distribution (left) and
  $p_T^Z$-distribution (right) of the $W^\pm Z$ pair after all
  selections. The shaded bands indicate the total statistical and
  systematic uncertainties of the MC prediction. The right-most bin
  contains overflow.}
\label{fig:ATLASWZ}
\end{figure*}

\begin{figure*}
\resizebox{0.49\textwidth}{!}{\includegraphics{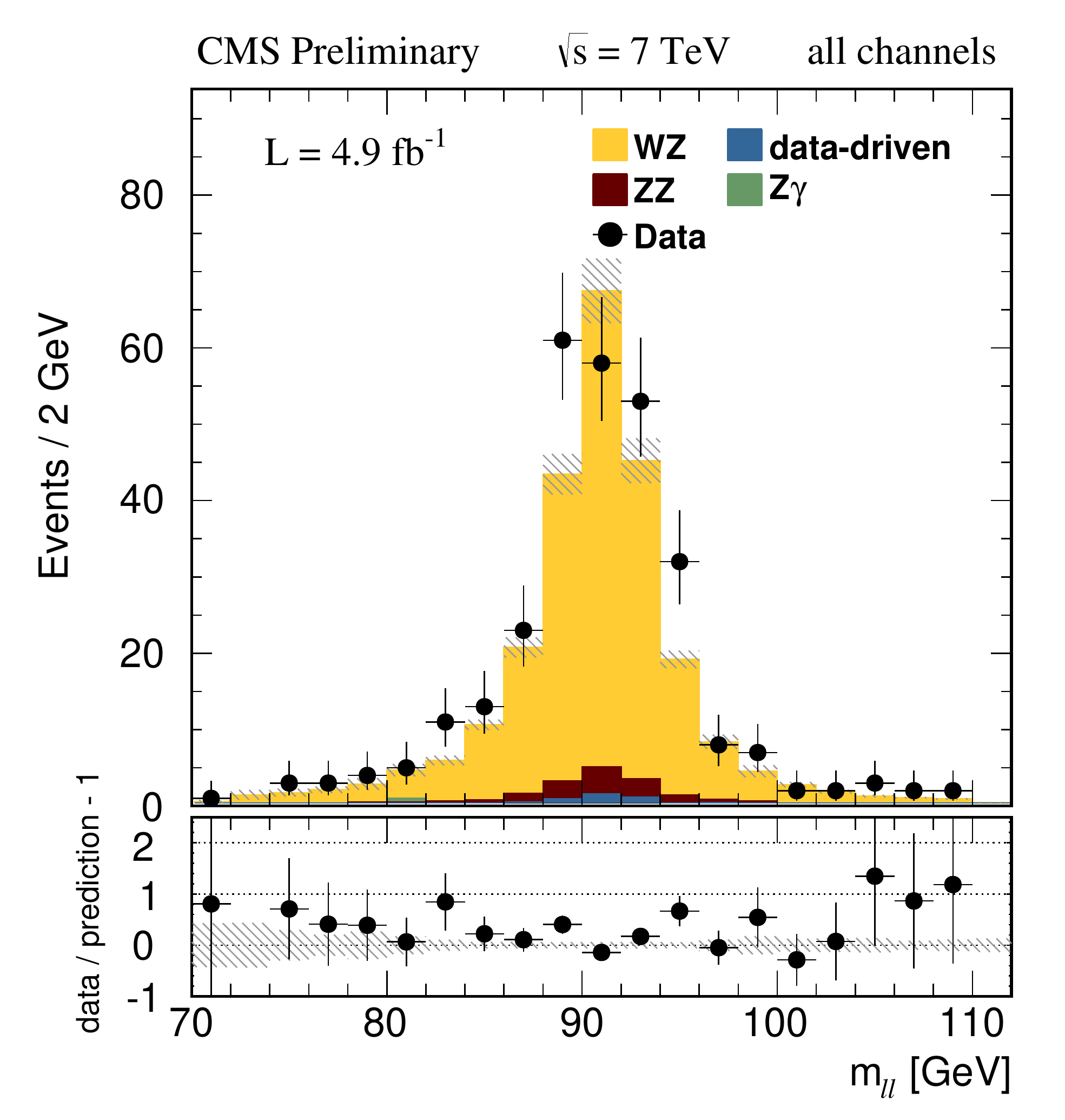}}
\resizebox{0.49\textwidth}{!}{\includegraphics{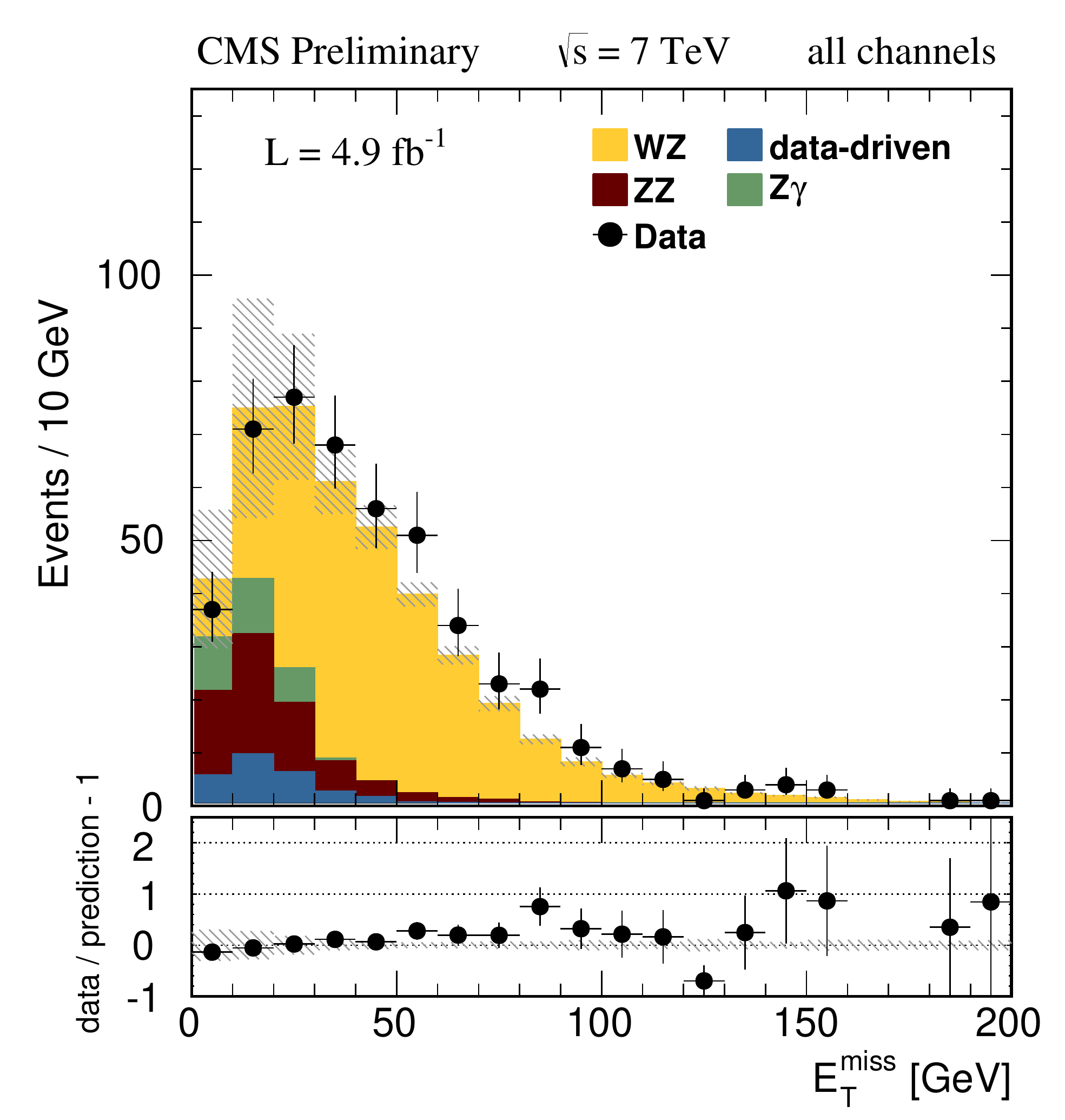}}
\caption{CMS: \Zboson boson candidate lepton invariant mass
  $m_{ll}$-distribution (left) and \MET distribution (right) for
  selected $W^\pm Z$ candidates (right). The shaded bands indicate the
  total statistical and systematic uncertainties of the MC
  prediction.}
\label{fig:CMSWZ}
\end{figure*}

Figure \ref{fig:ATLASWZ} shows the invariant mass of the $M_{WZ}$
system and the $p_T$ spectrum of the selected $Z$ boson in the signal
region for both data and MC for the ATLAS analysis. Figure
\ref{fig:CMSWZ} shows the invariant mass distribution of the $Z$ boson
candidate lepton pair and the observed \MET~spectrum for the CMS
analysis. A good description of the data by the MC prediction is
observed in both analyses.

\subsubsection{Cross Section Measurement}


The fiducial phase space in ATLAS at the generator level is defined
along the kinematic requirements and presented in
Tab. \ref{TAB:WZ:Selection}.  The momenta of photons within $\Delta
R<0.1$ of signal leptons are added to the respective lepton $p_T$.
The resulting acceptance factors $A_{WZ}$ for all decay channels range
from $0.330$ to $0.338$.  The differences are caused by final state radiation photons.
The corresponding uncertainties due to PDFs and momentum scales are
evaluated similarly to the ATLAS $WW$ analysis. The main uncertainty
is due to the limited knowledge of PDFs, which leads to an uncertainty
of $1.4\%$.

The detector correction factors $C_{WZ}$ range from $0.8$ in the
$\mu\mu\mu$ channel to $0.4$ in the $eee$ channel. The differences are
due to the lower electron identification efficiencies compared to the
muon channel. The primary uncertainties on $C_{WZ}$ are due to the
limited knowledge of the lepton reconstruction efficiencies and
momentum scales and resolution.

CMS did not publish a fiducial cross section number. The product of the
acceptance and efficiency correction factors $A_{WZ}\cdot C_{WZ}$ are
comparable to the ATLAS analyses.  The dominate experimental
uncertainties are due to \MET\ requirement and lepton reconstruction
and trigger efficiencies.


The fiducial and inclusive cross section is then calculated for each
channel separately.  A mass-cut of $66\,<m_{\ell^+\ell^-}<116\,\GeV$
is required for the inclusive cross section in order to define
an on-shell \Zboson boson.  The cross sections are calculated with
respect to the corresponding process $W^\pm Z\rightarrow \ell'^\pm \nu
\ell^+ \ell^-$, i.e. the contribution from $\tau$ lepton decays has to
be removed. This is done by multiplying the final cross-section by a
factor $(1-N_\tau^{MC}/N_{sig}^{MC})$ where $N_\tau^{MC}$ is the
number of reconstructed and selected $WZ$ where at least one
$\tau$-decay is present, and $N_{sig}^{MC}$ is the number of all reconstructed and
selected $WZ$ events.  The contribution from $\tau$ decays is in the
order of $4\%$.  The individual cross sections are then combined via a
maximum likelihood method, leading to the results shown in
Tab. \ref{Tab:WZ:Results}.  The main systematic uncertainties are due
to the data-driven background estimates and contribute $4\%$ to the
total systematic uncertainty of $4.6\%$ in the fiducial phase space.

The preliminary CMS results from \cite{CMS:2011wz} are also
presented. The combined cross section is obtained by combining the
four decay channels, requiring $71<m_{\ell\ell}<111$ GeV.


\begin{table}[h]
\tbl{Summary of measured inclusive cross sections for the $WZ$ process from the ATLAS and CMS collaborations.}  {
\begin{tabular}{lll}
\hline
\hline
 										& Measured				& Predicted						\\
\hline
ATLAS~~~~~~~~~~~~~~~~~~~~~~			&  							\\
$\sigma^{fid} (WZ\rightarrow \ell^+\ell^- \ell'\nu)$			& $92~\pm~7 \,(stat.)~\pm ~4\,(sys.)~\pm ~2 \,(lumi.) $ fb				&		\\
$\sigma^{incl.} (WZ\rightarrow\ell^+\ell^- \ell'\nu)$			& $19.0~\pm~1.4 \,(stat.)~\pm ~0.9\,(sys.)~\pm ~0.4 \,(lumi.)$ pb		&	$17.7 \pm 1.1$	 pb\\
\hline
CMS 											& 							\\
$\sigma^{incl.} (WZ\rightarrow e^+e^- e^\pm \nu)$			& $23.0~\pm~3.1 \,(stat.)~\pm ~1.4\,(sys.)~\pm ~~0.5 \,(lumi.)$ pb		&	$17.7 \pm 1.1$	pb\\
$\sigma^{incl.} (WZ\rightarrow e^+e^- \mu^\pm \nu)$		& $19.7~\pm~2.7 \,(stat.)~\pm ~1.5\,(sys.)~\pm ~~0.4 \,(lumi.)$ pb		&	$17.7 \pm 1.1$	pb\\
$\sigma^{incl.} (WZ\rightarrow \mu^+\mu^- e^\pm \nu)$		& $19.8~\pm~2.6 \,(stat.)~\pm ~1.6\,(sys.)~\pm ~~0.4 \,(lumi.)$ pb		&	$17.7 \pm 1.1$	pb\\
$\sigma^{incl.} (WZ\rightarrow \mu^+\mu^- \mu^\pm \nu)$		& $21.0~\pm~1.6 \,(stat.)~\pm ~1.5\,(sys.)~\pm ~~0.5 \,(lumi.)$ pb		&	$17.7 \pm 1.1$	pb\\
$\sigma^{incl.} (WZ\rightarrow \ell^+\ell^- \ell'^\pm \nu)$		& $20.8~\pm~1.3 \,(stat.)~\pm ~1.1\,(sys.)~\pm ~~0.5 \,(lumi.)$ pb		&	$17.7 \pm 1.1$	pb\\
\hline
\hline
\end{tabular}
\label{Tab:WZ:Results}
}  
\end{table}

\begin{figure*}
\resizebox{0.49\textwidth}{!}{\includegraphics{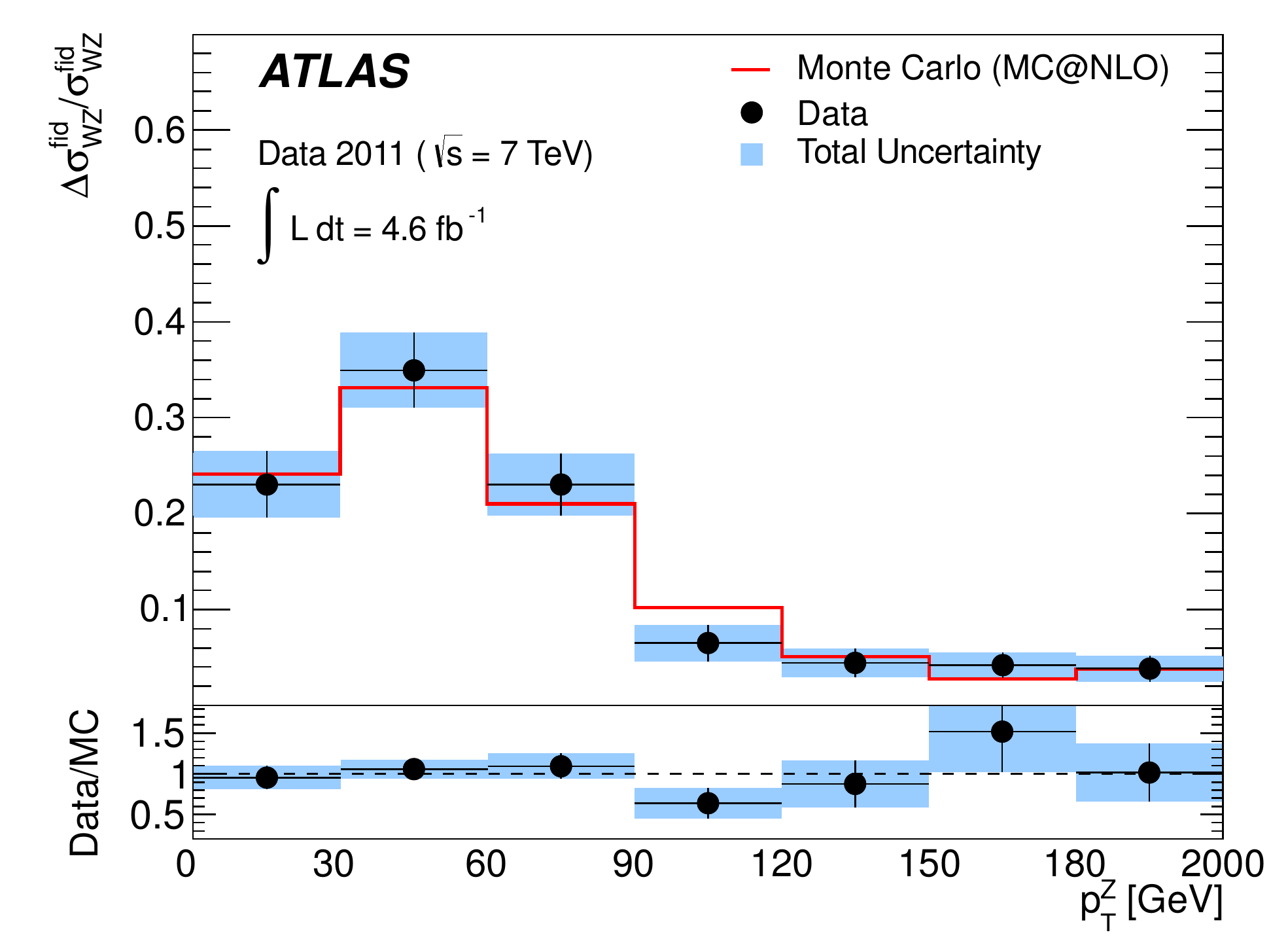}}
\resizebox{0.49\textwidth}{!}{\includegraphics{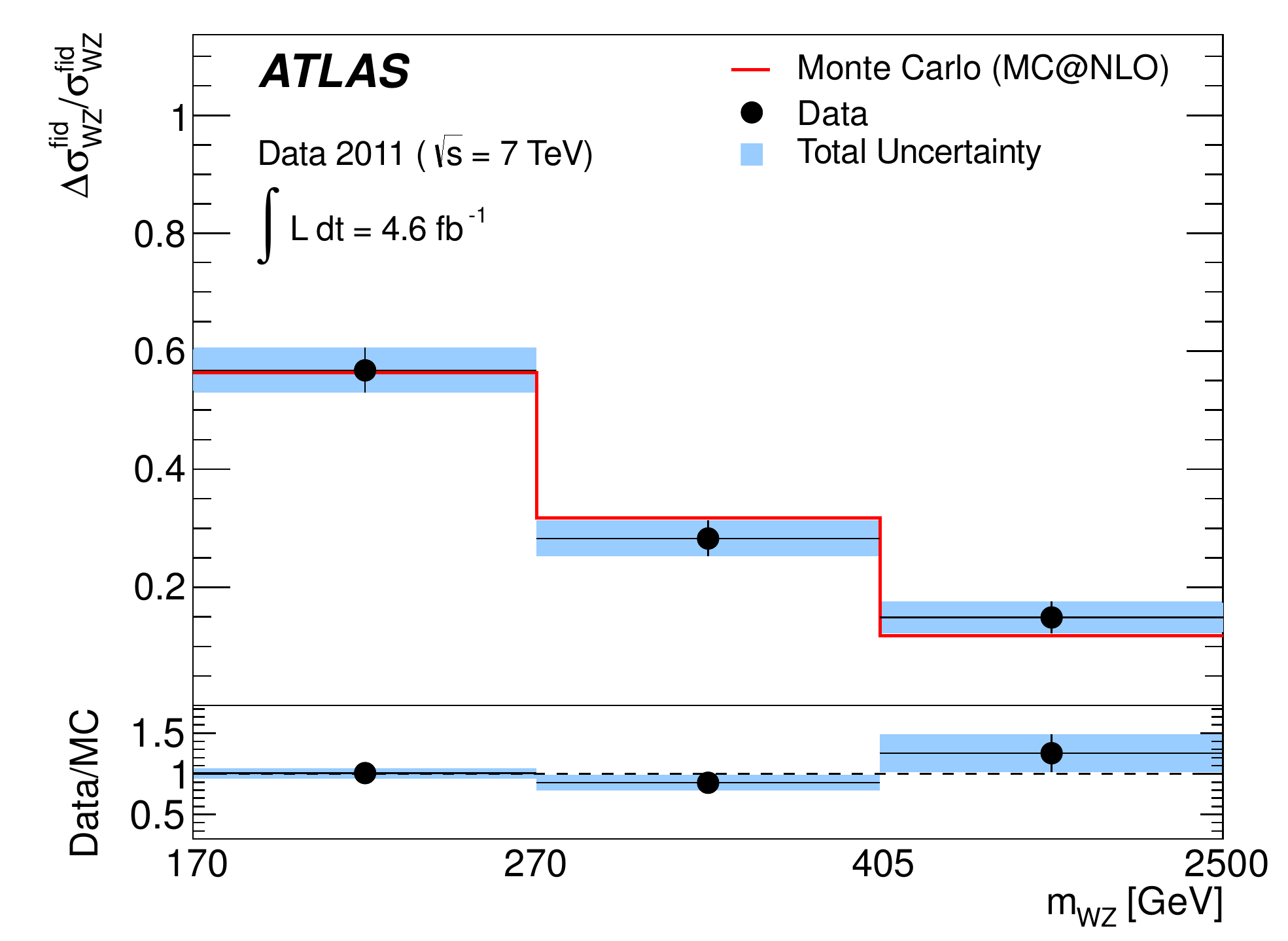}}
\caption{Normalized fiducial cross-sections in bins of $p_T^Z$ (left)
  and $m_{WZ}$ compared with the SM prediction. The full uncertainty
  contains statistical and systematic uncertainties.}
\label{fig:WZ:Unfolded}
\end{figure*}

 
Within the fiducial region, ATLAS also published normalized unfolded
distributions of $M_{WZ}$ and $p_T^Z$, shown in
Fig. \ref{fig:WZ:Unfolded}. These can be used to constrain aTGC
couplings without knowing details of detector effects and resolution.
A good agreement with the MC prediction is seen.

CMS measured the ratio of the production cross-sections of $W^+Z$ and
$W^-Z$, as they are not expected to be the same in a proton-proton
collider. Several systematic uncertainties cancel in the ratio, such
as the luminosity uncertainty. The resulting combined cross section
ratio is found to be $\sigma_{W^{+}Z}/\sigma_{W^{-}Z} = 1.94 \pm 0.25
(\mbox{stat.}) \pm 0.04 (\mbox{syst.})$, and is in agreement with the
NLO QCD calculation of $1.724\pm0.003$ using \MCFM and the MSTW2008
PDF set.

\subsubsection{Derived Results and Discussion}

For the combination of the inclusive $WZ$ cross section measurements
from both experiments, it is assumed that the uncertainties due to
integrated luminosity are fully correlated while all other
uncertainties are uncorrelated. The BLUE method yields a combined
cross section of
\[ \sigma ^{LHC}_{WZ} = 19.9 \pm 1.0 (\mbox{stat.}) \pm 0.8 (\mbox{syst.})~\text{pb}. \]
The measured cross section is in good agreement with the theoretical
prediction of $17.6\pm1.1\,pb$ (as shown in
Tab. \ref{tab:CrossSections}).  In contrast to the $WW$ measurement,
the statistical uncertainty is the dominant uncertainty in the
7 TeV dataset and is expected to be reduced by a factor of $\approx
2.5$ in the ongoing analyses of the 8 TeV dataset. However, the
overall relative precision of this measurement is better by $15\%$
compared to the $WW$ analyses and hence provides also an improved test
of the SM prediction of the diboson production.

A simplified fiducial cross section for $WZ$ events is defined by
requiring three charged leptons on the generator level with $p_T>20$ GeV
and $|\eta|<2.5$, one neutrino with $p_T>20$ GeV and an invariant mass
cut on the leptons from the \Zboson boson decay of $|m_Z-m_{\ell\ell}|<30$
GeV. The extrapolation factor from the ATLAS definition to the
simplified phase space is evaluated with \MCAtNLO and leads to
$1.07\pm0.02$, where uncertainties due to momentum scales and PDFs
are included.  The simplified cross section is found to be
\[ \sigma ^{fid,sim}_{WZ} = 98 \pm 7 (\mbox{stat.}) \pm 5 (\mbox{syst.}) \pm 2 (\mbox{lumi.})\,~{\text{fb}}.\]

The cross section within this simplified fiducial phase space but
requiring that the $p_T$ of the \Zboson boson in the range
$180\,\GeV<p_T^Z<2000\,\GeV$, results in $\sigma
^{fid,sim}_{WZ}(180\,\GeV<p_T^Z<2000\,\GeV)= 4.1 \pm 1.4$ fb, where
the corresponding extrapolation factor is again based on \MCAtNLO
predictions. This cross section can be used to test specifically the
TGC vertex of $WZ$ production.


\subsection{\label{sec:wgamma}$W\gamma$ Analysis}

\begin{figure}[bt]
    \begin{center}
        \includegraphics[width=1.0  \textwidth]{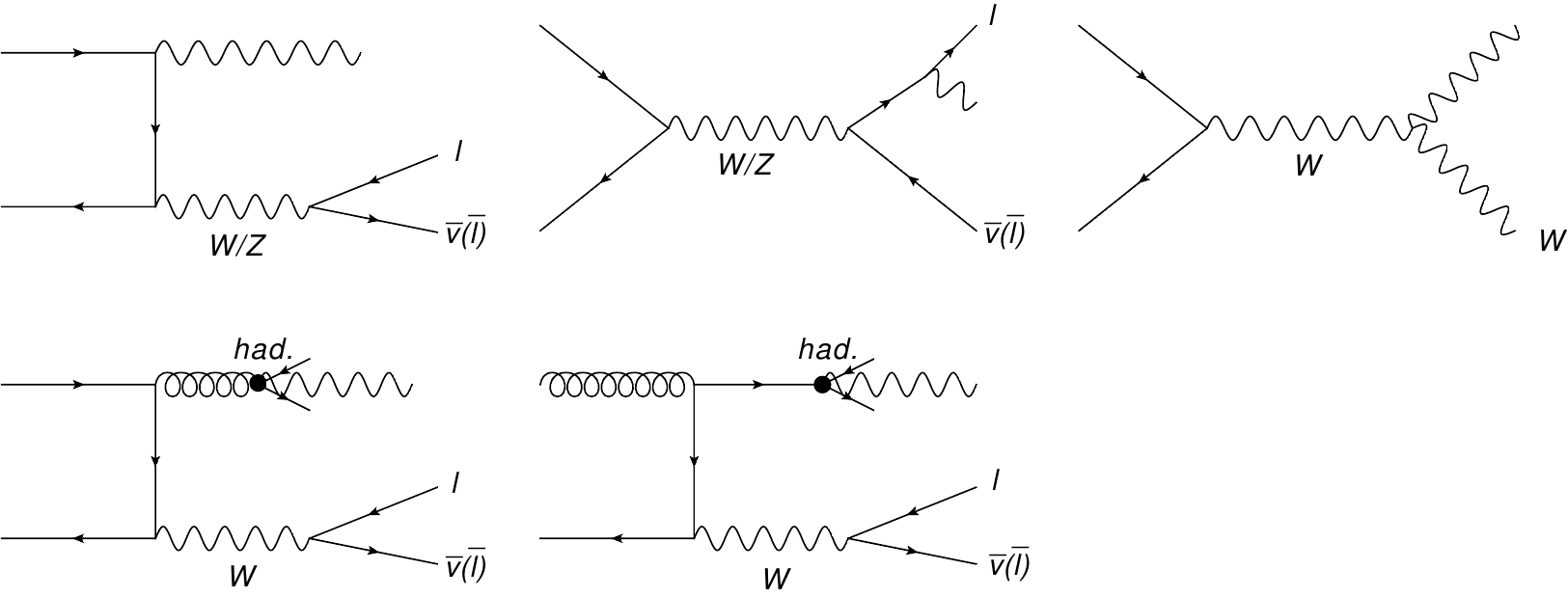}
        \caption{Feynman diagrams for $W\gamma$ and $Z\gamma$
          production via the $u$- and$t$-channels (upper left), final state radiation
          (upper middle), and the $s$-channel for $W\gamma$ production (upper right). Photons
          in the final state can also be produced in the
          fragmentation of final state quarks and gluons (lower row).}
        \label{Fig:WZGammaLOFeynman}
    \end{center}
\end{figure}

In the SM, the production process $pp \rightarrow W\gamma \rightarrow
\ell\nu\gamma (\ell=e, \mu)$ originates from $W$ production with
photons radiated from initial-state quarks or directly by the $W$
bosons. Other SM processes such as $W$ production with photon
bremsstrahlung from the charged leptons from the $W$ boson decays or
with photons from the fragmentation of secondary quarks and gluons
into isolated photons can result in the same final state (see
Fig. \ref{Fig:WZGammaLOFeynman}).  Both ATLAS and CMS collaborations
have measured the $W\gamma$ production cross section in the electron
and muon decay channels \cite{Aad:2012mr, Chatrchyan:2013fya}.

\subsubsection{Event Selection}

For the ATLAS $W\gamma \rightarrow \ell\nu\gamma$ analysis, candidate
events are selected by requiring exactly one lepton with $p_T>25$
\GeV, at least one isolated photon with $E_T>15$ \GeV, and
$E_T^{\text{miss}}>35$ \GeV. Leptons and photons are required to be
within the detector fiducial region. All selected leptons must satisfy
isolation requirements based on calorimeter (for electrons) or
tracking (for muons) information and must be consistent with
originating from the primary vertex.  Electromagnetic clusters without matching
tracks are directly classified as unconverted photon candidates, while
clusters that are matched to tracks that originate from reconstructed
conversion vertices in the tracker or to tracks consistent with coming
from a conversion are considered as converted photon candidates.
Tight requirements on the photon shower shapes are applied to suppress
the background from multiple showers produced in meson decays.  
Leptons and photons have the same isolation requirement which is
$\Delta R(\ell, \gamma)>0.7$
in order to suppress the contributions from FSR photons in the $W$
boson decays.  The $W\gamma$ candidates are also required to have the
transverse mass of the lepton-$E_T^{\text{miss}}$ system greater than
40 \GeV. A veto on reconstructed \Zboson boson events is applied in
the electron channel ($|m_{e\gamma}-m_Z|>15$ GeV) to reject
contributions from the $Z\gamma \rightarrow e^+e^-\gamma$ process
where one of the electrons is misidentified as an isolated photon.
ATLAS also measured the cross sections for events with and without the
requirement of zero jets reconstructed in each event.  Jets are
required to have $E_T>30$ GeV and $|\eta|<4.4$.

Similar event selection criteria are used for the CMS $W\gamma
\rightarrow \ell\nu\gamma$ analysis except that lepton candidates are
required to have $p_T>35$ \GeV\ and only photon candidates without
associated tracks in the pixel detector are considered. No selection
cut is applied on $E_T^{\text{miss}}$ and instead the transverse mass
of the lepton-$E_T^{\text{miss}}$ system is required to be greater
than 70 \GeV. A summary of the detailed selection cuts for both
experiments of shown in Table \ref{tab:Wgamma_selection}.

\begin{table}[h]
\tbl{Summary of the $W\gamma \rightarrow \ell\nu\gamma$ selection cuts
  used by the ATLAS and CMS collaborations.}  {
\begin{tabular}{lll}
\hline
\hline
        & ATLAS & CMS \\
\hline
$W\gamma \rightarrow \ell \nu \gamma$    &  Combined muons with $p_T>25$ GeV  & Muons with $p_T>35$ GeV\\
$(\ell'=e, \mu)$                                           &  Combined electrons with $E_T>25$ GeV & Electrons with $p_T>35$ GeV \\
                                                                    & Photon $E_T>15$ GeV and $E_T^{\Delta R<0.3}<6$ GeV & Photon $E_T>15$ GeV, isolation and \\ 
                                                                    &  & $~~~~~$ shower-shape requirements  \\
                                                                    & $E_T^{\text{miss}}>35$ GeV, $m_T(\ell\nu)>40$ GeV & $m_T(\ell\nu)>70$ GeV \\
                                                                    & $|m_{e\gamma}-m_Z|>15$ GeV and only one lepton & only one lepton \\
                                                                    & $\Delta R(\ell, \gamma)>0.7$ & $\Delta R(\ell, \gamma)>0.7$  \\
\hline
\hline
\end{tabular}
\label{tab:Wgamma_selection}
}
\end{table}

\subsection{Background Estimation}
The primary backgrounds to the $\ell\nu\gamma$ signal come from the
$W+$jets, $Z \rightarrow \ell\ell$, $\gamma+$ jets, and $t\bar{t}$,
single top quark and $WW$ processes.

Data is used to estimate the background from the $W+$ jets production
when photons come from the decays of mesons produced in jet
fragmentation.  ATLAS used a two-dimensional sideband method which
considered the distribution of signal and background events in a phase
space defined by two uncorrelated variables for which the signal and
background have different shapes. The phase space is partitioned into
four regions with one signal region and three control regions in which
one or both selections are reversed. The ratio of the number of
background events in two control regions provides the transfer factor
which is combined with events in the third control region to estimate
the background contribution in the signal region.  The two
discriminating variables are the photon isolation and the photon
identification based on the shower shape in the ATLAS analysis.  CMS
used a template method which relies on a maximum-likelihood fit to the
data distribution of the photon energy-weighted width in
pseudorapidity ($\sigma_{\eta\eta}$).  The fit makes use of the
expected distributions (``templates") for genuine photons and
misidentified jets. The distribution in $\sigma_{\eta\eta}$ is very
narrow and symmetric, and the templates are obtained from simulated
$W\gamma$ events. The $\sigma_{\eta\eta}$ distribution for backgrounds
is asymmetric with a slow falloff at large value, and the templates
are defined by events in a background-enriched isolation sideband of
data. The maximum-likelihood fit is performed in several photon $p_T$
bins. The estimated background contribution has been crosschecked
with an alternative method.
The prediction from the alternative approach is found to be consistent
with the template method, and their difference is treated as an
additional source of systematic uncertainty in the analysis.

Further background is due to the $Z/\gamma^* \rightarrow \ell\ell$
production, when one of the leptons is misidentified as a photon or is
not identified and the photon originates from initial state radiation
or from bremsstrahlung from a decay lepton. ATLAS used MC simulation
for its estimation, and various studies of the probability to lose one
lepton from $Z$ decay due to acceptance and the modeling of
$E_T^{\text{miss}}$ in $Z+\gamma$ and $Z+$ jets events have been
performed and checked with collision data.  ATLAS also used a
two-dimensional sideband method to estimate the $\gamma+$ jets
background, with lepton isolation and $E_T^{\text{miss}}$ as the
independent variables.  CMS used a data-driven method to estimate the
$Z/\gamma^* \rightarrow \ell\ell$ background. The background in the
$e\nu\gamma$ decay channel is dominated by $Z+$ jets events, and the
contribution is obtained from a fit to the invariant mass distribution
of the photon and electron candidates. To determine the background in
the $\mu\nu\gamma$ decay channel where an electron is misidentified as a
photon, a new sample is selected with events passing all event
selection criteria except that the presence of a track in the pixel
detector associated with the photon candidate is ignored. The
contribution from genuine electrons misidentified as photons can
therefore be derived from the total number of events in this new
sample and the probability for an electron not to have a matching
track. This probability can be measured from $Z \rightarrow ee$ data
by requiring stringent electron identification criteria on one
electron and checking how often the other electron passes the full
photon selection criteria, including the requirement of having no
associated track in the pixel detector.

All other backgrounds are estimated using MC simulations for both
ATLAS and CMS analyses. A summary of the expected signal and
background events in given in Table \ref{tab:Wgamma}. The primary
background contribution and also the dominant associated uncertainty
is due to the $W+$jets process for both experiments. The signal to
background ratio is significantly better for the ATLAS analysis, which
can be explained by the superior performance of the ATLAS
electromagnetic calorimeter.

\subsubsection{Fiducial Cross Section Results}
\begin{figure}[h]
\begin{center}
\resizebox{0.99\textwidth}{!}{\includegraphics{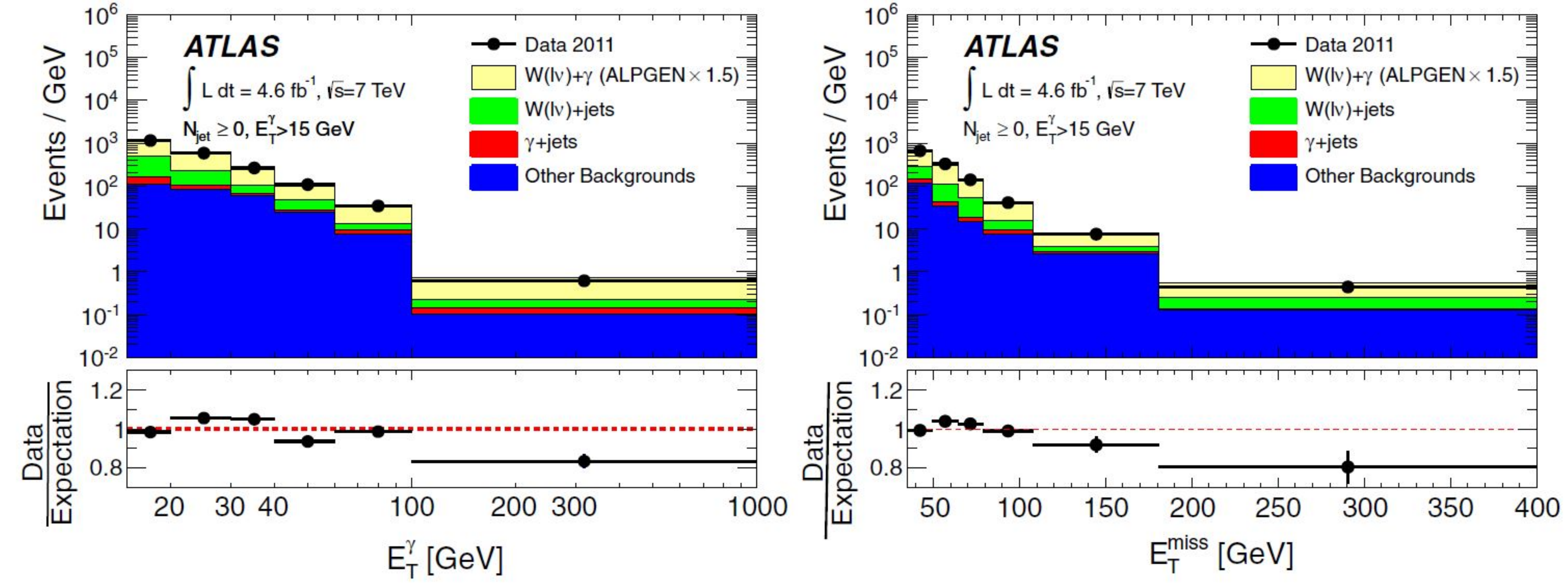}}
\caption{\label{fig:ATLAS_WGamma} Combined distributions for $\ell\nu\gamma$ candidate events in the electron and muon channels of the photon transverse energy (left) and the missing transverse energy (right) for the ATLAS analysis.}
\end{center}
\end{figure}

\begin{figure}[h]
\begin{center}
\resizebox{0.9\textwidth}{!}{\includegraphics{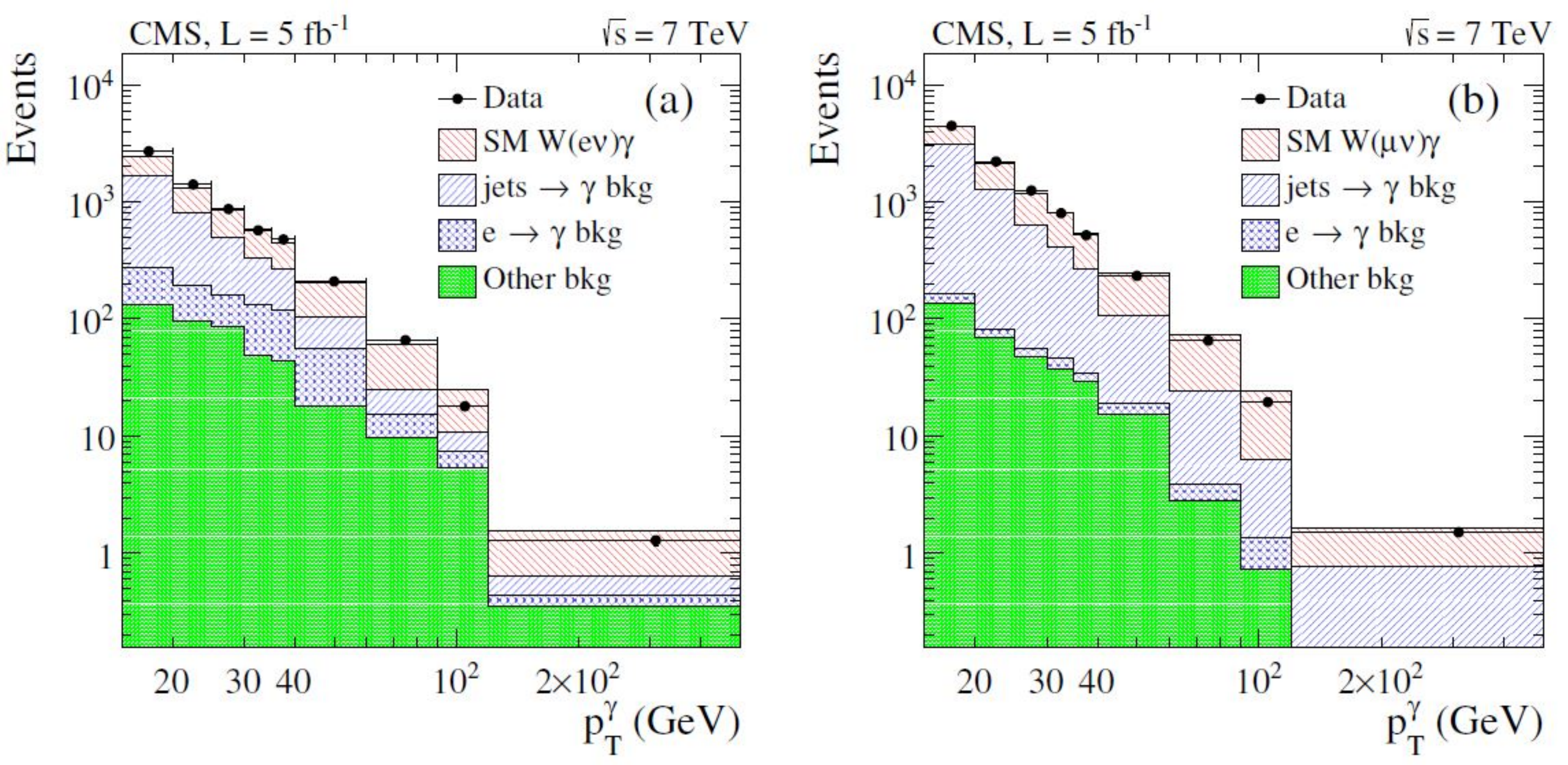}}
\caption{\label{fig:CMS_WGamma} Distributions in $p^\gamma_T$ for
  $\ell\nu\gamma$ candidate events in data: signal and background MC
  simulation contributions to $W\gamma \rightarrow e\nu\gamma$ (left) and 
$W\gamma \rightarrow \mu\nu\gamma$ (right). The same channels are
  shown for comparison for the CMS analysis.}
\end{center}
\end{figure}

The expected and observed event yields after applying all selection
criteria are shown in Tab.~\ref{tab:Wgamma} for all decay channels.
The photon transverse energy and \ETMiss\ distributions for
all selected $\ell\nu\gamma$ candidate events are shown in
Fig.~\ref{fig:ATLAS_WGamma} for the ATLAS analysis, and the photon
transverse energy distribution is shown separately for the electron
and muon decay channels in Fig.~\ref{fig:CMS_WGamma} for the CMS
analysis.  Reasonable agreement between the data and expected signal
and background contributions are observed from each experiment.

\begin{table}[h]
\tbl{Summary of observed $W\gamma$ candidates in the data, 
  background estimates, and expected signal for the individual decay
  modes. The first uncertainty is statistical while the second is
  systematic.}  {
\begin{tabular}{lccccr}
\hline
\hline
        & \multicolumn{2}{c}{ATLAS} & \multicolumn{2}{c}{CMS}  & \\
        & $e\nu\gamma$ & $\mu\nu\gamma$ & $e\nu\gamma$ & $\mu\nu\gamma$ \\
\hline
$W+$jets  & $1240 \pm 160 \pm 210$  & $2560 \pm 270 \pm 580$  & $3180 \pm 50 \pm 300$  & $5350 \pm 60 \pm 510$ \\
$Z(\rightarrow \ell\ell) +X$ & $678 \pm 18 \pm 86$ & $779 \pm 19 \pm 93$ & $690 \pm 20 \pm 50$ & $91 \pm 1 \pm 5$ \\
$\gamma+$jets & $625 \pm 80 \pm 86$ & $184 \pm 9 \pm 15$ & N/A & N/A \\
$t\bar{t}$ & $320 \pm 8 \pm 28$ & $653 \pm 11 \pm 57$ & N/A & N/A \\
Other backgrounds & $141 \pm 16 \pm 13$ & $291 \pm 29 \pm 26$ & $410 \pm 20 \pm 30$ & $400 \pm 20 \pm 30$ \\
Signal & $4390 \pm 200 \pm 250$ & $6440 \pm 300 \pm 590$ & $3200 \pm 100 \pm 320$ & $4970 \pm 120 \pm 530$ \\ \hline
Observed & 7399 & 10914 & 7470 & 10809 \\
\hline
\hline
\end{tabular}
\label{tab:Wgamma}
}
\end{table}

ATLAS measured the production cross section in an extended fiducial
region defined as charged lepton $p_T>25$ \GeV\ and $|\eta|<2.47$,
neutrino $p_T>35$ \GeV, and photon $E_T>15$ \GeV\ and $\Delta R(\ell,
\gamma)>0.7$. Cross sections for events with and without the
requirement of zero jets are listed separately. The jets are required
to have $E_T>30$ GeV and $|\eta|<4.4$.  CMS measured the production
cross section in an extended fiducial region defined as photon
$E_T>15$ \GeV\ and $\Delta R(\ell, \gamma)>0.7$. The measured and
predicted cross sections are listed in
Tab.~\ref{tab:Wgamma_xsection}. ATLAS also measured the unfolded
differential cross section as a function of the photon $E_T$ of the
$W\gamma \rightarrow \ell\nu\gamma'$ process. The results are shown
separately for the inclusive ($N_{jet} \ge 0$) and exclusive
($N_{jet}=0$) fiducial regions.

\begin{table}[h]
\tbl{Summary of the measured and predicted $W\gamma \rightarrow
  \ell\nu\gamma$ cross sections from the ATLAS and CMS
  collaborations. Different fiducial regions are defined by the ATLAS
  and CMS collaborations.}  {
\begin{tabular}{lll}
\hline
\hline
        & Measured [pb] & Predicted [pb] \\
\hline
ATLAS & & \\
$~~~~~$ $\sigma^{fid}_{W\gamma \rightarrow \ell \nu \gamma}$ ($N_{jet} \ge 0$)   &  $2.77 \pm 0.03$ (stat.) $\pm 0.33$ (syst.) $\pm 0.14$ (lumi.)  & $1.96 \pm 0.17$ \\ [1.5ex]
$~~~~~$ $\sigma^{fid}_{W\gamma \rightarrow \ell \nu \gamma}$ ($N_{jet} = 0$)   &  $1.76 \pm 0.03$ (stat.) $\pm 0.21$ (syst.) $\pm 0.08$ (lumi.)  & $1.39 \pm 0.13$ \\ [1.5ex]
CMS & & \\
$~~~~~$ $\sigma^{fid}_{W\gamma \rightarrow \ell \nu \gamma}$  &  $37.0 \pm 0.8$ (stat.) $\pm 4.0$ (syst.) $\pm 0.8$ (lumi.)  & $31.8 \pm 1.8$ \\ [1.5ex]
\hline
\hline
\end{tabular}
\label{tab:Wgamma_xsection}
}
\end{table}

\begin{figure}[h]
\begin{center}
\includegraphics[height=0.54\textwidth, width=0.48\textwidth]{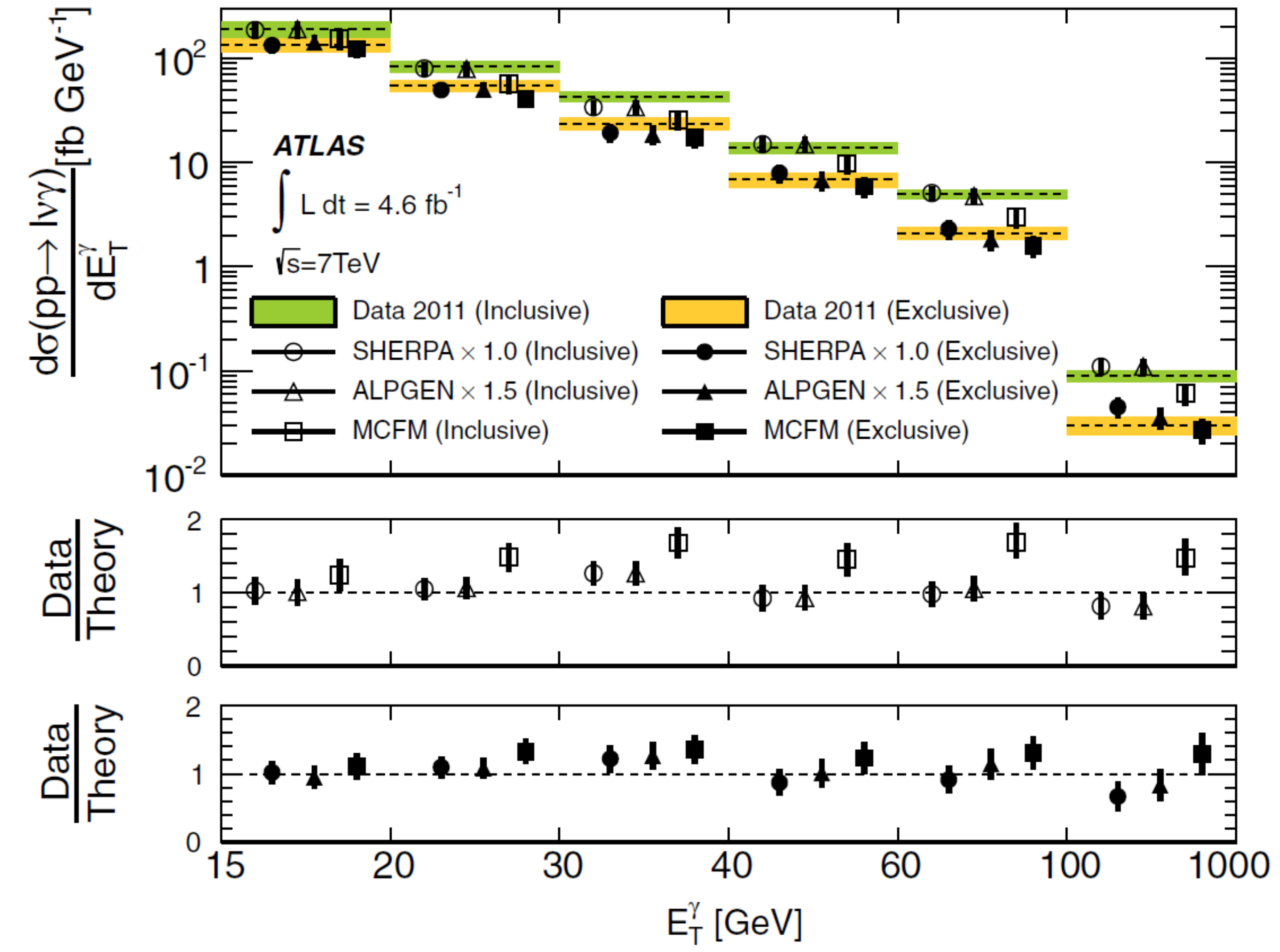}
\includegraphics[height=0.54\textwidth, width=0.48\textwidth]{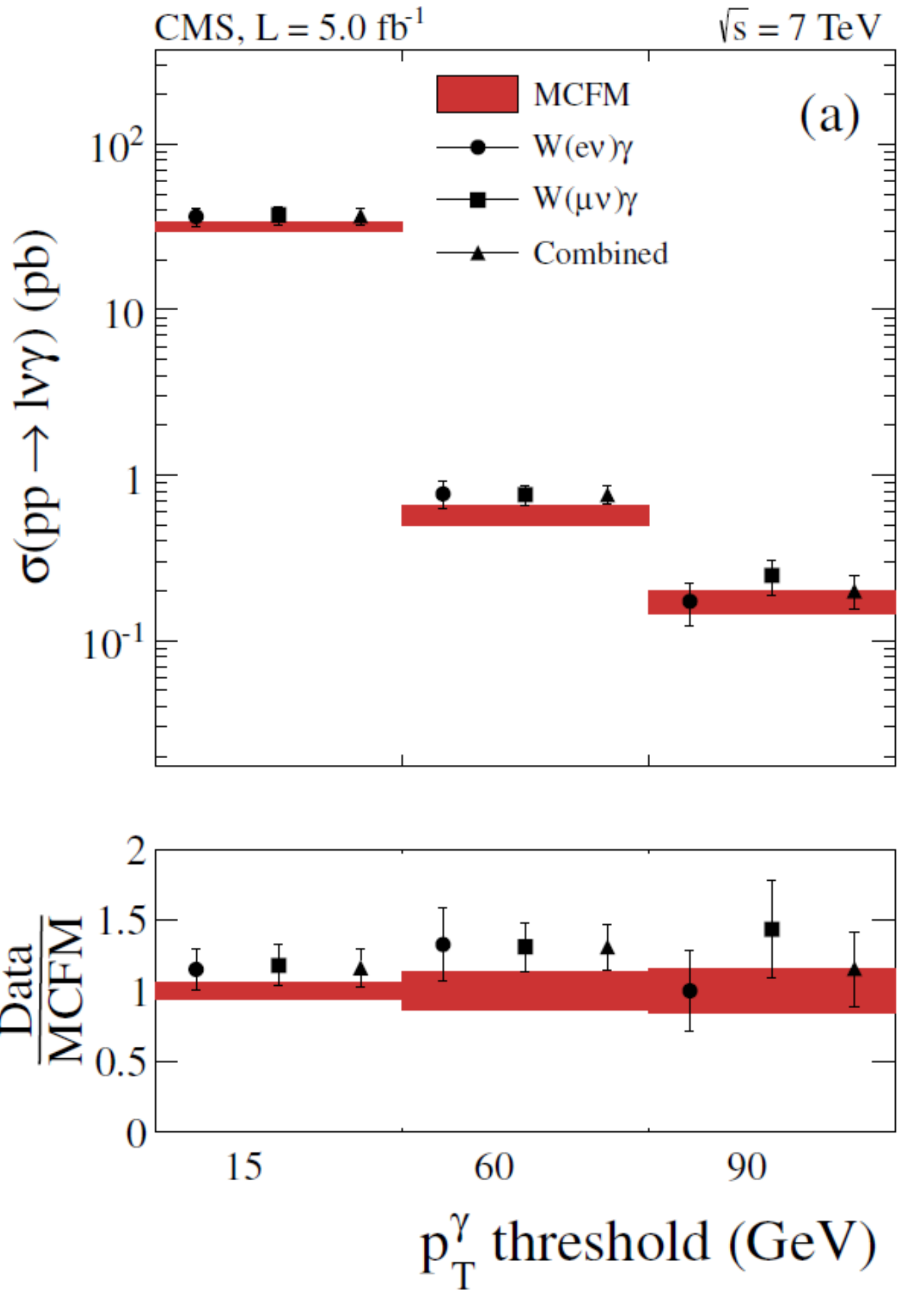}
\caption{\label{fig:ATLAS_WGamma_Unfolded} Left: Measured photon $E_T$
  differential cross section of the $W\gamma \rightarrow
  \ell\nu\gamma$ process, using combined electron and muon
  measurements in the inclusive ($N_{jet} \ge 0$) and exclusive
  ($N_{jet}=0$) fiducial regions. Right: Measured $W\gamma$ cross
  sections for three photon $E_T$ thresholds, compared to SM
  predictions.}
\end{center}
\end{figure}

ATLAS also measured the differential cross section as a function of
the photon $E_T$ using combined electron and muon measurements in the
inclusive ($N_{jet} \ge 0$) and exclusive ($N_{jet}=0$) fiducial
regions. CMS measured the $W\gamma$ production cross section for three
different photon $E_T$ thresholds (15 GeV, 60 GeV, and 90 GeV).  The
measured cross sections are compared with several theoretical
predictions as shown in Fig.~\ref{fig:ATLAS_WGamma_Unfolded}.

\subsubsection{Derived Results and Discussion}
Since ATLAS and CMS measured the $W\gamma$ cross section in different
phase spaces we adopt the extended fiducial region defined by the CMS
collaboration in order to combine the cross sections. The CMS
predicted theoretical cross section number is used and the ATLAS
measured inclusive cross section number is scaled up by $16.2\pm0.4$.  The
statistical uncertainties for both measurements are treated
independently. The uncertainties on the integrated luminosity are
treated as fully correlated. Other systematic uncertainties are
dominated by uncertainties related to the $W+$jets and $Z+$jets
data-driven background estimation, lepton, photon and
$E_T^{\text{miss}}$ identification, and energy and resolution. These
uncertainties are treated to be completely uncorrelated between the
two experiments. Using the BLUE method, the combined inclusive cross
section is found to be 
\[ \sigma(pp \rightarrow W\gamma) = 39.6 \pm 0.6 (stat.) \pm 3.5 (syst.)~{\text{fb}}, \] 
which is consistent with a theoretical prediction of $31.8 \pm 1.8$ fb.


\section{\label{sec:zzgammavertex}Studies of the $ZZ^{(*)}$ and $Z\gamma$ final states}

In contrast to the decay channels discussed in the
previous section, the $ZZ^{(*)}$ and $Z\gamma$ final states can occur only
via $t$- and $u$-channels within the SM. The study of their
production could therefore test the existence of an anomalous $ZZZ$,
$ZZ\gamma$ or $Z\gamma\gamma$ vertex. Similar to the previous
measurements, the event selection, the background estimation methods
and results of both experiments for the production cross section
measurements are summarized in this section. We also give
special emphasis on the differences between the two experiments and
derive combinations of the measured cross sections.

\subsection{$ZZ^{(*)}$ Analysis}
In the SM, non-resonant $ZZ^{(*)}$ production proceeds via the $t$-
and $u$-channel $q\bar{q}$ interactions and via gluon-gluon fusion.
Both ATLAS and CMS collaborations have measured the $ZZ$ production
cross section using events that are consistent with two $Z$ bosons
decaying to electrons or muons ($ZZ \rightarrow \ell\ell\ell'\ell'$
with $\ell, \ell'=e, \mu$) \cite{Aad:2012awa, Chatrchyan:2012sga}. ATLAS also lowered the mass requirement
for the second dilepton pair to measure the production cross section
for the $ZZ^* \rightarrow \ell\ell\ell'\ell'$ process \cite{Aad:2012awa}. In addition,
CMS measured the production cross section for events with one $Z$
boson decaying to $e^+e^-$ or $\mu^+\mu^-$ and a second $Z$ boson
decaying to $\tau^+\tau^-$ in four possible final states:
$\tau_h\tau_h$, $\tau_e\tau_h$, $\tau_\mu\tau_h$ and $\tau_e\tau_\mu$,
where $\tau_h$ represents a $\tau$ decaying hadronically, while
$\tau_e$ and $\tau_\mu$ indicate taus decaying into an electron or a
muon, respectively\cite{Chatrchyan:2012sga}.  The ATLAS collaboration also measured the $ZZ
\rightarrow \ell \ell \nu \nu$ production cross section using events
with one $Z$ boson decaying to electrons or muons and a second $Z$
boson decaying to neutrinos \cite{Aad:2012awa}.

\subsubsection{Event Selection}
\begin{figure}[bt]
    \begin{center}
        \includegraphics[width=1.0  \textwidth]{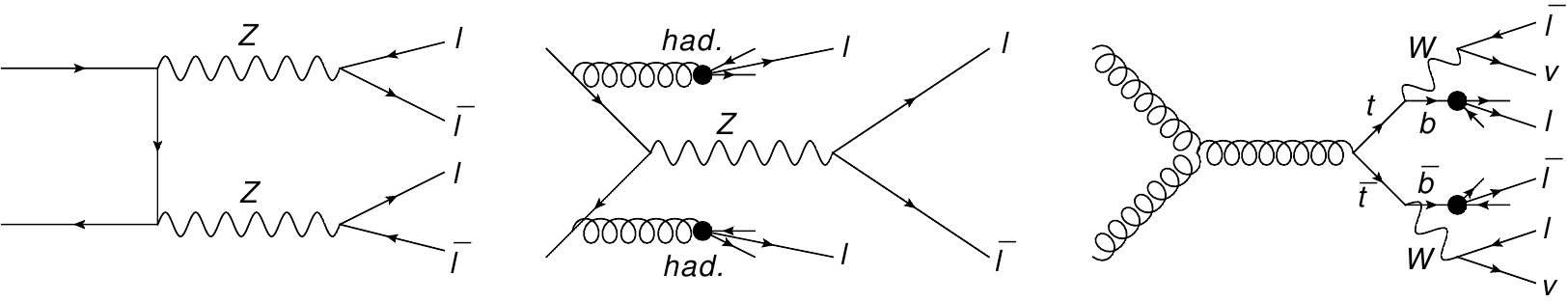}
        \caption{Feynman diagrams for the $t$-channel $ZZ$ signal
          process (left) and dominant background processes (middle:
          $Z$+jets, right: $t\bar t$ pair production).}
        \label{Fig:ZZLOFeynman}
    \end{center}
\end{figure}

The presence of four high-$p_T$ leptons in the final state for the
$ZZ^{(*)} \rightarrow \ell\ell\ell'\ell'$ process provides a clean
signature with only a small contribution from other SM background
processes (see Fig. \ref{Fig:ZZLOFeynman}).  Loose lepton selection
criteria are thus applied to maximize the overall signal acceptance
times efficiency.
The $ZZ \rightarrow \ell\ell\nu\nu$ process is characterized by large
\ETMiss\ and two high-$p_T$ isolated electrons or muons.  It has a
branching ratio six times larger than the purely leptonic decay processes
$ZZ^{(*)} \rightarrow \ell\ell\ell'\ell'$. However larger SM
background from the $Z+$jets process is expected and thus tighter
selection criteria are used.

For the ATLAS $ZZ^{(*)} \rightarrow \ell\ell\ell'\ell'$ selection,
highly energetic electrons are required (i.e. $E_T>20$ GeV) 
satisfying the loose identification criterion, while muons are
required only to fulfill a rather loose momentum cut of $p_T>7$ GeV.
All selected leptons must satisfy isolation requirements based on
calorimetric and tracking information and must be consistent with
originating from the primary vertex.
Forward spectrometer and calorimeter-tagged muons as well as
calorimeter-only electrons passing the tight identification
requirements are also considered in order to increase the overall
signal acceptance.

Selected events are required to have at least one highly energetic
electron (muon) with $p_T>25 (20)$ GeV and to contain two pairs of
same-flavor, oppositely-charged leptons.  In the $4e$ and $4\mu$
channels, ambiguities are resolved by choosing the
combination which produces the smallest sum of relative differences
between the dilepton mass and the $Z$ boson mass from each pair.  At
least one lepton pair is required to have an invariant mass within the
$Z$ mass window ($66<m_{\ell\ell}<116$ GeV).  If the second $Z$
candidate has $66<m_{\ell\ell}<116$ GeV the event is classified as a
$ZZ$ event; if the second candidate satisfies $m_{\ell\ell}>20$ GeV,
the event is classified as a $ZZ^*$ event.

For the CMS $ZZ \rightarrow \ell\ell\ell'\ell'$ selection electrons
are selected with a threshold of 7 GeV and muons are selected with a
threshold of 5 GeV. Events are required to have at least one $Z
\rightarrow \ell\ell$ candidate with $60<m_{\ell\ell}<120$ GeV.
The leading lepton must have $p_T>20$ GeV.  Lepton isolation
requirements depend on the $ZZ$ decay mode.

The $\tau$ leptons are required to fulfill the kinematic requirements
of $p_T>20$ GeV and $|\eta|<2.3$. Since hadronically decaying $\tau$
leptons have much larger misidentification rates than the other
leptons, tighter electron and muon isolation criteria are applied for
the electrons and muons in the final states $\tau_e\tau_h$ and $\tau_\mu \tau_h$.

For the ATLAS $ZZ \rightarrow \ell\ell\nu\nu$ selection, 
leptons are required to have $p_T>20$ GeV. Selected candidate events
are required to have exactly two same-flavor leptons with
$76<m_{\ell\ell}<116$ GeV.  Events with a third electron or muon with
$p_T>10$ GeV are rejected to reduce the $WZ \rightarrow
\ell\nu\ell'\ell'$ background.  Events containing at least one
well-reconstructed jet with $p_T>25$ GeV
are vetoed to reduce top quark background.  To remove the dominant
$Z+$jets background, a minimal cut on the so-called
axial-$E_T^{\text{miss}}$ of 75\,\GeV is
applied \footnote{Axial-$E_T^{\text{miss}}$ is defined as the
  projection of the $E_T^{\text{miss}}$ along the direction opposite
  to the $Z$ candidate direction in the transverse plane.} .
The two $Z$ bosons from the $pp \rightarrow ZZ \rightarrow
\ell\ell\nu\nu$ process tend to have similar transverse momenta. To
exploit this tendency, the fractional difference between
$E_T^{\text{miss}}$ and $p_T^Z$, $|E_{T}^{\text{miss}} -
p_T^Z|/p_T^Z$, is required to be less than 0.4.

The details of the selection criteria used for all $ZZ^{(*)}$ analyses
are summarized in Tab.~\ref{tab:ZZ_selection}.

\begin{table}[h]
\tbl{Summary of $ZZ^{(*)}$ selection cuts used by the ATLAS and CMS
  collaborations.}  {
\begin{tabular}{lll}
\hline
\hline
        & ATLAS & CMS \\
\hline
$ZZ^{(*)} \rightarrow $    					&  Combined, forward or 							& Muons with $p_T>5$ GeV, $|\eta|<2.4$\\
$\ell^+\ell^-\ell'^+\ell'^-$  					&  calorimeter-tagged muons   						& Electrons with $E_T>7$ GeV, $|\eta|<2.5$\\ 
$(\ell, \ell'=e, \mu)$ 	                            			&  Combined or calorimeter-only electrons 			& $\ge$ one lepton with $p_T>20$ GeV\\ 
                                                               			& $\ge$ one lepton with $p_T^{e(\mu)}>25 (20)$ GeV 	& First $Z$ with $60<m_{\ell\ell}<120$ GeV \\
                                                               			& First $Z$ with $66<m_{\ell\ell}<116$ GeV 			& Second $Z$ with $60<m_{\ell\ell}<120$ GeV \\
                                                               			& Second $Z$ with $66<m_{\ell\ell}<116$ GeV ($ZZ$) 	&  \\
                                                               			& Second $Z$ with $m_{\ell\ell}>20$ GeV ($ZZ^*$) 		& \\
\hline
$ZZ \rightarrow$ 				   		&  &  $\ge$ one lepton from $Z$ with \\
$\ell^+\ell^-\tau^+\tau^-$    				&  &  $p_T^{e(\mu)}>20$ GeV  \\
$(\ell =e, \mu)$                                     			&  & $\tau_{e, \mu}$ lepton with $p_T^{e(\mu)}>10$ GeV \\
                                                              			&  & $\tau_h$ lepton with $p_T>20$ GeV\\
                                                              			&  & and $|\eta|<2.3$ \\
                                                              			&  & $30<m_{\tau\tau}^{vis}<80$ GeV for $\tau_h\tau_{e, \mu}$ \\
                                                              			&  & $30<m_{\tau\tau}^{vis}<90$ GeV for $\tau_e\tau_{\mu}$ \\
\hline
$ZZ \rightarrow $   &  Combined muons (electrons)  & \\
$\ell^+\ell^-\nu\nu$                          & $76<m_{\ell^+\ell^-}<106$ GeV & \\
 $(\ell=e, \mu)$				& Axial-$\met>75$ GeV & \\
                                                   & $|\met-p_T^Z|/p_T^Z<0.4$ & \\
                                                   & Third lepton veto, jet veto & \\
\hline
\hline
\end{tabular}
\label{tab:ZZ_selection}
}
\end{table}

\subsubsection{Background Estimation}
For the $ZZ \rightarrow \ell\ell\ell'\ell'$ channel, the major
contributions to the background are due to $Z$ and $WZ$ production in
association with jets or $t\bar{t}$. In all of these cases a jet or
non-isolated lepton is misidentified as an isolated lepton.  The
relative contribution of each background source depends on the final
state. The rate for loosely-isolated objects to be misidentified as
isolated ones is estimated with events with one selected $Z$ boson and
an additional probe lepton. No isolation requirement is applied to the
probe lepton. The misidentification rate is defined as the ratio of
the number of probe candidates that pass the isolation requirements to
the initial number of probe candidates. This rate is then applied to
events which pass all selection requirements but requiring the probe
candidate not to be isolated to estimate the number of background
events in the signal region.  The uncertainties on $Z+$jets, $WZ+$jets
and $t\bar{t}$ backgrounds reflect the uncertainties of the
misidentification rates and the limited quantity of data in the
control region, and amount to $30-50\%$ depending on the decay
channel.

For the $ZZ \rightarrow \ell\ell\nu\nu$ channel, the major
contributions to the background are due to $Z \rightarrow \tau\tau$,
$WW$, $WZ$, $Wt$ and $t\bar{t}$ production that result in two true
isolated leptons with missing transverse energy. Production of a $Z$
(or $W$) boson in association with jets can also give similar final
states if the jet momenta are mismeasured or if one jet is
misidentified as an isolated lepton.  The contributions from $Z
\rightarrow \tau\tau$, $Wt$, $WW$ and $t\bar{t}$ processes are
measured by extrapolating from a control sample of events with one
electron and one muon which otherwise satisfy the full $ZZ \rightarrow
\ell\ell\nu\nu$ selection. The extrapolation from the $e\mu$ channel
to the $ee$ or $\mu\mu$ channel uses the relative branching fractions
as well as the ratio of the efficiencies $\epsilon_{ee}$ or
$\epsilon_{\mu\mu}$ of the $ee$ or $\mu\mu$ selections to the
efficiency $\epsilon_{e\mu}$ of the $e\mu$ selection, which differs
from unity due to differences in the electron and muon
efficiencies. The contribution from the $WZ$ process is estimated
using the simulated samples and checked using a control region with
three high-$p_T$ isolated leptons. The contribution from the $Z+$jets
process is estimated using events with a high-$p_T$ photon and jets as
a template, since the mechanism for large missing transverse energy is
the same as in $Z+$jets events.  The events are reweighted such that
the photon $p_T$ matches the observed $Z$ boson $p_T$ and are
normalized to the observed $Z+$jets yield. The procedure is repeated
in bins of $Z$ boson $p_T$ in order to obtain the $p_T$ distribution
of the $Z+$jets backgrounds. The background from processes in which
one of the two leptons comes from the decays of a $W$ or $Z$ boson and
the second lepton from non-prompt leptons or misidentified mesons or
conversions is found to be negligible.

\subsubsection{Inclusive and Fiducial Cross Section Results}
\begin{figure}[h]
\begin{center}
\resizebox{0.48\textwidth}{!}{\includegraphics{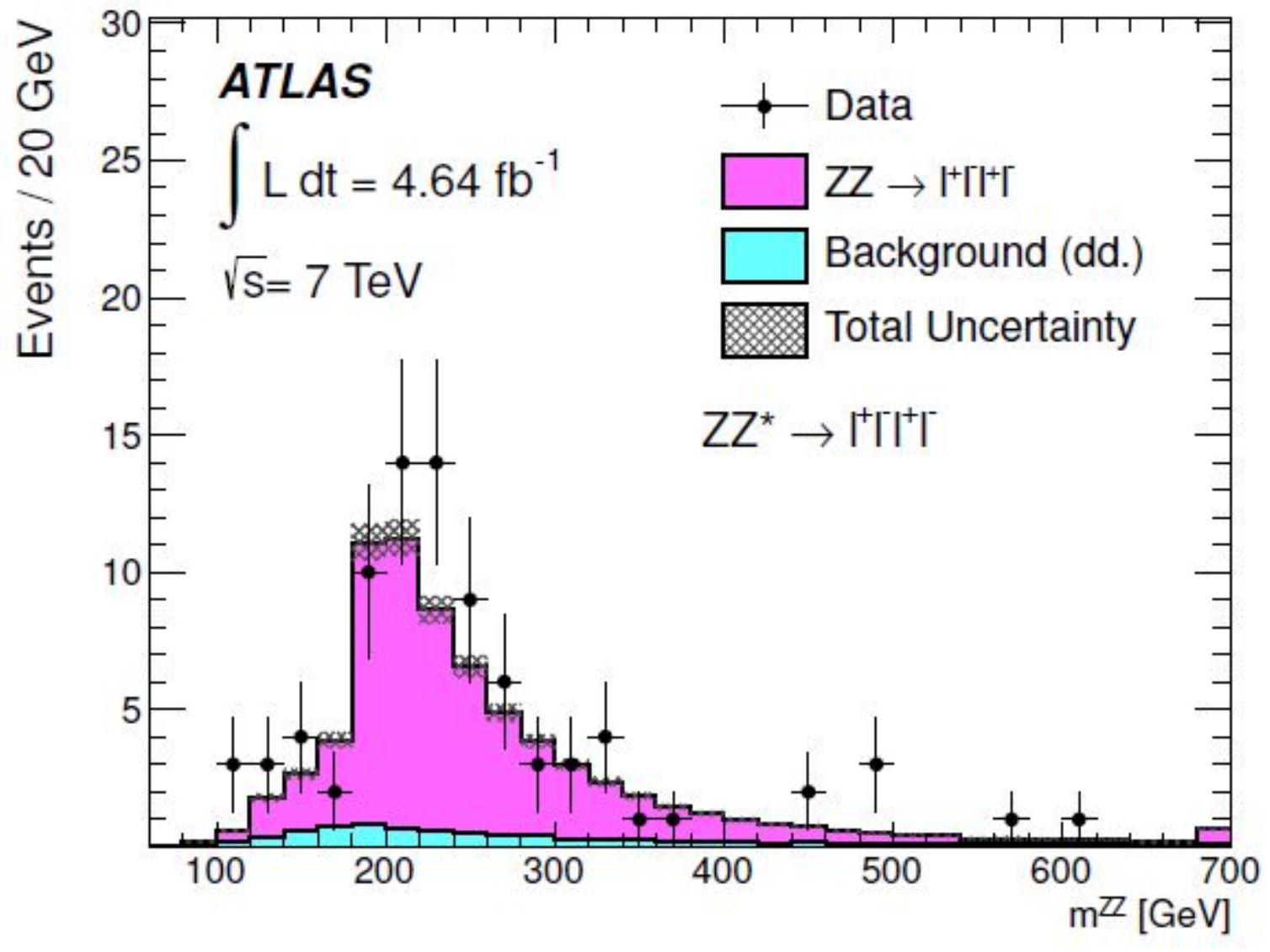}}
\resizebox{0.48\textwidth}{!}{\includegraphics{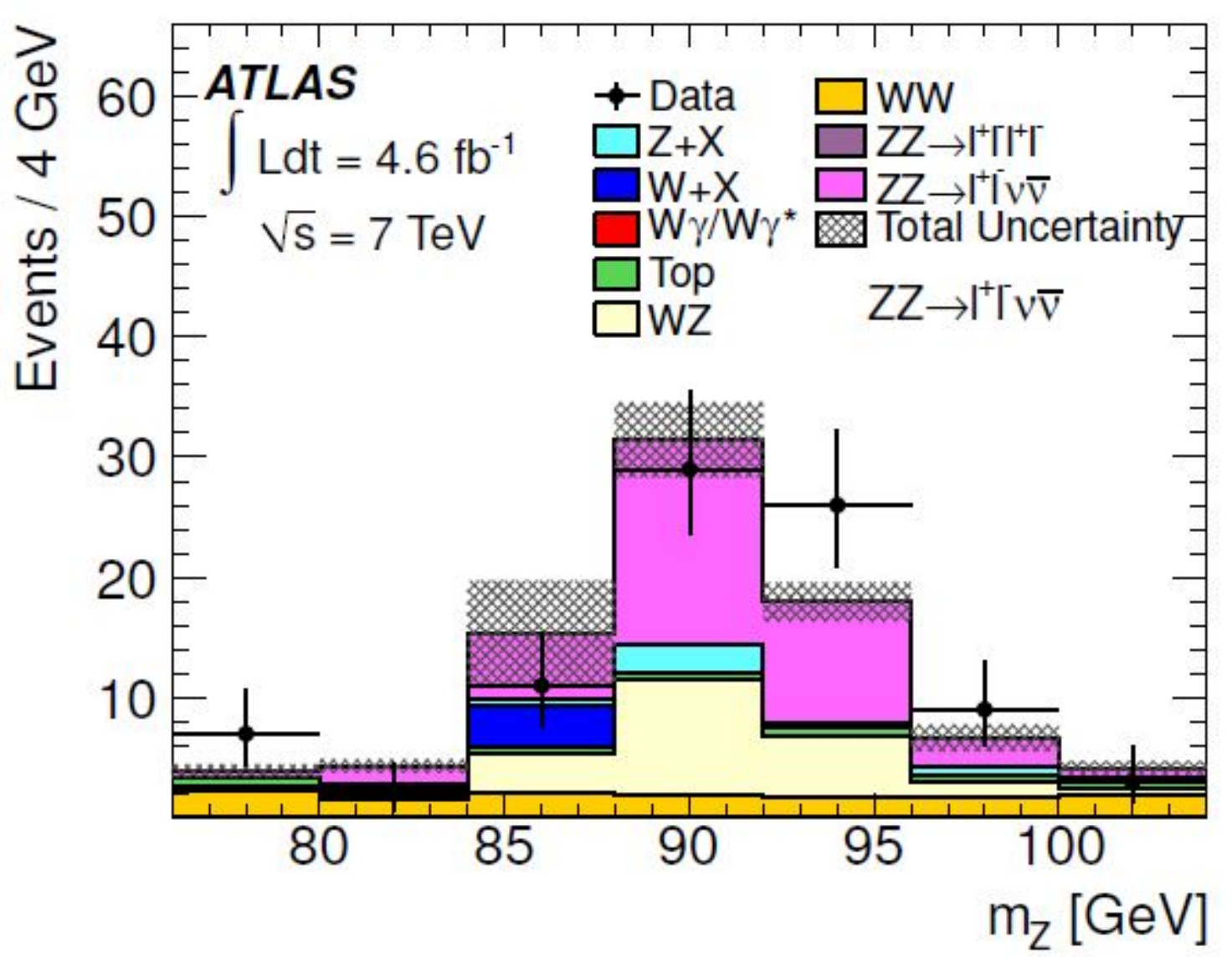}}
\caption{\label{fig:ATLAS_ZZ} Invariant mass $m^{ZZ}$ of the
  four-lepton system for the $ZZ^{(*)} \rightarrow \ell\ell\ell'\ell'$
  selection (Left) and mass $m_Z$ of the two-charged-lepton system for
  the $ZZ \rightarrow \ell\ell\nu\nu$ selection for the ATLAS
  analysis.}
\end{center}
\end{figure}

\begin{figure}[h]
\begin{center}
\resizebox{0.96\textwidth}{!}{\includegraphics{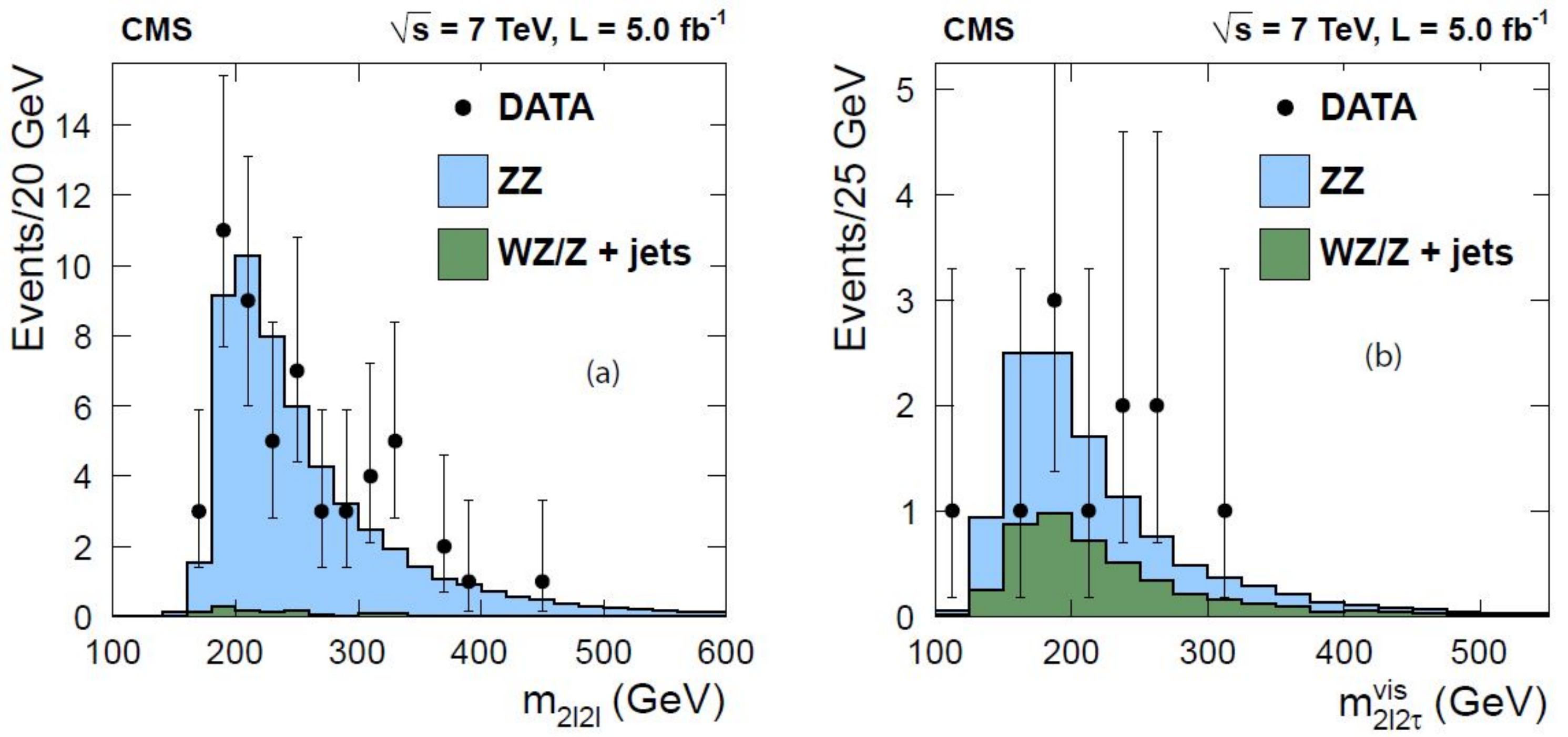}}
\caption{\label{fig:CMS_ZZ} Distributions of the four-lepton
  reconstructed mass for the sum of the $4e$, $4\mu$, and $2e2\mu$
  channels (Left) and the sum of the $2\ell 2\tau$ channels (Right)
  for the CMS analysis.}
\end{center}
\end{figure}

The expected and observed event yields after applying all selection
criteria are shown in Tab.~\ref{tab:ZZ} for all decay channels.  Good
agreement between data and SM expectations can be observed in all decay channels.
Figure~\ref{fig:ATLAS_ZZ} shows the invariant mass of the four-lepton
system for the $ZZ^* \rightarrow \ell\ell\ell'\ell'$ selection and
the mass of the two-charged-lepton system for the $ZZ \rightarrow
\ell\ell\nu\nu$ selection.  Distributions of the four-lepton
reconstructed mass for the sum of the $4e$, $4\mu$, and $2e2\mu$
channels and the sum of the $2\ell 2\tau$ channels for the CMS
analysis are shown in Fig.~\ref{fig:CMS_ZZ}.

\begin{table}[h]
\tbl{Summary of observed $ZZ \rightarrow \ell\ell\ell'\ell'$, $ZZ^*
  \rightarrow \ell\ell\ell'\ell'$, $ZZ \rightarrow \ell\ell\nu\nu$ and
  $ZZ \rightarrow \ell\ell\tau\tau$ candidates in the data, total
  background estimates and expected signal for the individual decay
  modes. The first uncertainty is statistical while the second is
  systematic. The uncertainty on the integrated luminosity (3.9\%) is
  not included for ATLAS numbers.}  {
\begin{tabular}{lccccr}
\hline
\hline
        									& Expected signal 			& Background  			& Observed \\
\hline
ATLAS & & & \\
$~~~~$ $ZZ \rightarrow \ell\ell\ell'\ell'$		& $80.9 \pm 1.1 \pm 0.7$ 		& $0.9 \pm 1.1 \pm 0.7$  & 66 \\ 
$~~~~$ $ZZ^* \rightarrow \ell\ell\ell'\ell'$		&  $64.4 \pm 0.4 \pm 4.0$  	&  $9.1 \pm 2.3 \pm 1.3$ & 84 \\
$~~~~$ $ZZ \rightarrow \ell\ell\nu\nu$                & $39.3 \pm 0.4 \pm 3.7$		&$46.9 \pm 4.8 \pm 1.9$ 	& 87 \\
 \hline
CMS & & & \\
$~~~~$ $ZZ \rightarrow \ell\ell\ell'\ell'$		& $53.15 \pm 0.12 \pm 2.96$ 	& $1.35 \pm 0.34 \pm 0.35$  & 54 \\
$~~~~$ $ZZ \rightarrow \ell\ell\tau\tau$		& $7.05 \pm 0.04 \pm 0.20$ 	& $4.37 \pm 0.80 \pm 0.29$  & 11 \\
\hline
\hline
\end{tabular}
\label{tab:ZZ}
}
\end{table}

To include all the final states in the calculation of the cross
section, a simultaneous fit to the numbers of observed events in all
the decay channels is performed.  The fit is constrained by the
requirement that all the measurements come from the same initial state
via different decay modes.

The joint likelihood is a combination of the likelihoods for the
individual channels, which include the signal and background
hypotheses.  The resulting combined cross sections are compared with
theoretical predictions in Tab.~\ref{tab:ZZ_xsection}.  The integrated
fiducial $ZZ^{(*)}$ cross sections for different final states are also
listed from the ATLAS collaboration.  The predicted cross
sections are calculated with {\sc powhegbox} and {\sc gg2zz} \cite{Binoth:2006mf} for both
$Z$ bosons in the mass range $66<m_Z<116$ GeV. The CMS predictions are
calculated with MCFM at NLO for $q\bar{q} \rightarrow ZZ$ and LO for
$gg \rightarrow ZZ$ for both $Z$ bosons in the mass range $60<m_Z<120$
GeV.


\begin{table}[h]
\tbl{Summary of measured and predicted inclusive and fiducial cross
  sections for the $ZZ^{(*)}$ process from the ATLAS and CMS
  collaborations.}  {
\begin{tabular}{lll}
\hline
\hline
        & Measured (fb)& Predicted (fb) \\
\hline
ATLAS & & \\
$~~~~$ $\sigma^{fid}_{ZZ \rightarrow \ell^+\ell^-\ell'^+\ell'^-}$    &  $25.4 ^{+3.3}_{-3.0}$ (stat.) $^{+1.2}_{-1.0}$ (syst.) $\pm 1.0$ (lumi.)  & $20.9 \pm 0.1$ (stat.) $^{+1.1}_{-0.9}$ (theory) \\ [1.5ex]
$~~~~$ $\sigma^{fid}_{ZZ^* \rightarrow \ell^+\ell^-\ell'^+\ell'^-}$   &  $29.8 ^{+3.8}_{-3.5}$ (stat.) $^{+1.7}_{-1.5}$ (syst.) $\pm 1.2$ (lumi.)  & $25.6 \pm 0.1$ (stat.) $^{+1.3}_{-1.1}$ (theory) \\ [1.5ex]
$~~~~$ $\sigma^{fid}_{ZZ \rightarrow \ell^+\ell^-\nu\nu}$  &  $12.7 ^{+3.1}_{-2.9}$ (stat.) $\pm 1.7$ (syst.) $\pm 0.5$ (lumi.)  & $12.5 \pm 0.1$ (stat.) $^{+1.0}_{-1.1}$
 (theory) \\ [1.5ex]
$~~~~$ $\sigma^{tot}_{ZZ}$   &  $6700 \pm 700$ (stat.) $^{+400}_{-300}$ (syst.) $\pm 300$ (lumi.)  & $5890$ $^{+220}_{-180}$ \\ [1.5ex]
CMS & & \\
$~~~~$ $\sigma^{tot}_{ZZ}$  &  $6240 ^{+860}_{-800}$ (stat.) $^{+410}_{-320}$ (syst.) $\pm 140$ (lumi.)  & $6300 \pm 400$ \\ [1.5ex]
\hline
\hline
\end{tabular}
\label{tab:ZZ_xsection}
}  
\end{table}

\begin{figure}[h]
\begin{center}
\resizebox{0.48\textwidth}{!}{\includegraphics{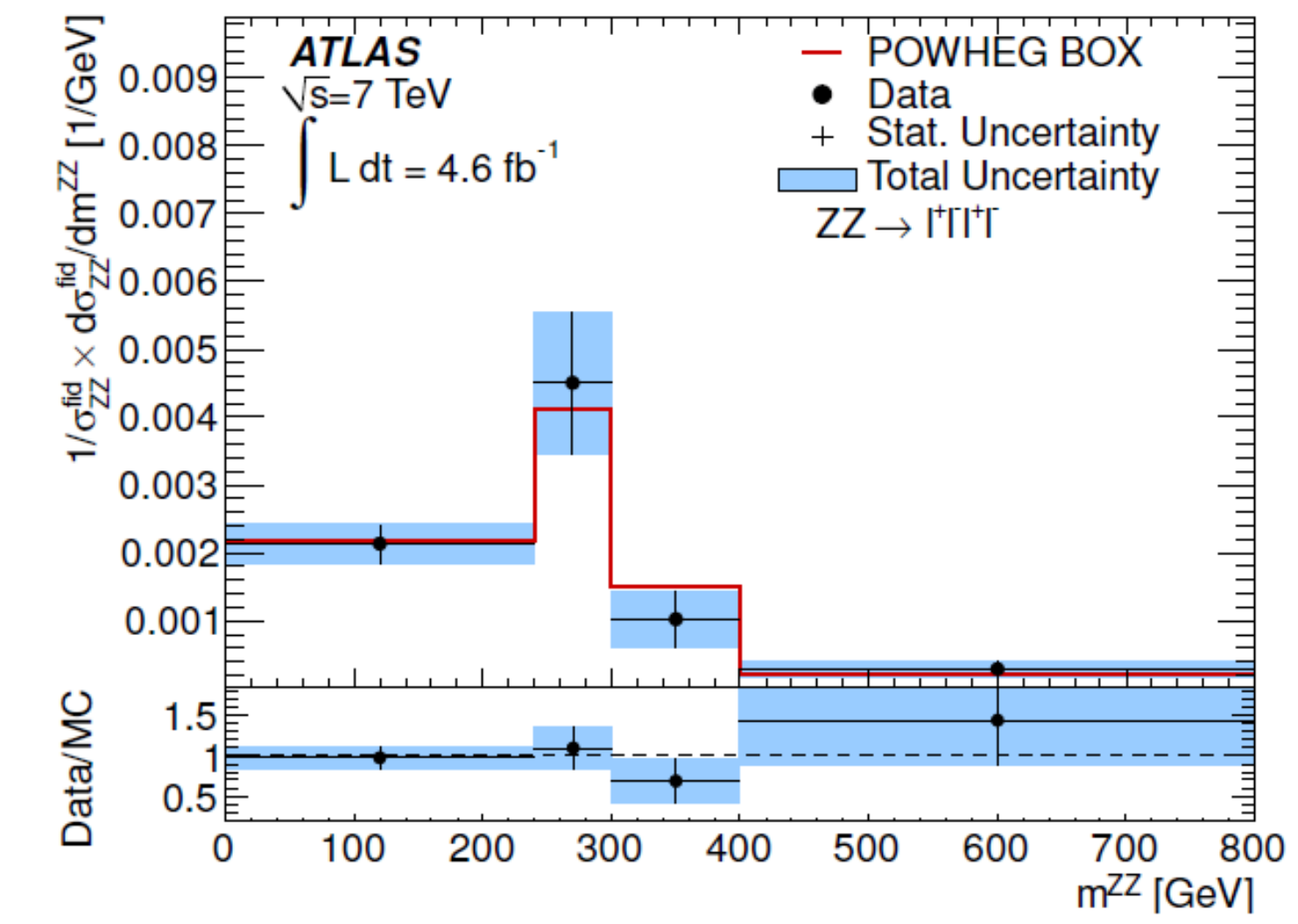}}
\resizebox{0.48\textwidth}{!}{\includegraphics{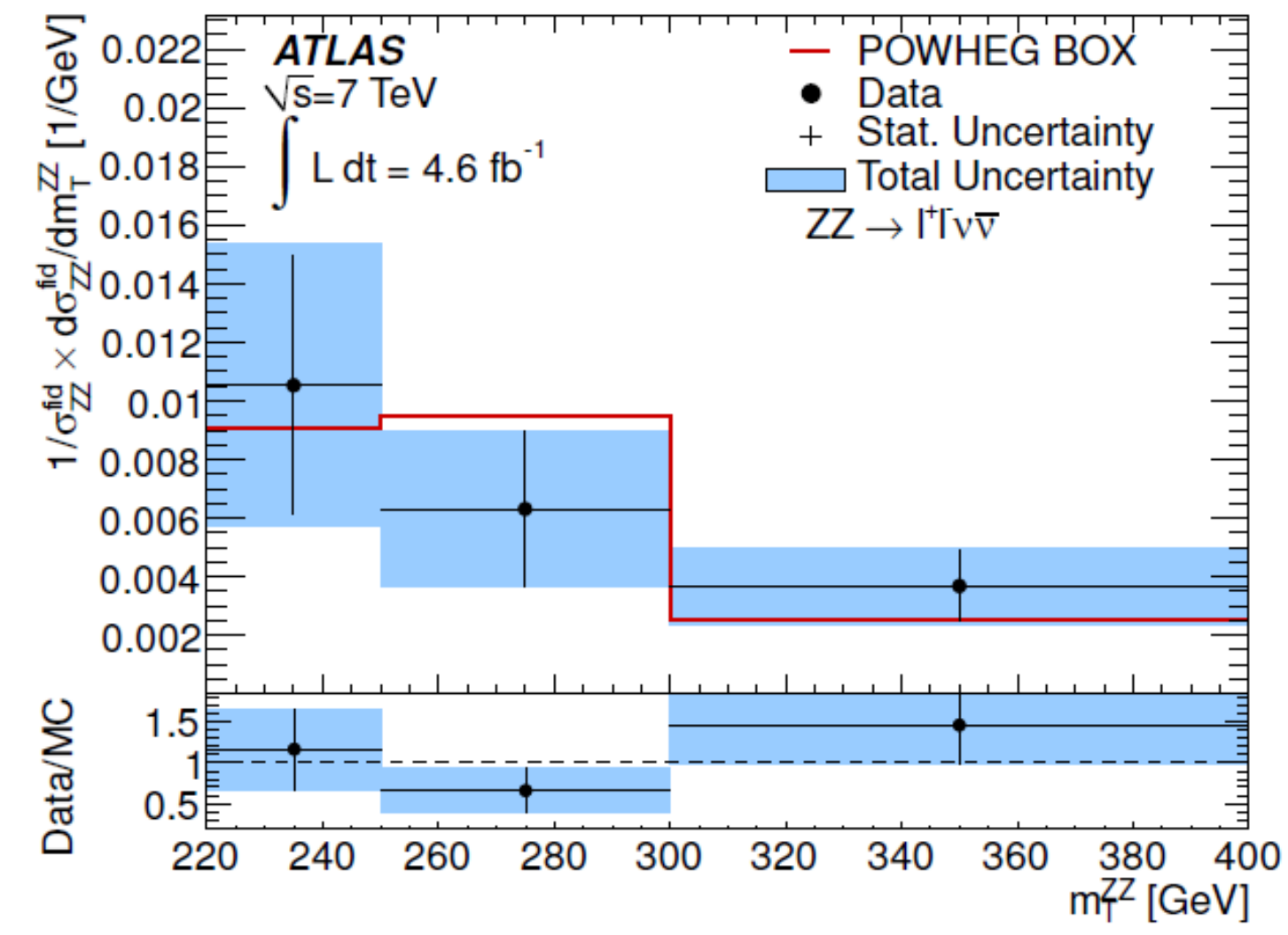}}
\caption{\label{fig:ZZ_unfolded} Unfolded $ZZ$ fiducial cross sections
  in bins of $m_{ZZ}$ for the $ZZ \rightarrow \ell\ell\ell'\ell'$
  selection and $m_T^{ZZ}$ for the $ZZ \rightarrow \ell\ell\nu\nu$
  selection.}
\end{center}
\end{figure}

ATLAS also measured the unfolded differential cross section as a
function of the four-lepton invariant mass $m_{ZZ}$ of the $ZZ
\rightarrow \ell\ell\ell'\ell'$ process and transverse mass $m_T^{ZZ}$
of the $ZZ \rightarrow \ell\ell\nu\nu$ process. The results are shown
in Fig.~\ref{fig:ZZ_unfolded}.

\subsubsection{Derived Results and Discussion}
Since ATLAS and CMS used different $Z$ mass requirements for the final
reported inclusive cross section measurement, for the combination, we
adapt the ATLAS result to the phase space defined by the CMS
collaboration (both $Z$ bosons must have $60<m_Z<120$ GeV).  The CMS
predicted theoretical cross section number is used and the ATLAS
measured inclusive cross section number is scaled up by 7\%. The
overall uncertainty is dominated by statistical uncertainty for both
measurements which are treated independently. The uncertainties on the
integrated luminosity are fully correlated between the two
experiments. Other major contributors to systematic uncertainties are
those due to lepton and jet identification, jet-energy and jet-resolution, and
also in the $Z+$jets data-driven background estimation. These
uncertainties are treated as uncorrelated between the two
experiments. Using the BLUE method, the combined inclusive cross
section is found to be
\[ \sigma(pp \rightarrow ZZ) = 6727 \pm 682 (stat.) \pm 400 (syst.) ~{\text{fb}},\] 
which is consistent with a theoretical prediction of $6300 \pm 400$
fb.  The fiducial cross section measurements for $ZZ^{(*)} \rightarrow
\ell\ell\ell'\ell'$ and $ZZ \rightarrow \ell\ell\nu\nu$ are not
combined into a common fiducial phase space due to their very different
detector signatures.

\subsection{$Z\gamma$ Analysis}
In the SM, the $Z$ boson cannot directly emit a photon. Hence the study of the $pp \rightarrow Z\gamma$ process allows a precise test of this 
prediction. Both ATLAS and CMS collaborations have measured the
production cross section in the $Z \rightarrow \ell\ell (\ell=e, \mu)$
and $Z \rightarrow \nu\bar{\nu}$ decay channels \cite{Aad:2012mr, Chatrchyan:2013fya}.

\subsubsection{Event Selection}
For the ATLAS $Z\gamma \rightarrow \ell\ell\gamma$ analysis, lepton
and photon identification criteria are exactly the same as used in the
$W\gamma$ analysis. The $Z\gamma$ candidates are selected by requiring
exactly two oppositely charged same-flavor leptons with an invariant
mass greater than 40 GeV and one isolated photon with $E_T>15$
GeV. The $\nu\bar{\nu}$ candidates are selected by requiring one
isolated photon with $E_T>100$ GeV and $E_T^{\text{miss}}>90$ GeV. The
reconstructed photon, $E_T^{\text{miss}}$, and jets are required to be
well separated in the transverse plane in order to reduce the
$\gamma+$ jet background. Events with identified electrons and muons
are vetoed to reject $W+$ jets and $W\gamma$ background. The selection
criteria to identify the electrons and muons are the same as in the
$Z(\ell\ell)\gamma$ analysis.
ATLAS also measured the cross sections for events with and without the
requirement of zero jets in each event. Jets are required to have
$E_T>30$ GeV and $|\eta|<4.4$.

Similar event selection criteria are used for the CMS $Z\gamma
\rightarrow \ell\ell\gamma$ analysis except lepton candidates are
required to have $p_T>20$ GeV and the invariant mass of the two
leptons is required to satisfy $m_{\ell\ell}>50$ GeV.  The
$\nu\bar{\nu}$ candidates are selected by requiring one isolated
photon with $E_T>145$ GeV and $E_T^{\text{miss}}>130$ GeV.  Only
photons reconstructed in the barrel EM calorimeter region
($|\eta|<1.4$) are used.  Events are vetoed if they contain other
particles of significant energy or momentum.

\begin{table}[h]
\tbl{Summary of the $Z\gamma \rightarrow \ell\ell\gamma$ and $Z\gamma
  \rightarrow \nu\nu\gamma$ selection cuts used by the ATLAS and CMS
  collaborations.}  {
\begin{tabular}{lll}
\hline
\hline
        & ATLAS & CMS \\
\hline
$Z\gamma \rightarrow \ell \ell \gamma$    		&  Combined muons with $p_T>25$ GeV  				& Muons with $p_T>20$ GeV\\
$(\ell'=e, \mu)$                                           		&  Combined electrons with $E_T>25$ GeV 				& Electrons with $p_T>20$ GeV \\
                                                                    		& Photon $E_T>15$ GeV and 						 	& Photon $E_T>15$ GeV, isolation and \\ 
                                                                    		&  $~~$ $E_T^{\Delta R<0.3}<6$ GeV					& $~~$ shower-shape requirements  \\
                                                                    		& $m_{\ell\ell}>40$ GeV 								& $m_{\ell\ell}>50$ GeV \\
                                                                    		& $\Delta R(\ell, \gamma)>0.7$	 						& $\Delta R(\ell, \gamma)>0.7$  \\ 
\hline 
$Z\gamma \rightarrow \nu \nu \gamma$    	&  Photon $E_T>100$ GeV and						 &  Photon $E_T>145$ GeV, $|\eta|<1.4$,  isolation \\ 
                                                                    		&  $~~$  $E_T^{\Delta R<0.3}<6$ GeV					& $~~$ and shower-shape requirements  \\
                                                                    		& $E_T^{\text{miss}} > 90$ GeV 						& $E_T^{\text{miss}} > 130$ GeV \\
                                                                    		& $\Delta \phi(E_T^{\text{miss}}, \gamma) > 2.6$ and		  & \\
									& $~~$  $\Delta \phi(E_T^{\text{miss}}, jet)>0.4$			& \\
                                                                    		& Veto events with rec. leptons 						& Veto events with rec. leptons and jets \\
\hline
\hline
\end{tabular}
\label{tab:Zgamma_selection}
}
\end{table}

\subsubsection{Background Estimation}
The main background to the $\ell\ell\gamma$ signal originates from
events with $Z+$ jets where jets are misidentified as photons.
Background to the $\nu\bar{\nu}\gamma$ signal originates mainly from
$W(e\nu)$ events when the electron is misidentified as a photon,
$Z(\nu\bar{\nu})+$ jets and multijet events when one of the jets in
the event is misidentified as a photon, and $\tau\nu\gamma$ and
$\ell\nu\gamma$ events from $W\gamma$ production when the $\tau$
decays into hadrons or when the electron or muon from $\tau$ or $W$
decay is not reconstructed, and $\gamma+$ jets events when large
$E_T^{\text{miss}}$ is created by a combination of real
$E_T^{\text{miss}}$ from neutrinos in heavy quark decays and
mismeasured jet energy.

For the $\ell\ell\gamma$ analysis, ATLAS used a sideband method
similar to the one used for the $W\gamma$ analysis to estimate the
$Z+$ jets contamination.
For the $\nu\bar{\nu}\gamma$ analysis, the sideband method is also
used to estimate the $Z(\nu\bar{\nu})+$ jets and multijet
background. A similar method using the probability of electrons to be
misidentified as photons is used to derive the $W(e\nu)$
background. The $\gamma+$ jets contamination is estimated using a
semi-data-driven method. A sample is selected by applying all signal
region selection criteria except for requiring $\Delta
\phi(E_T^{\text{miss}}) < 0.4$. By requiring the $E_T^{\text{miss}}$
direction to be close to the jet direction, the selected events in the
control region are dominated by $\gamma+$ jets background. The yield
of $\gamma+$ jets events obtained in control regions is then scaled by an
extrapolation factor to predict the $\gamma+$ jets background yield in
the signal region, where the extrapolation factor is taken from a
$\gamma+$ jets MC sample. The $\tau\nu\gamma$ and $\ell\nu\gamma$
background is estimated using MC simulation.

For the $\ell\ell\gamma$ analysis, CMS also chose to perform a
template fit to the photon $\sigma_{\eta\eta}$ distribution in data to
determine the $Z+$ jets contamination.  For the $\nu\bar{\nu}\gamma$
analysis, the background from misidentified photons originating in jet
fragmentation and decay processes is estimated by constructing a
control data sample enriched with multijet events. This sample is then
used to calculate a misidentification ratio, defined as the number of
events where the photon candidate satisfies the signal selection
criteria to the number of events where the photon candidate satisfies
looser selection criteria but fails the isolation condition. The
background contribution due to misidentified jets is then estimated
from the multiplication of the misidentification ratio and the number
of events in the signal data sample that pass the photon selection
criteria but fail the isolation requirements. The $W(e\nu)$ background
is estimated using a similar method to that used by ATLAS. All other
backgrounds are estimated from MC simulation.

\subsection{Cross Section Results}
\begin{figure}[h]
\begin{center}
\resizebox{0.48\textwidth}{!}{\includegraphics{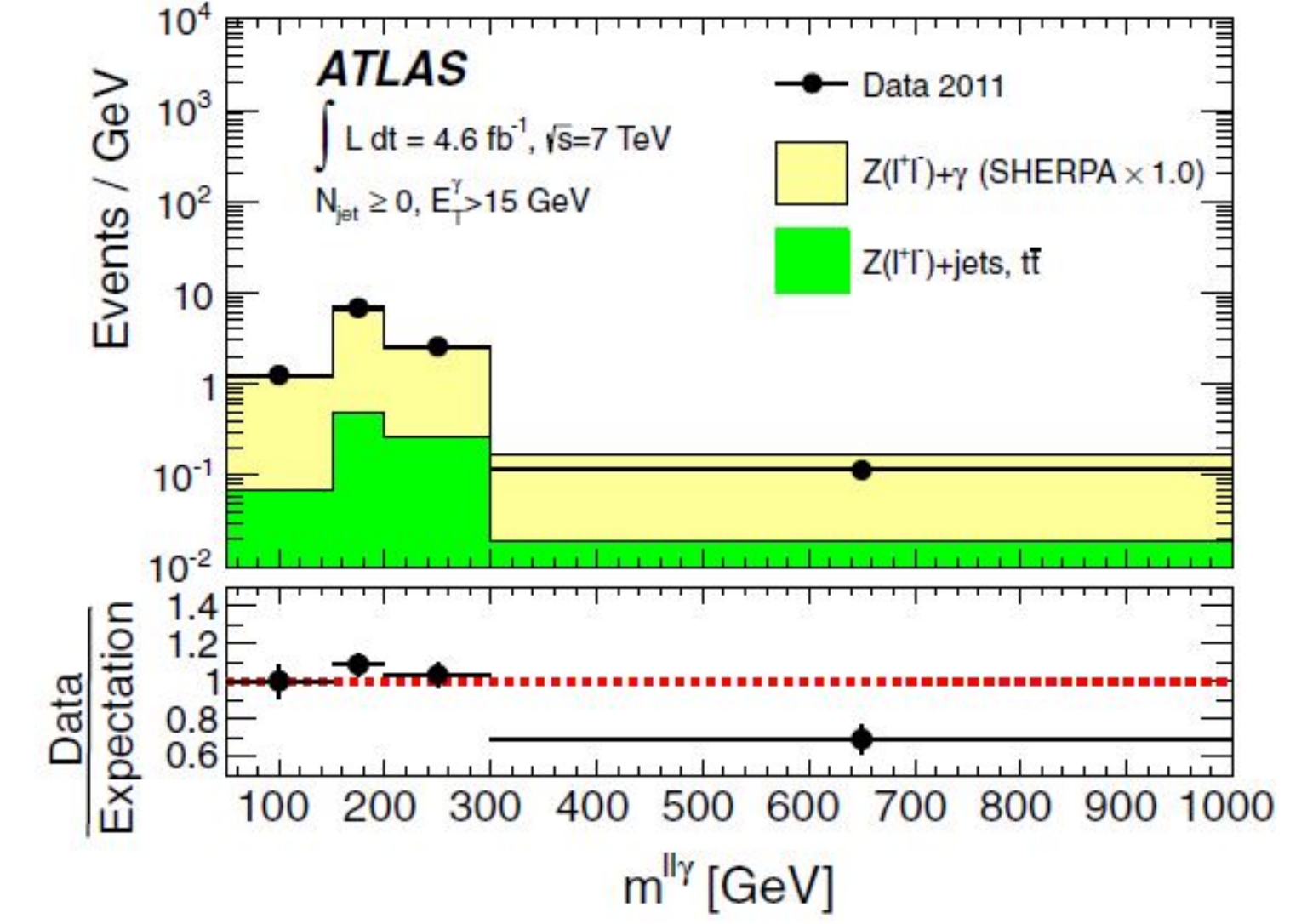}}
\resizebox{0.48\textwidth}{!}{\includegraphics{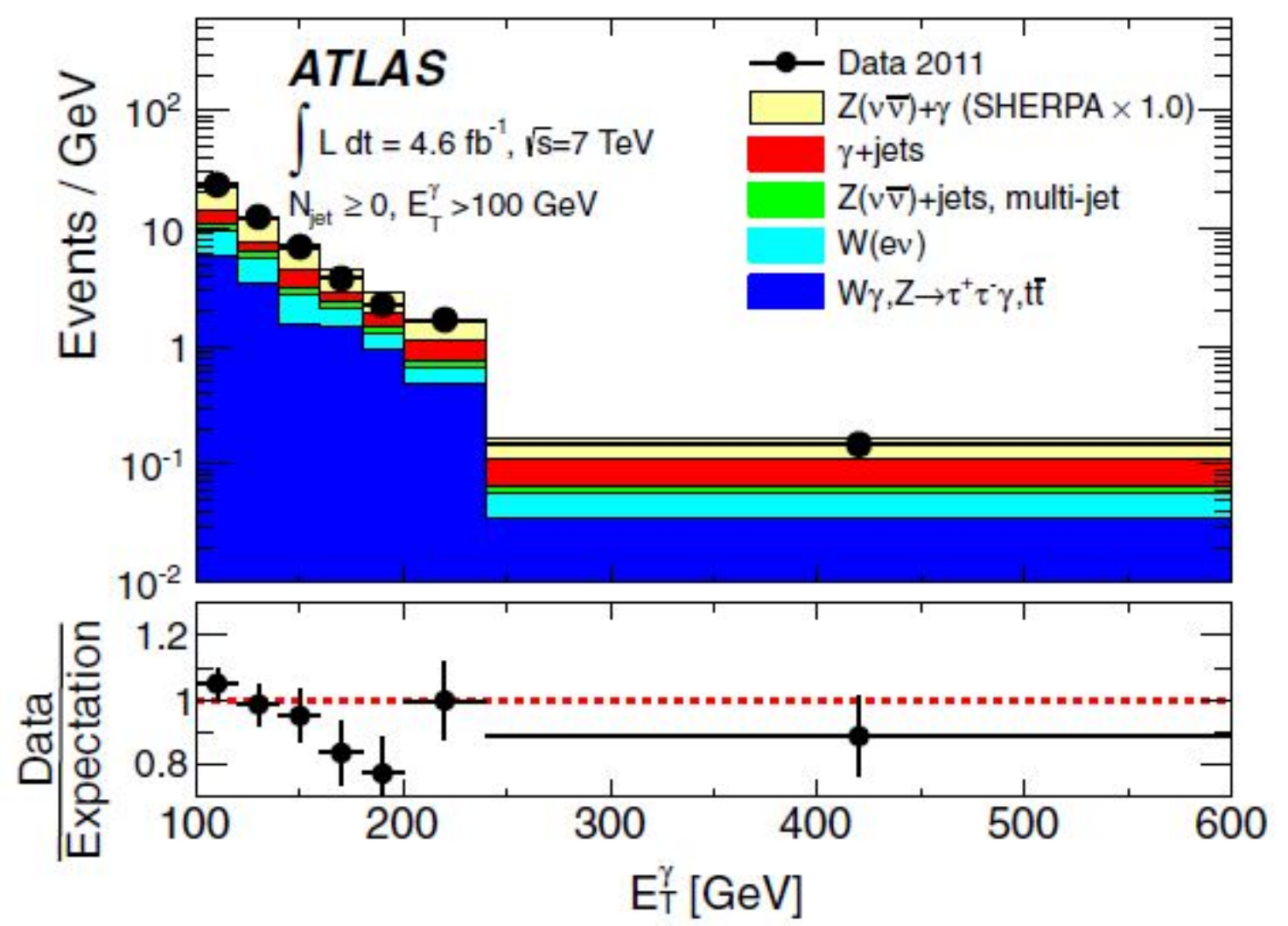}}
\caption{\label{fig:ATLAS_ZGamma} Combined distributions for
  $\ell\ell\gamma$ candidate events in the electron and muon channels
  of the $\ell\ell\gamma$ invariant mass (Left) for the ATLAS $pp
  \rightarrow Z\gamma \rightarrow \ell\ell\gamma$ analysis and the
  photon transverse energy for the ATLAS $pp \rightarrow Z\gamma
  \rightarrow \nu\bar{\nu}\gamma$ analysis.}
\end{center}
\end{figure}

\begin{figure}[h]
\begin{center}
\resizebox{0.9\textwidth}{!}{\includegraphics{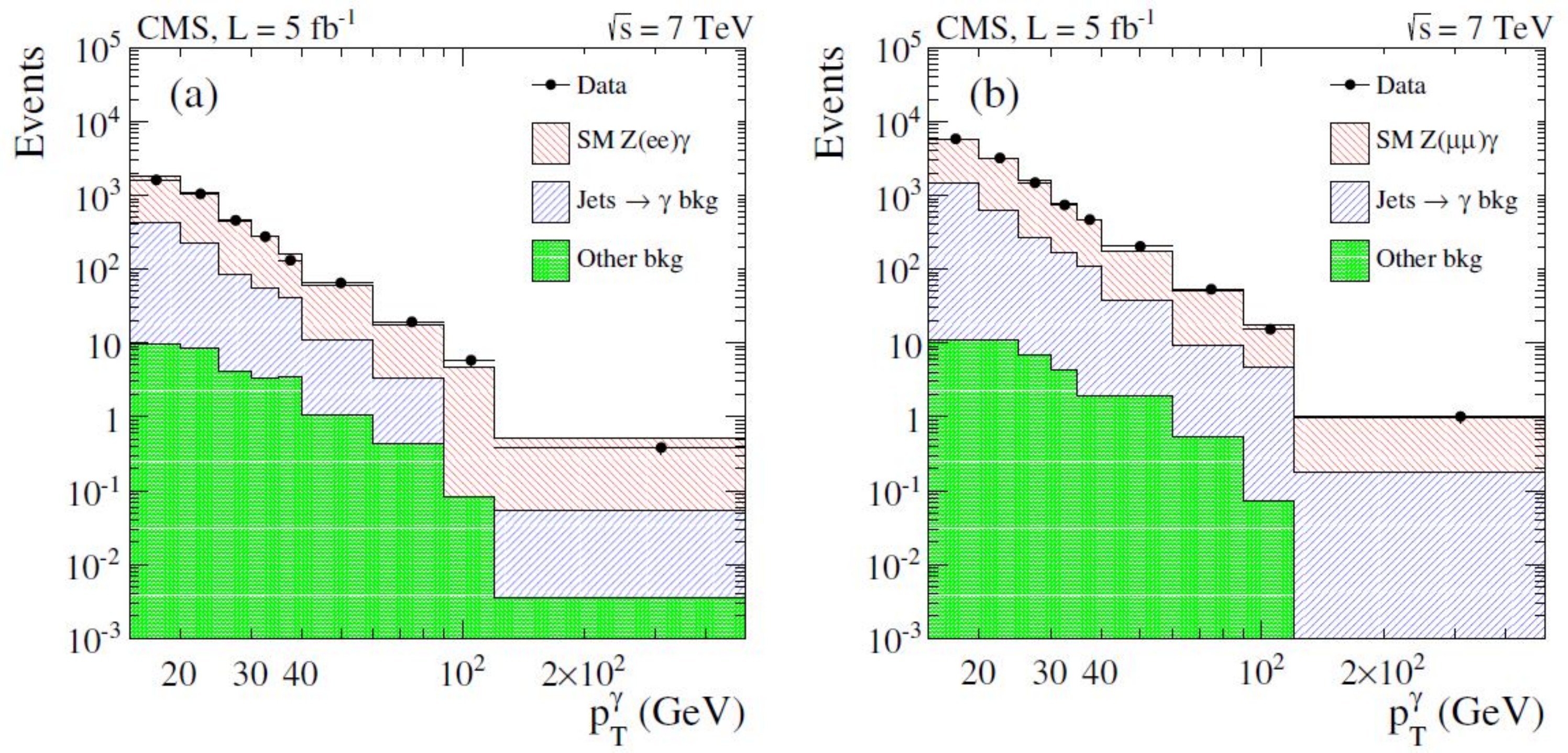}}
\caption{\label{fig:CMS_ZGamma} CMS distributions of $p^\gamma_T$ for
  $\ell\ell\gamma$ candidate events in data, with signal and
  background MC simulation contributions to $Z\gamma \rightarrow
  ee\gamma$ (Left) and $Z\gamma \rightarrow \mu\mu\gamma$ (Right).}
\end{center}
\end{figure}

The expected and observed event yields after applying all selection
criteria are shown in Tab.~\ref{tab:Zgamma} for all decay channels.
The combined photon $\ell\ell\gamma$ invariant mass distribution for
all selected $\ell\ell\gamma$ candidate events and the photon
transverse energy distribution for all selected $\nu\bar{\nu}\gamma$
candidate events are shown in Fig.~\ref{fig:ATLAS_ZGamma} for the
ATLAS analysis.  The photon transverse energy distribution is shown
separately for the electron and muon decay channels in
Fig.~\ref{fig:CMS_ZGamma} for the CMS $\ell\ell\gamma$ analysis.
Reasonable agreement between the data and expected signal and
background contributions are observed from each experiment.

\begin{table}[h]
\tbl{Summary of observed $Z\gamma$ candidates in the data, total
  background estimates and expected signal for the individual decay
  modes. The first uncertainty is statistical while the second is
  systematic.}  {
\begin{tabular}{lccccr}
\hline
\hline
        & Decay channel & Expected signal & Background  & Observed \\
\hline
ATLAS      & $ee\gamma$    & $1600 \pm 71 \pm 68$  & $311 \pm 57 \pm 68$ & 1908  \\
                & $\mu\mu\gamma$    & $2390 \pm 97 \pm 73$  & $366 \pm 83 \pm 73$ & 2756  \\
                & $\nu\bar{\nu}\gamma$    & $420 \pm 42 \pm 60$  & $670 \pm 27 \pm 60$ & 1094  \\ \hline
CMS         & $ee\gamma$    & $3160 \pm 80 \pm 90$  & $950 \pm 50 \pm 40$ & 4108  \\
                & $\mu\mu\gamma$    & $5030 \pm 100 \pm 1210$  & $1424 \pm 60 \pm 80$ & 6463  \\
                & $\nu\bar{\nu}\gamma$    & $45.3 \pm 6.9$  & $30.2 \pm 6.5$ & 73  \\ 
\hline
\hline
\end{tabular}
\label{tab:Zgamma}
}
\end{table}

For the $Z\gamma \rightarrow \ell\ell\gamma$ analysis, ATLAS measured
the production cross section in an extended fiducial region defined as
two leptons with $p_T>25$ GeV and $|\eta|<2.47$, dilepton invariant
mass $m_{\ell\ell}>40$ GeV, photon $E_T>15$ GeV, and $\Delta R(\ell,
\gamma)>0.7$.  CMS measured the production cross section in an
extended fiducial region defined as photon $p_T>15$ GeV, $\Delta
R(\ell, \gamma)>0.7$ and $m_{\ell\ell}>50$ GeV.  For the $Z\gamma
\rightarrow \nu\bar{\nu}\gamma$ analysis, ATLAS measured the
production cross section in an extended fiducial region defined as
$p_T^{\nu\bar{\nu}}>90$ GeV and photon $E_T>100$ GeV, while CMS
measured the production cross section in an extended fiducial region
defined as photon $E_T>145$ GeV and $|\eta|<1.4$.  The measured and
predicted cross sections are listed in
Tab.~\ref{tab:Zgamma_xsection}. ATLAS also measured the unfolded
differential cross section as a function of the photon $E_T$ of the
$Z\gamma \rightarrow \ell\ell\gamma'$ process. The results are shown
separately for the inclusive ($N_{jet} \ge 0$) and exclusive
($N_{jet}=0$) fiducial regions.

\begin{table}[h]
\tbl{Summary of the measured and predicted $W\gamma \rightarrow
  \ell\nu\gamma$ cross sections from the ATLAS and CMS
  collaborations. Different fiducial regions are defined by the ATLAS
  and CMS collaborations.}  {
\begin{tabular}{lll}
\hline
\hline
        & Measured (pb) & Predicted (pb) \\
\hline
ATLAS & & \\
$~~~~~$ $\sigma^{fid}_{Z\gamma \rightarrow \ell \ell \gamma}$ ($N_{jet} \ge 0$)   &  $1.31 \pm 0.02$ (stat.) $\pm 0.11$ (syst.) $\pm 0.05$ (lumi.)  & $1.18 \pm 0.05$ \\ [1.5ex]
$~~~~~$ $\sigma^{fid}_{Z\gamma \rightarrow \nu \nu \gamma}$ ($N_{jet} \ge 0$)   &  $0.133 \pm 0.013$ (stat.) $\pm 0.020$ (syst.) $\pm 0.005$ (lumi.)  & $0.156 \pm 0.012$ \\ [1.5ex]
$~~~~~$ $\sigma^{fid}_{Z\gamma \rightarrow \ell \ell \gamma}$ ($N_{jet} = 0$)   &  $1.05 \pm 0.02$ (stat.) $\pm 0.10$ (syst.) $\pm 0.04$ (lumi.)  & $1.06 \pm 0.05$ \\ [1.5ex]
$~~~~~$ $\sigma^{fid}_{Z\gamma \rightarrow \nu \nu \gamma}$ ($N_{jet} = 0$)   &  $0.116 \pm 0.010$ (stat.) $\pm 0.013$ (syst.) $\pm 0.004$ (lumi.)  & $0.115 \pm 0.009$ \\ [1.5ex]
CMS & & \\
$~~~~~$ $\sigma^{fid}_{Z\gamma \rightarrow \ell \ell \gamma}$  &  $5.33 \pm 0.08$ (stat.) $\pm 0.25$ (syst.) $\pm 0.12$ (lumi.)  & $5.45 \pm 0.27$ \\ [1.5ex]
$~~~~~$ $\sigma^{fid}_{Z\gamma \rightarrow \nu \nu \gamma}$  &  $0.0211 \pm 0.0042$ (stat.) $\pm 0.0043$ (syst.) $\pm 0.0005$ (lumi.)  & $0.0219 \pm 0.0011$ \\ [1.5ex]
\hline
\hline
\end{tabular}
\label{tab:Zgamma_xsection}
}
\end{table}

\begin{figure}[h]
\begin{center}
\includegraphics[height=0.4\textwidth, width=0.42\textwidth]{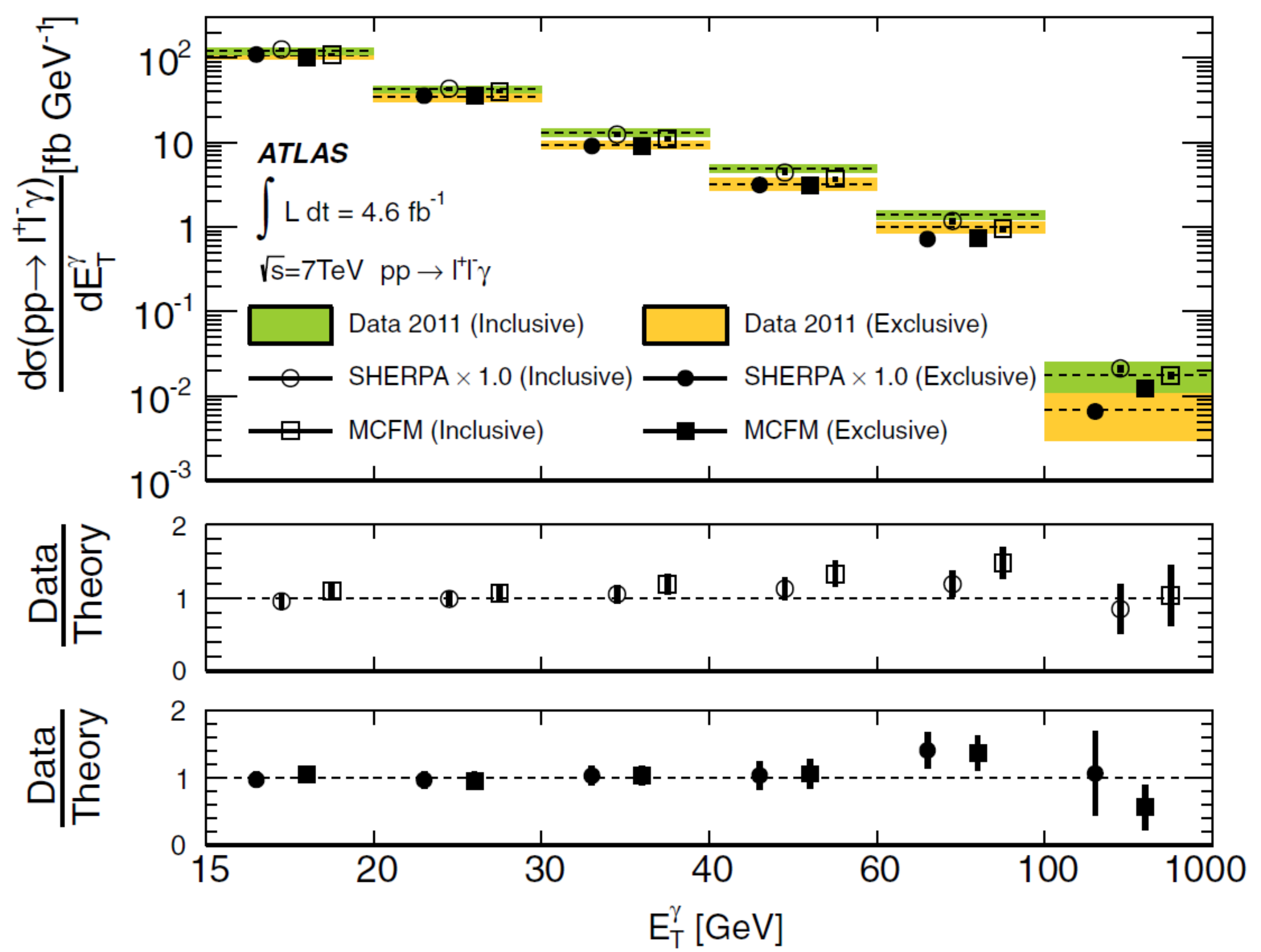}
\includegraphics[height=0.4\textwidth, width=0.36\textwidth]{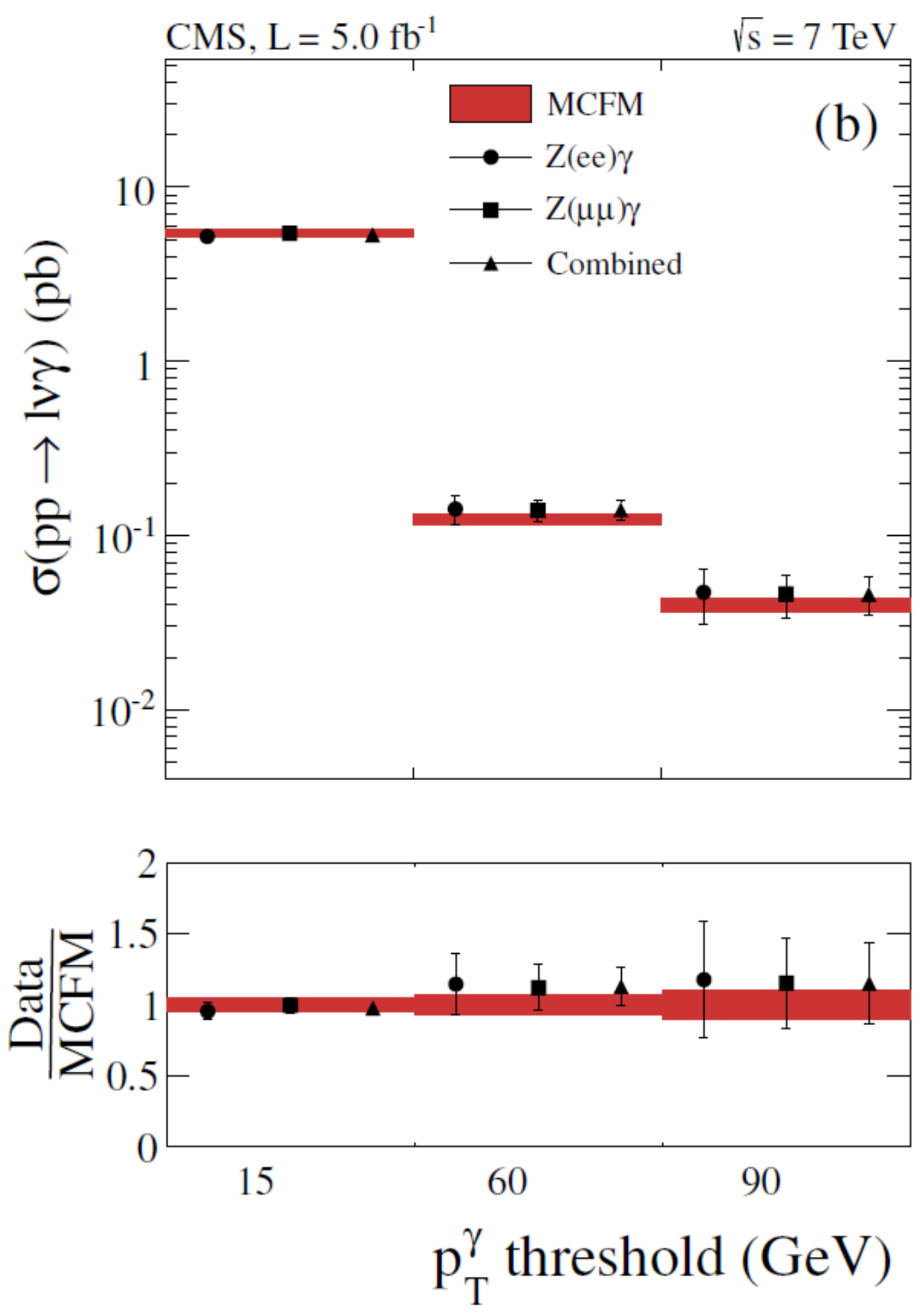}
\caption{\label{fig:ATLAS_ZGamma_Unfolded} Left: Measured photon $E_T$
  differential cross section of the $Z\gamma \rightarrow
  \ell\ell\gamma$ process, using combined electron and muon
  measurements in the inclusive ($N_{jet} \ge 0$) and exclusive
  ($N_{jet}=0$) fiducial regions. Right: Measured $Z\gamma$ cross
  sections for three photon $E_T$ thresholds, compared to SM
  predictions.}
\end{center}
\end{figure}

ATLAS measured the differential cross section as a function of the
photon $E_T$ using combined electron and muon measurements in the
inclusive ($N_{jet} \ge 0$) and exclusive ($N_{jet}=0$) fiducial
regions.  CMS measured the $Z\gamma$ production cross section for
three different photon $E_T$ thresholds (15 GeV, 60 GeV, and 90
GeV). The measured cross sections are compared with several
theoretical predictions as shown in
Fig.~\ref{fig:ATLAS_ZGamma_Unfolded}.

\subsubsection{Derived Results and Discussion}
Since ATLAS and CMS measured the $Z\gamma \rightarrow \ell\ell\nu\gamma$ and
$Z\gamma \rightarrow \nu\nu\gamma$ cross sections in different phase
spaces, for the cross section combination, we choose the phase space
defined by the CMS collaboration for $Z\gamma \rightarrow \ell\ell\nu\gamma$
and the phase space defined by the ATLAS collaboration for $Z\gamma
\rightarrow \nu\nu\gamma$.  The ATLAS measured $Z\gamma \rightarrow
\ell\ell\gamma$ cross section is scaled up by $4.6\pm0.1$, and the CMS
measured $Z\gamma \rightarrow \nu\nu\gamma$ cross section value is
scaled up by $7.1\pm0.2$.  The statistical uncertainties for both
measurements are treated independently. The uncertainties on the
integrated luminosity are treated as fully correlated. Other
systematic uncertainties are dominated by uncertainties related to the
$Z+$jets data-driven background estimation, lepton and photon
identification, and energy and resolution. These uncertainties are
treated as fully uncorrelated between the two experiments. Using the
BLUE method, the combined inclusive cross section is found to be
\[\sigma(pp \rightarrow Z\gamma \rightarrow \ell\ell\gamma) = 5.45 \pm 0.07\,\mbox{(stat.)} \pm 0.27\,\mbox{(sys.)}\,\fb, \] 
and 
\[\sigma(pp \rightarrow Z\gamma \rightarrow \nu\nu\gamma) = 0.137 \pm 0.011\,\mbox{(stat.)} \pm 0.021\,\mbox{(sys.)} \,\fb.\]
The corresponding predicted cross section in each fiducial region is $5.45 \pm 0.27$ fb and $0.156 \pm 0.012$ fb respectively.


\section{\label{sec:interpretation}Interpretation of Results and Outlook}

\subsection{Inclusive Cross Section Measurements}

\begin{figure}[bt]
    \begin{center}
        \includegraphics[width=1.0  \textwidth]{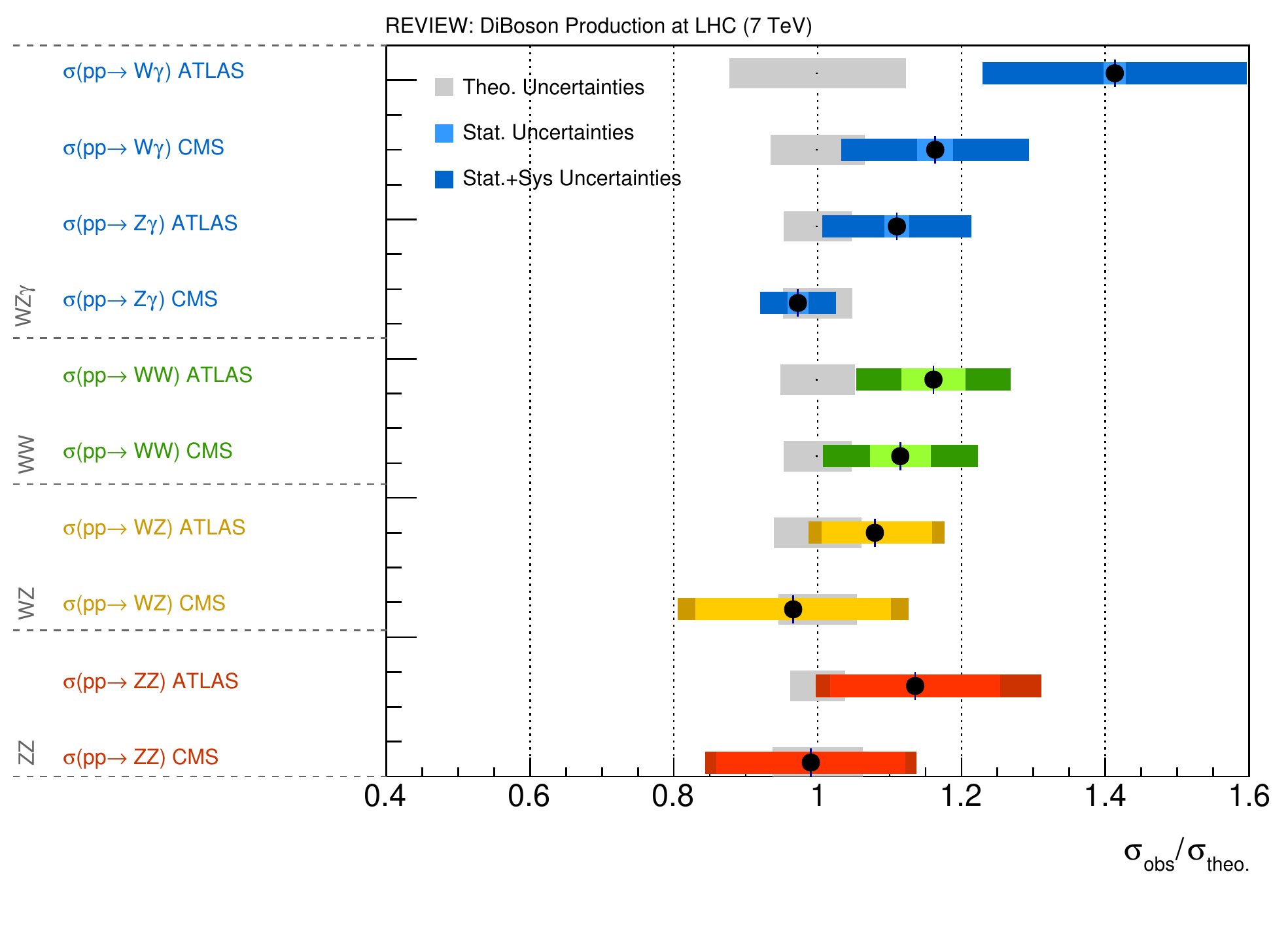}
        \caption{Summary of cross sections for different diboson production processes. Shown is the ratio between data and predictions. The grey bands represent the theory uncertainties, the light colored bands are the statistical uncertainties and the dark colored bands the combined statistical and systematic uncertainties. }
        \label{fig:SummaryCrossSection}
    \end{center}
\end{figure}

Several inclusive cross section measurements are not longer statistically limited and the associated systematic uncertainties are at percent level. The measured cross sections are compatible between the two experiments. Their combination, which have been derived in the Sect. \ref{sec:wwvvertex} and \ref{sec:zzgammavertex}, lead to a reduction of the overall experimental uncertainties by $20-30\%$. On the theoretical side, the SM predictions of the diboson cross sections are known to NLO and NNLO in $\alpha_s$ (see Sect. \ref{sec:eventgen}). A comparison between theoretical predictions and measured and combined cross sections from both experiments is illustrated in Fig. \ref{fig:SummaryCrossSection}. All results are compatible with the theoretical expectation within the statistical and systematic uncertainties. 

In addition to the inclusive cross-section measurements, the ATLAS collaboration also published normalized differential cross section of various diboson final states. Also here, a good agreement between theoretical predictions and the measurements can be observed. However, the measured differential cross sections are still dominated by statistical uncertainties and a stringent test of these distributions is yet to come.

In summary, the good agreement between theory and experiment for the inclusive and differential diboson production cross sections is an impressive confirmation of the perturbative QCD calculations rather than a precise test of the electroweak gauge group structure, as deviations from the latter only affect the high energetic tails of the measured phase space.


\subsection{Limits on Anomalous Triple Gauge Couplings}

The theoretical basis of aTGCs and their effect on the diboson production cross sections is introduced in Sect. \ref{sec:theoryatgc}. In the presence of non-zero aTGCs, 
the production cross section in the $s$-channel is enhanced. This enhancement increases with the center-of-mass energy, leading to an 
increase in the inclusive production cross section, and also to a change of the shape of the differential cross section. 
Observables which probe high center-of-mass energies are therefore of particular importance for the study of aTGCs.

\begin{figure*}
\begin{minipage}{0.49\textwidth}
\resizebox{1.0\textwidth}{!}{\includegraphics{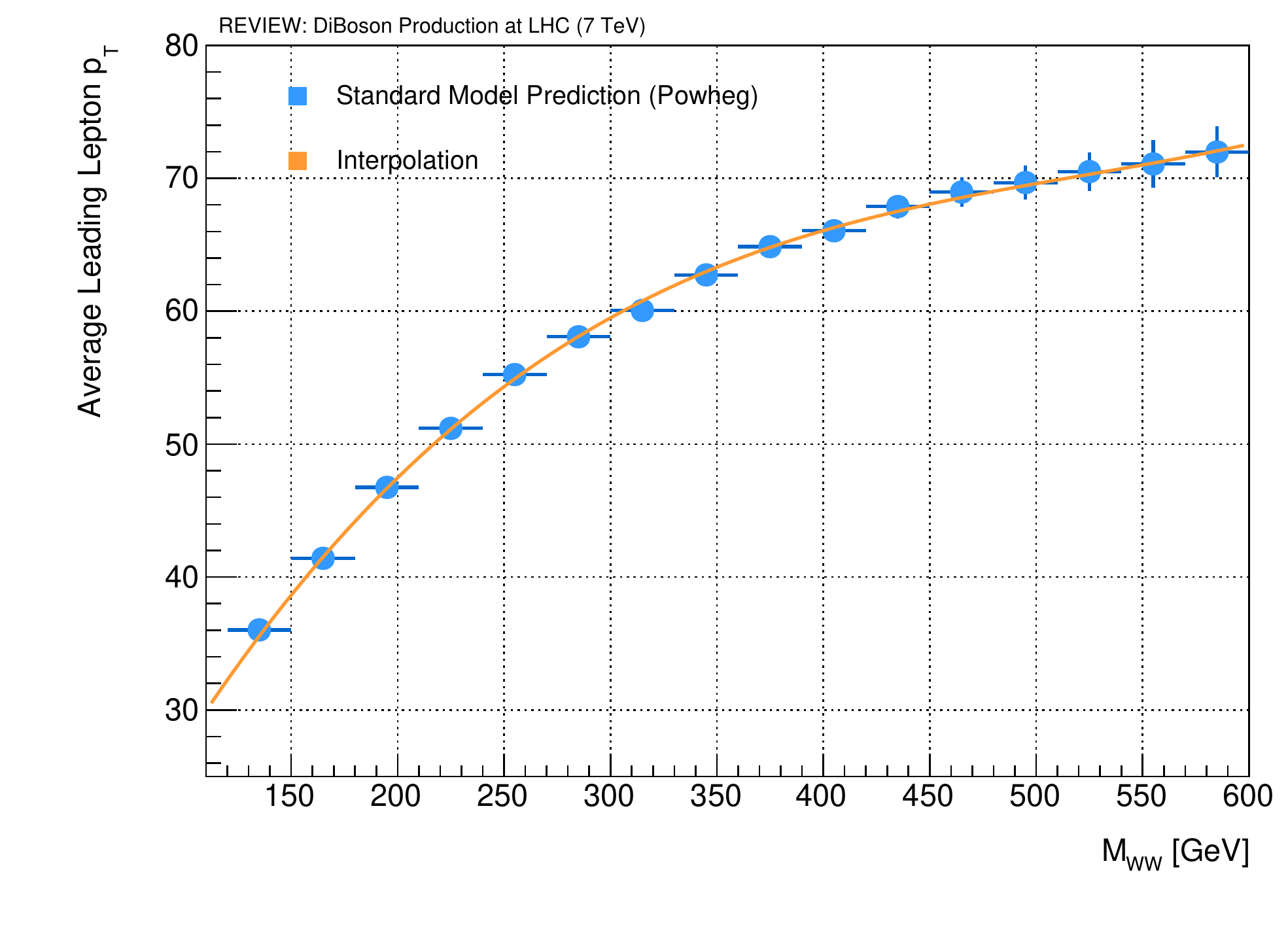}}
\caption{\label{Fig:WWCorelatin} Correlation between the invariant mass of the $W^+W^-$ system and the leading lepton transverse momentum.\vspace{1.35cm}}
\end{minipage}
\hspace{0.0cm}
\begin{minipage}{0.49\textwidth}
\resizebox{1.0\textwidth}{!}{\includegraphics{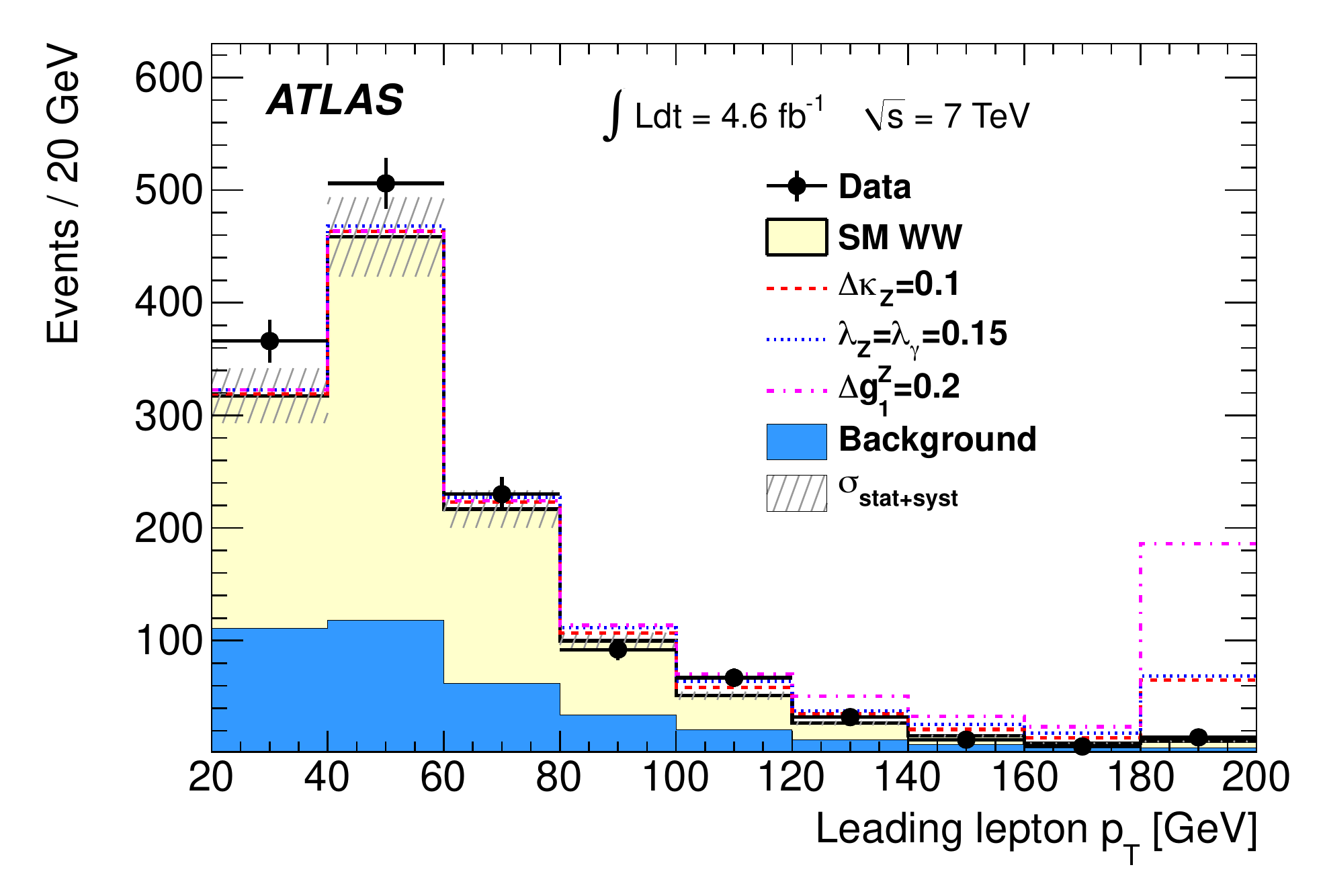}}
\caption{\label{Fig:wwaTGCExample} The reconstructed leading lepton $p_T$ spectrum in data and sum of MC signal and background for the SM prediction and for three different anomalous TGC predictions. The rightmost bin shows the sum of all events with the leading lepton $p_T$ above 180 GeV.}
\end{minipage}
\end{figure*}

If available, the full invariant mass of the diboson system can be taken as observable for aTGC studies. However, this is not available for final 
states which involve a $W$ boson as the decay neutrino does not allow for a full kinematic reconstruction. The transverse momentum of the leading (highest $p_T$) lepton and the reconstructed transverse momentum of the $W$ boson can be used as sensitive variables in case of the $WW$ and $WZ$ analyses, respectively. The correlations between the leading lepton $p_T$ in the $WW$ final state with the center-of-mass energy $\hat s = m_{WW}$ is shown in Fig. \ref{Fig:WWCorelatin}. 

Since no deviations from SM expectations were observed, the measured distributions have been used to derive limits on aTGCs. This has been done on reconstruction level where the expected signal yields are compared to data. MC event generators such as MC@NLO or BHO \cite{Baur:1993ir} are used to generate events for a chosen set of aTGCs. These MC samples are then used in a second step to reweight a fully simulated and reconstructed standard model MC sample. As an example, the reconstructed leading lepton transverse momentum of a $WW$ process is shown in Figure \ref{Fig:wwaTGCExample}. Several predictions of this distributions for different aTGCs scenarios are also shown. 

A maximum likelihood fit is performed in order to derive a 95\% confidence limit on aTGCs. The likelihood $L$ is defined as a product of Poisson probability distribution functions for the observed number of events $N^i_{obs}$ and the number of expected signal and background events ($N^i_{exp}$) for each bin $i$ of the reconstructed distribution:
\[  
L = \Pi_{i=1}^{n} \frac{e^{-N^i_{exp}} (N^i_{exp})^{N^i_{obs}}}{N^i_{obs}!} 
\]
This likelihood is evaluated for one one varying aTGC parameter, while all other aTGC parameters are kept at zero, and the corresponding 95\% confidence limits are derived. CMS published aTGC limits without any usage of form factors, while ATLAS derived limits with and without form factors. The cut-off scale in the ATLAS analyses is 
chosen to match the energy at which the unitarity is still preserved. 

\begin{figure*}
\resizebox{0.49\textwidth}{!}{\includegraphics{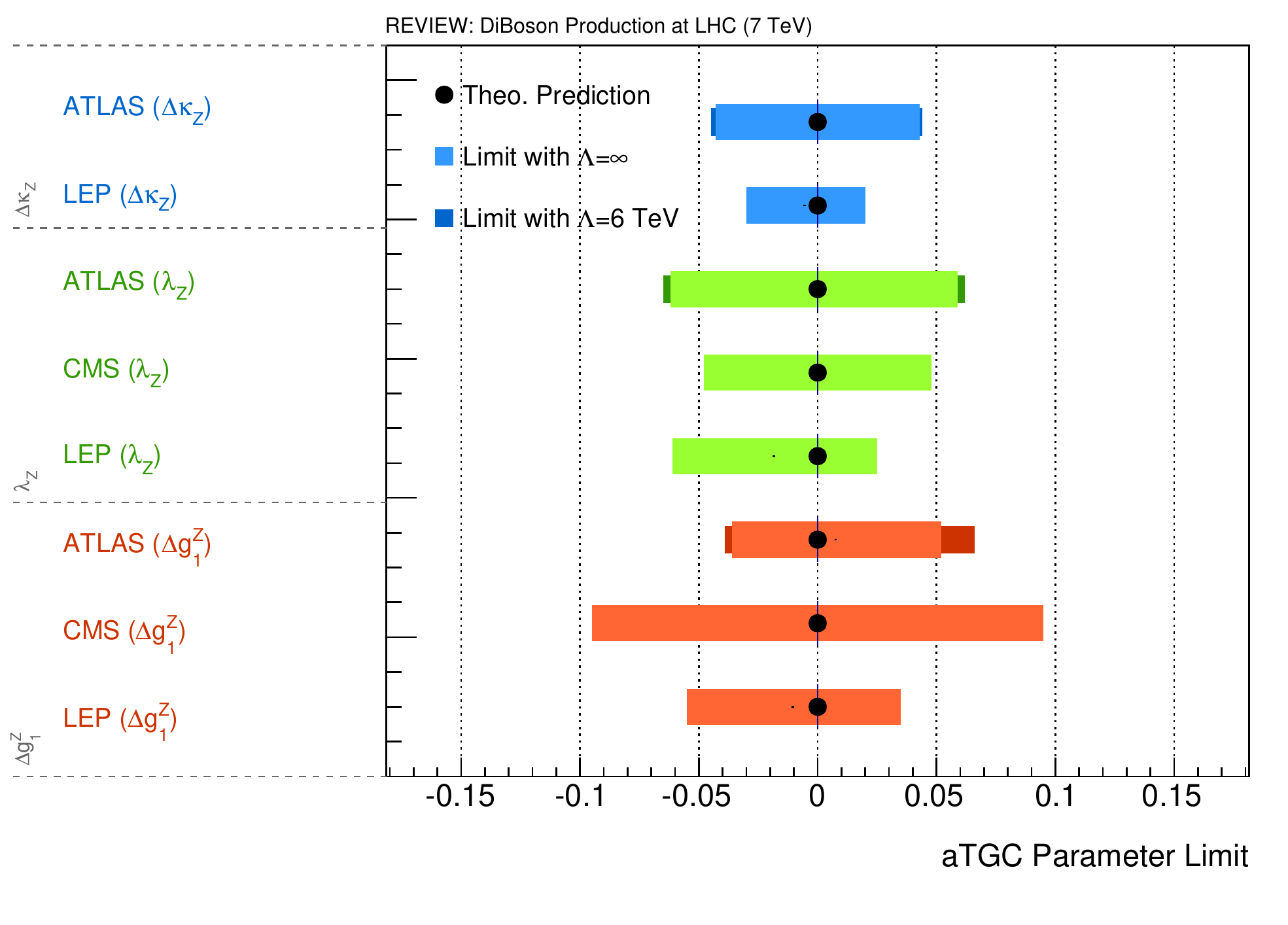}}
\resizebox{0.49\textwidth}{!}{\includegraphics{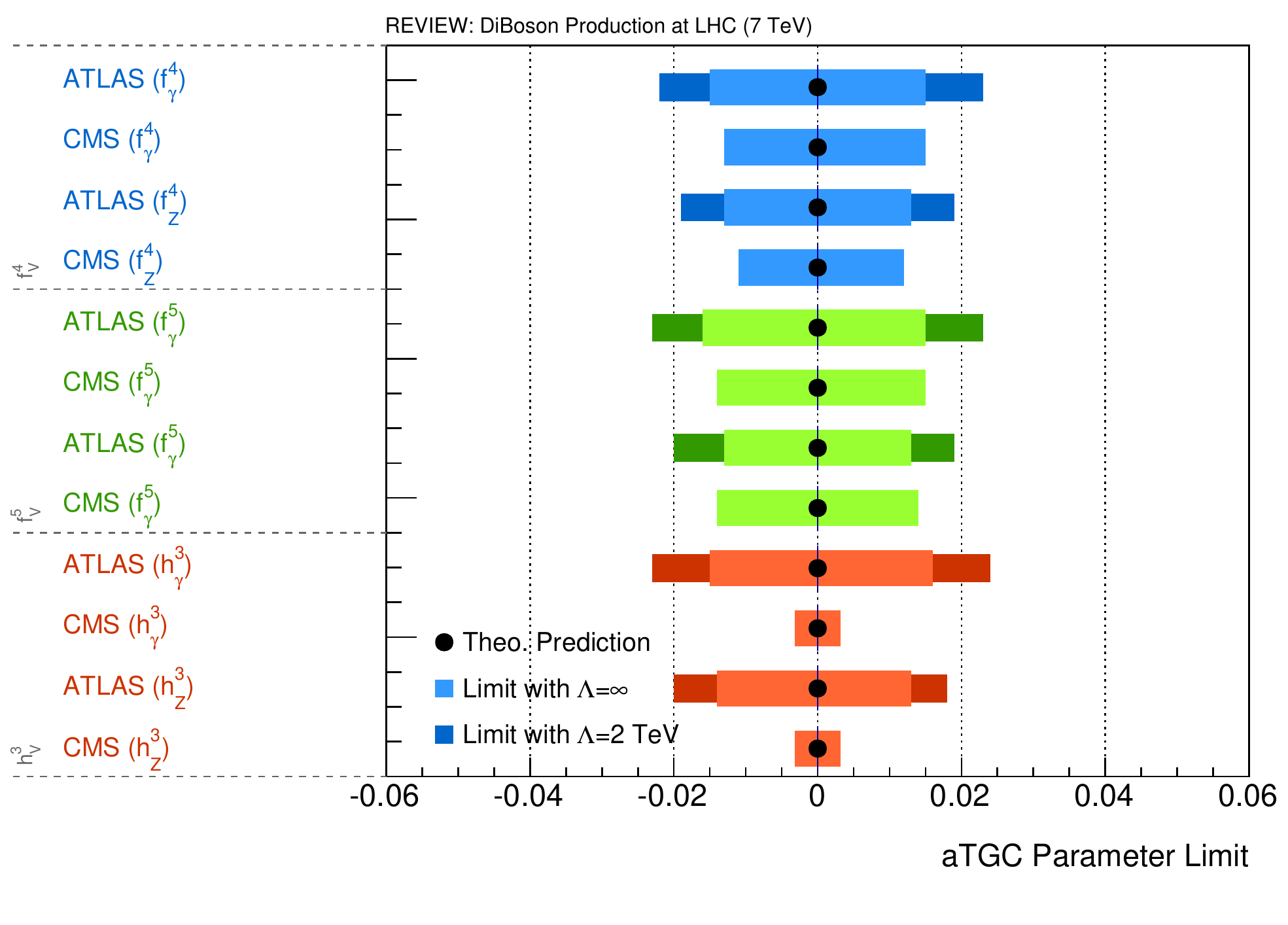}}
\caption{Illustration of the observed 95\%  confidence exclusion limits on anomalous couplings on the $WWV$ (left), the $ZZV$ and the $Z\gamma V$ vertex (right) from ATLAS, CMS and the LEP experiments. The limits on the $WWV$ are based on the LEP scenario. Except for the coupling under study, all other anomalous couplings are set to zero. The results are shown with an energy cut-off scale $\Lambda=\infty$ (i.e. no form factor), and if applicable with $\Lambda = 2-3 \GeV$. The limits on $h^4_V$ are too small to be shown in this Figure.}
\label{Fig:SumWWV}
\end{figure*}

As previously discussed, the $WWZ$ and $WW\gamma$ vertices can be studied by the $WW$, $WZ$ and $W\gamma$ final states. Both ATLAS and CMS experiments derived aTGC limits using the transverse momentum of the leading lepton in the $WW$ final state. CMS published limits only based on the LEP scenario \cite{Chatrchyan:2013yaa}, while ATLAS provided limits assuming the LEP scenario, the equal coupling assumptions, the HISZ assumptions and no further constraints \cite{ATLAS:2012mec}. 
Limits on aTGCs via the $WZ$ analysis are available from the ATLAS collaboration \cite{Aad:2012twa} and are based on the study of the transverse momentum spectrum of the $Z$ boson. Both experiments derived limits via the $W\gamma$ final state \cite{Aad:2012mr, Chatrchyan:2013fya} using the measured energy of the final state photon. 
A summary on aTGC parameters involving the $WWZ$ and $WW\gamma$ without form factors is given in Tab. \ref{tab:SummaryZZVATGCs} and illustrated in Fig. \ref{Fig:SumWWV}. The results from the $WW$ analyses provide the most sensitive limits assuming defined relationships between the couplings. For unconstrained limits, the $WZ$ and $W\gamma$ analyses lead to significant improvements on the relevant coupling parameters. The effect of a form factor on the observed limits varies from a degradation of 10\% for the $WW$ analysis with a cut-off scale of $\Lambda = 6\,\TeV$, to $30\%$ for the $WZ$-analysis with $\Lambda = 2\,\TeV$.


\begin{table}[h]
\tbl{Summary of the observed 95\% confidence exclusion limits on anomalous couplings on the $WWV$ vertex, assuming the LEP scenario and also with no further assumptions  for the ATLAS and CMS experiment. Except for the coupling under study, all other anomalous couplings are set to zero. The results are shown for an energy cut-off scale $\Lambda=\infty$, i.e. no form factor has been used. The limits marked with (*) are derived by the authors by assuming perfect error ellipses, that describe the published results of the experiments.}
{
\begin{tabular}{lcccc}
\hline
							& ATLAS 				& CMS 			& Final State	& Assumptions  \\
\hline
$\Delta \kappa _Z$				&	[-0.043, 0.043]		& ([-0.091,0.091]*)	& $WW$			& LEP	\\
							&	[-0.078, 0.092]		& 				& $WW$			& -	\\
							&	[-0.37, 0.57] 		& 				& $WZ$			& -	\\
$\lambda_Z$					&	[-0.062, 0.059]		& [-0.048,0.048]	& $WW$			& LEP	\\
							&	[-0.074, 0.073] 		& 				& $WW$			& -	\\
							&	[-0.046, 0.047] 		& 				& $WZ$			& -	\\
$\lambda_\gamma$				&	[-0.062, 0.059]		& [-0.048,0.048]	& $WW$			& LEP	\\
							&	[-0.152, 0.146]		& 				& $WW$			& -	\\
							&	[-0.060,0.060]		& [-0.050, 0.037] 	& $W\gamma$		& -	\\
$\Delta g^Z_1$					&	[-0.039, 0.052]		& [-0.095, 0.095]	& $WW$			& LEP	\\
							&	[-0.373, 0.562]		& 				& $WW$			& -	\\
							&	[-0.057, 0.093] 		& 				& $WZ$			& -	\\
$\Delta \kappa _\gamma$			&	([-0.20,  0.20]*)		& [-0.21, 0.22]		& $WW$			& LEP	\\
							&	[-0.135, 0.190]		& 				& $WW$			& -	\\
							&	[-0.33,0.37]		& [-0.38, 0.29]		& $W\gamma$		& -	\\
\hline
\end{tabular}
\label{tab:SummaryZZVATGCs}
}  
\end{table}

The $\gamma ZZ$ and the $\gamma\gamma Z$ vertex are forbidden in the SM and are studied using the $ZZ$ and $Z\gamma$ final states, respectively.
In the $ZZ$ analyses, CMS used the invariant mass of the four-lepton system \cite{Chatrchyan:2012sga}, while ATLAS uses the transverse momentum of the $Z$ boson to 
derive limits on aTGCs \cite{Aad:2012awa}. Both experiments used the transverse energy of the photon in the $Z\gamma$ channel for the limit extraction \cite{Aad:2012mr, Chatrchyan:2013fya}. 
A summary of the derived limits assuming no form factors is given in Tab. \ref{tab:SummaryZZVATGCs} and illustrated in Fig. \ref{Fig:SumWWV}. The given limits 
degrade by $30\%$ to $150\%$ with $\Lambda = 3 \TeV$ for the $ZZ$ analyses and with $\Lambda=2 \TeV$ for the $Z\gamma$ analyses, respectively. 
The limits on the $ZZV$ and $Z\gamma V$ vertices are partly comparable between the two experiments, where the limits on the $ZZ\gamma$ and $Z\gamma\gamma$ 
couplings are driven by the analyses of the $Z\gamma\rightarrow \nu\nu\gamma$ process. However, CMS achieves significantly more stringent exclusion limits on $h^3_V$ and $h^4_V$, which is due to an optimized binning of the sensitive signal region.

\begin{table}[h]
\tbl{The summary of the observed 95\% C.L. on anomalous couplings on the $ZZV$ and $Z\gamma V$ vertex for the ATLAS and CMS experiment. Except for the coupling under study, all other anomalous couplings are set to zero. The results are shown an energy-cut of scale $\Lambda=\infty$, i.e. no form factor has been used.}
{
\begin{tabular}{lccc}
\hline
							& ATLAS 				& CMS					& Final State		\\
\hline
$f^4_\gamma$					&	[-0.015, 0.015]		& [-0.013, 0.015]			&	$ZZ$				\\
$f^4_Z$						&	[-0.013, 0.013]		& [-0.011, 0.012]			&	$ZZ$				\\
$f^5_\gamma$					&	[-0.016, 0.015]		& [-0.014, 0.015]			&	$ZZ$				\\
$f^5_Z$						&	[-0.013, 0.013]		& [-0.014, 0.014]			&	$ZZ$				\\
\hline
$h^3_\gamma$				&	[-0.015,0.016]		& [-0.0032, 0.0032]			&	$Z\gamma$		\\
$h^3_Z$						&	[-0.013,0.015]		& [-0.0032, 0.0032]			&	$Z\gamma$		\\
$h^4_\gamma$				&	[-0.000094,0.000092]	& [-0.000016, 0.000016]		&	$Z\gamma$		\\
$h^4_Z$						&	[-0.000087,0.000087]	& [-0.000014, 0.000014]		&	$Z\gamma$		\\
\hline
\end{tabular}
\label{tab:SummaryATGCs2}
}  
\end{table}

As discussed in Sect. \ref{sec:theoryatgc}, the present limits depend on various theoretical assumptions, thus it will be difficult to reinterpret those limits in the future 
when new theoretical models are proposed. One possibility to overcome those constraints is the publication of the production cross sections in sensitive regions of 
the phase space, which in turn can be easily used to derive limits on various theoretical scenarios. Those published cross sections are model independent to a high extend and do not require any further knowledge on the detector. 
Examples of such cross sections are derived in Sect. \ref{sec:wwvvertex} and \ref{sec:zzgammavertex}. 

In order to test the usefulness of this approach, we compare the derived limits on the $WWV$ vertex based on the reconstructed lepton $p_T$ distribution to the limits obtained using the published differential cross sections. In a first step, we use the \textsc{MCFM} generator to predict the production cross sections for one aTGC parameter ($\lambda_Z$ in this case) in the fiducial phase space defined in Sect. \ref{sec:WWDerived} for the leading lepton $p_T$ between $140\,\GeV$ and $350\,\GeV$. All other couplings are set to their SM values. In a second step, we compare the predicted cross sections for different $\lambda_Z$ with the measured fiducial cross section in this phase space region of $24\pm10$ fb. This leads to a limit of $[-0.13, 0.13]$, which is weaker by approximately $60 \%$. This degradation is due to the fact that the ATLAS limit derivation is based on the full reconstructed spectra, i.e. uses the full leading lepton $p_T$ shape information. However, the advantages of the limit derivation via fiducial cross sections remain and should be taken as a valid alternative option for future measurements.


\subsection{\label{sec:QGC}Outlook on Quartic Gauge Coupling Measurements}

We have concentrated on TGC studies in the above sections. Even though
the available data from the LHC run in 2010 and 2011 are not
sufficient to measure quartic gauge couplings (QGCs), first
sensitivity studies showing the potential of the upcoming LHC results
have been made available. The SM predicts four-point vertices between
electroweak gauge bosons, described in Eqn. \ref{EQN:SMLagrangian} by
the term $\LC_{WWVV}$. The highest cross section at the LHC is
expected for the $WW\gamma\gamma$ vertex, which can be studied in the
reaction $pp \rightarrow p W^+ W^- p$ with a subsequent leptonic decay
of the $W$ boson. A schematic sketch of the involved processes is
shown in Fig. \ref{Fig:PhotonFusion}.

\begin{figure*}[b]
\begin{minipage}{0.57\textwidth}
\resizebox{0.33\textwidth}{!}{\includegraphics{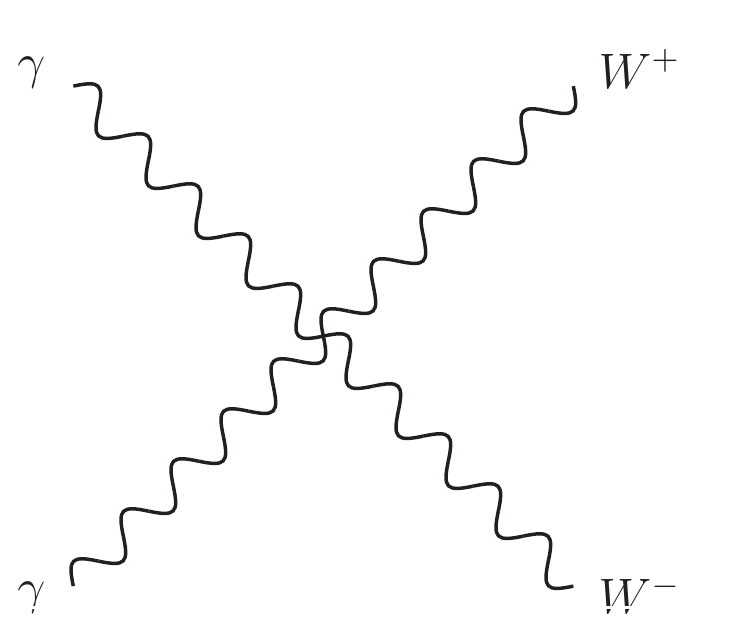}}
\resizebox{0.28\textwidth}{!}{\includegraphics{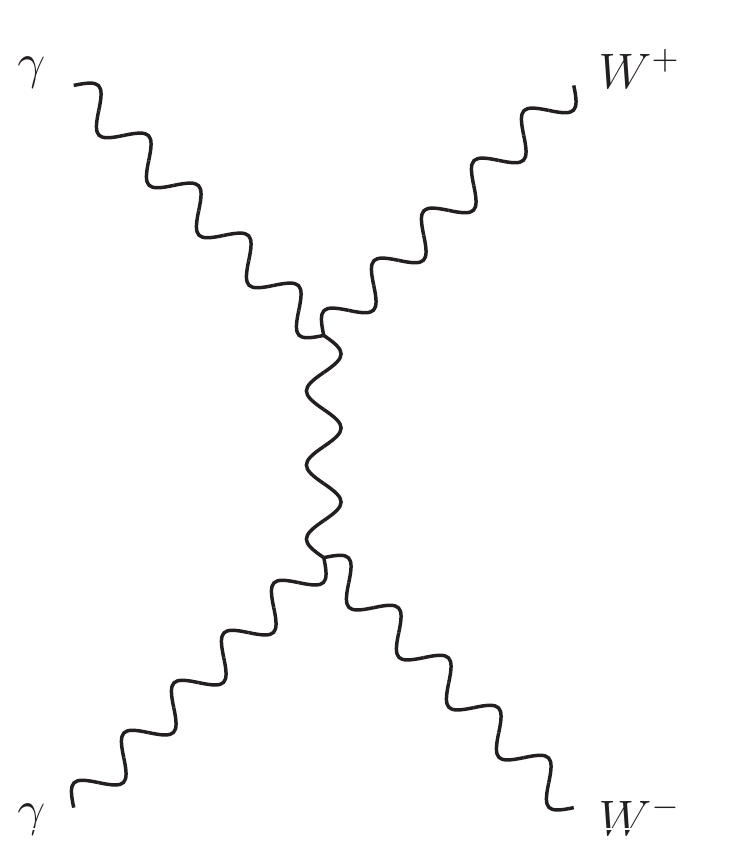}}
\resizebox{0.33\textwidth}{!}{\includegraphics{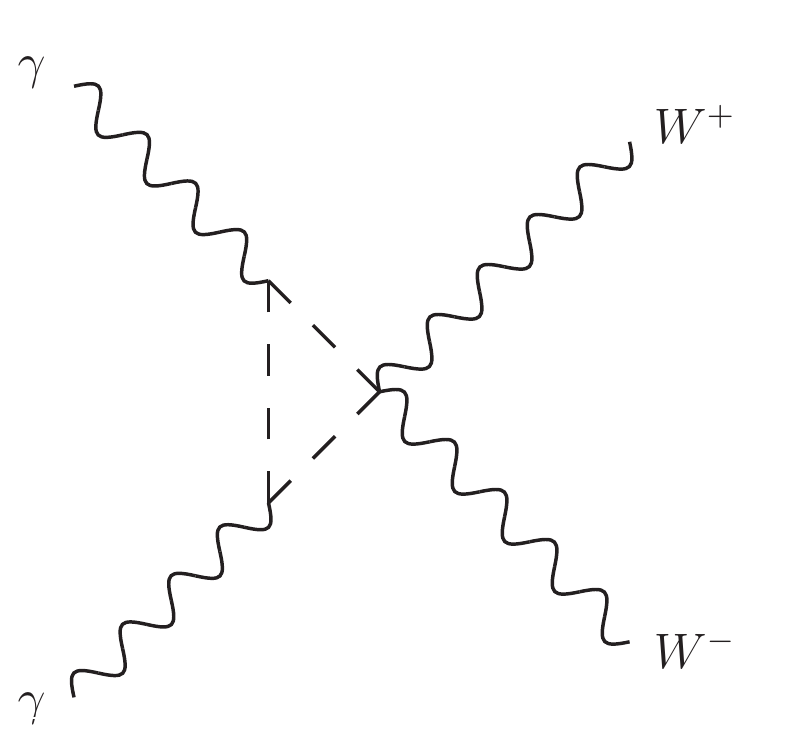}}
\caption{\label{Fig:WWPP:CMS:2}LO Feynman diagrams for the $\gamma
  \gamma \rightarrow W^+W^-$ production.}
\end{minipage}
\begin{minipage}{0.42\textwidth}
\resizebox{.8\textwidth}{!}{\includegraphics{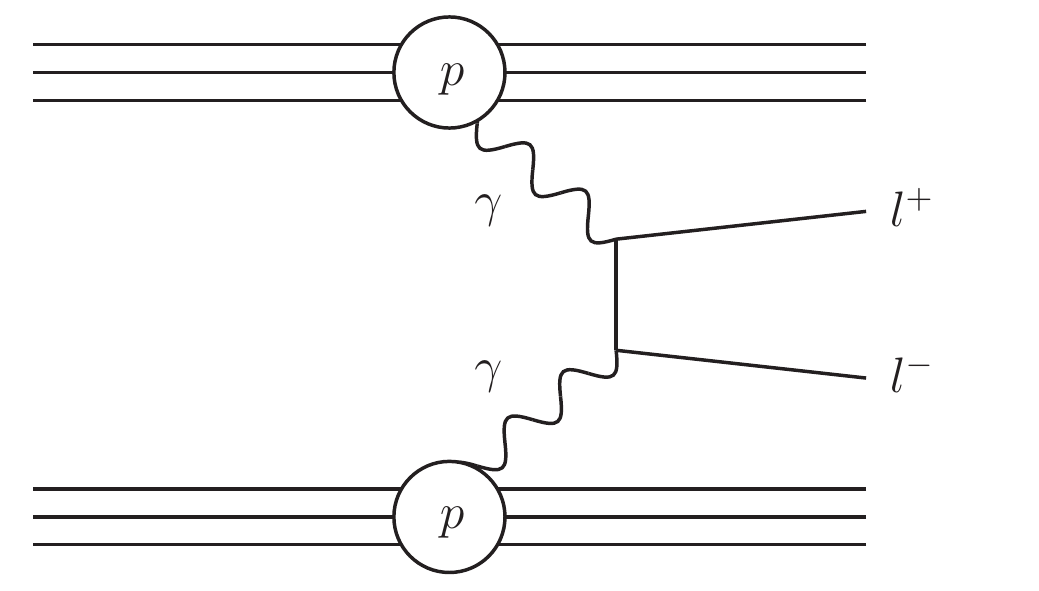}}
\caption{\label{Fig:WWPP:CMS:3}LO Feynman diagram for
  $\gamma\gamma\rightarrow \ell^+\ell^-$.}
\end{minipage}

\end{figure*}

The fully exclusive (``elastic") scattering is theoretically well
understood and can be modeled for example by the {\sc calhep}
\cite{Pukhov:2004ca} program.  The production cross section is 1.2 fb,
taking into account the leptonic branching ratios of the \Wboson
boson. The same process with a similar final state can also be induced
by inelastic scattering, in which one or even both protons dissociate
into a low-mass system which also escapes the detection. The inelastic
scattering is theoretically less well known due to strong interactions
between the proton remnants (rescattering) which produce additional
hadronic activity.  The LO diagrams for the $\gamma\gamma \rightarrow
W^+ W^-$ interactions are shown in Fig. \ref{Fig:WWPP:CMS:2}.  Two TGC
vertices are present in the $t$- and $u$-channels, while the quartic
couplings only appear in the $s$-channel.

The experimental signature of these processes are two high energetic, charged 
leptons from the \Wboson boson decay and two forward scattered protons beyond 
the acceptance of the detector. The dilepton system is expected to have  
large invariant mass and large transverse momentum. In addition, only two tracks 
from a common vertex are expected. Additional tracks would indicate further 
hadronic activity which is not expected in the $\gamma\gamma \rightarrow W^+W^-$ process.

The main background processes are the inclusive $pp\rightarrow W^+W^-$
production, the inclusive $Z$ boson production with a
leptonic decay, and the two photon process $\gamma \gamma \rightarrow
\ell^+ \ell^-$ as shown in Fig. \ref{Fig:WWPP:CMS:3}.  While the first
two are expected to have additional tracks associated with the
production vertex, the latter process is produced in a similar
way to the signal.
The two photon processes have an order of magnitude times larger cross
section than the signal process.

The CMS experiment has analyzed 5 fb$^{-1}$ of data to search for the
evidence of the $\gamma\gamma \rightarrow W^+W^-$ 
process \cite{CMS:2011wzexcl}. In order to achieve a clean signal selection, only the
$e^\pm\mu^\mp$ decay channel has been used. The requirement of two
different flavor leptons greatly reduces the background contribution
from the $\gamma \gamma \rightarrow \ell^+ \ell^-$ process.
Signal candidates are selected by requiring two oppositely-charged
leptons ($e, \mu$) with $p_T>20\,\GeV$ and $|\eta|<2.4$. The two
leptons must origin from a common vertex and with no other reconstructed
tracks associated to the production vertex. The dilepton system is
required to have $m_{e\mu}>20\,\GeV$ and $p_T(e\mu)>30\,\GeV$.  The
corresponding signal selection efficiency is $10.6\%$.

\begin{figure*}[tb]
\begin{minipage}{0.62\textwidth}
\resizebox{1.0\textwidth}{!}{\includegraphics{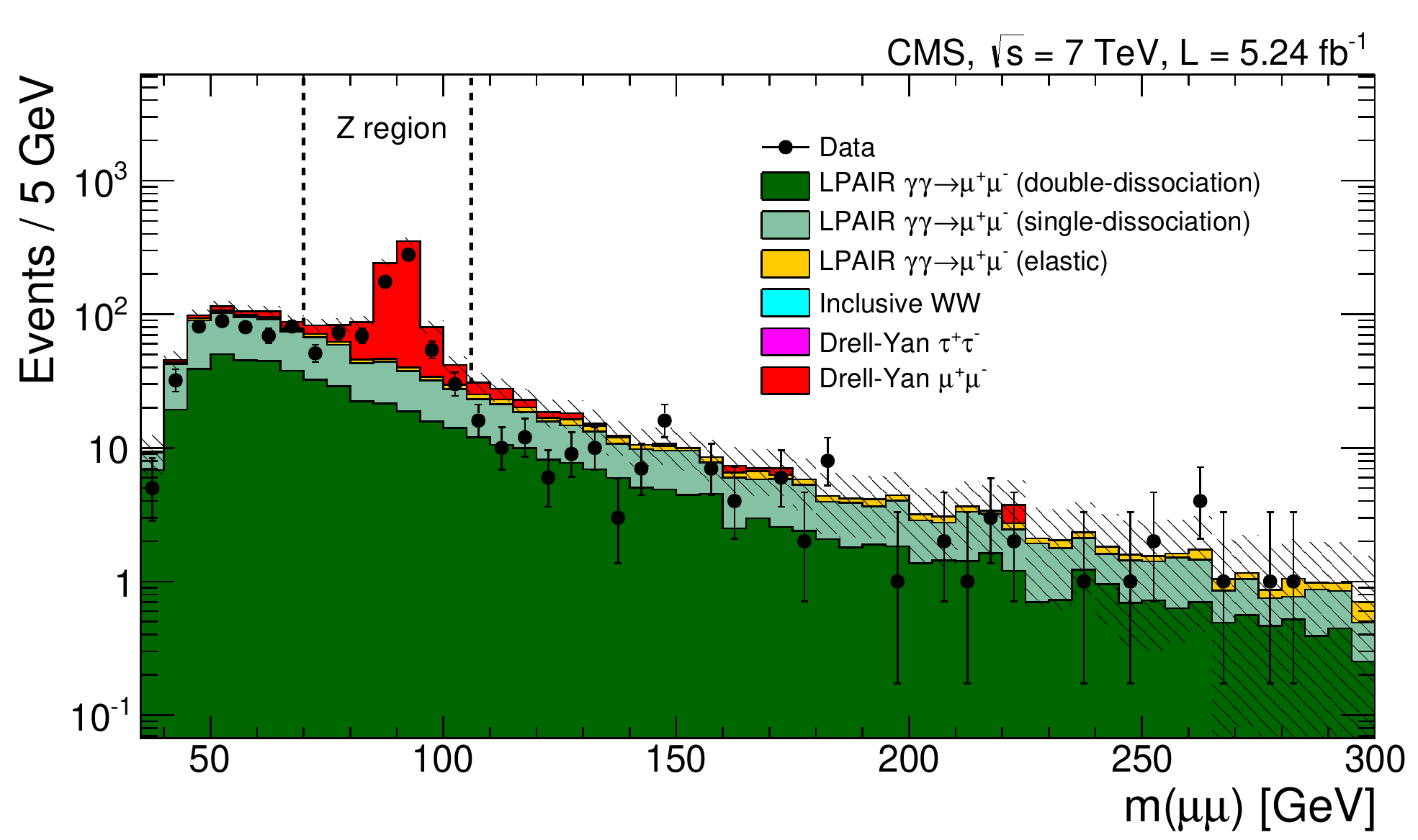}}
\caption{\label{Fig:CMSQT1} Invariant mass distribution of the muon
  pairs for the dissociation selection. The dashed lines indicate the
  $Z$-peak region. The hatched bands indicate the statistical
  uncertainty in the simulation.\vspace{0.0cm}}
\end{minipage}
\begin{minipage}{0.37\textwidth}
\resizebox{1.0\textwidth}{!}{\includegraphics{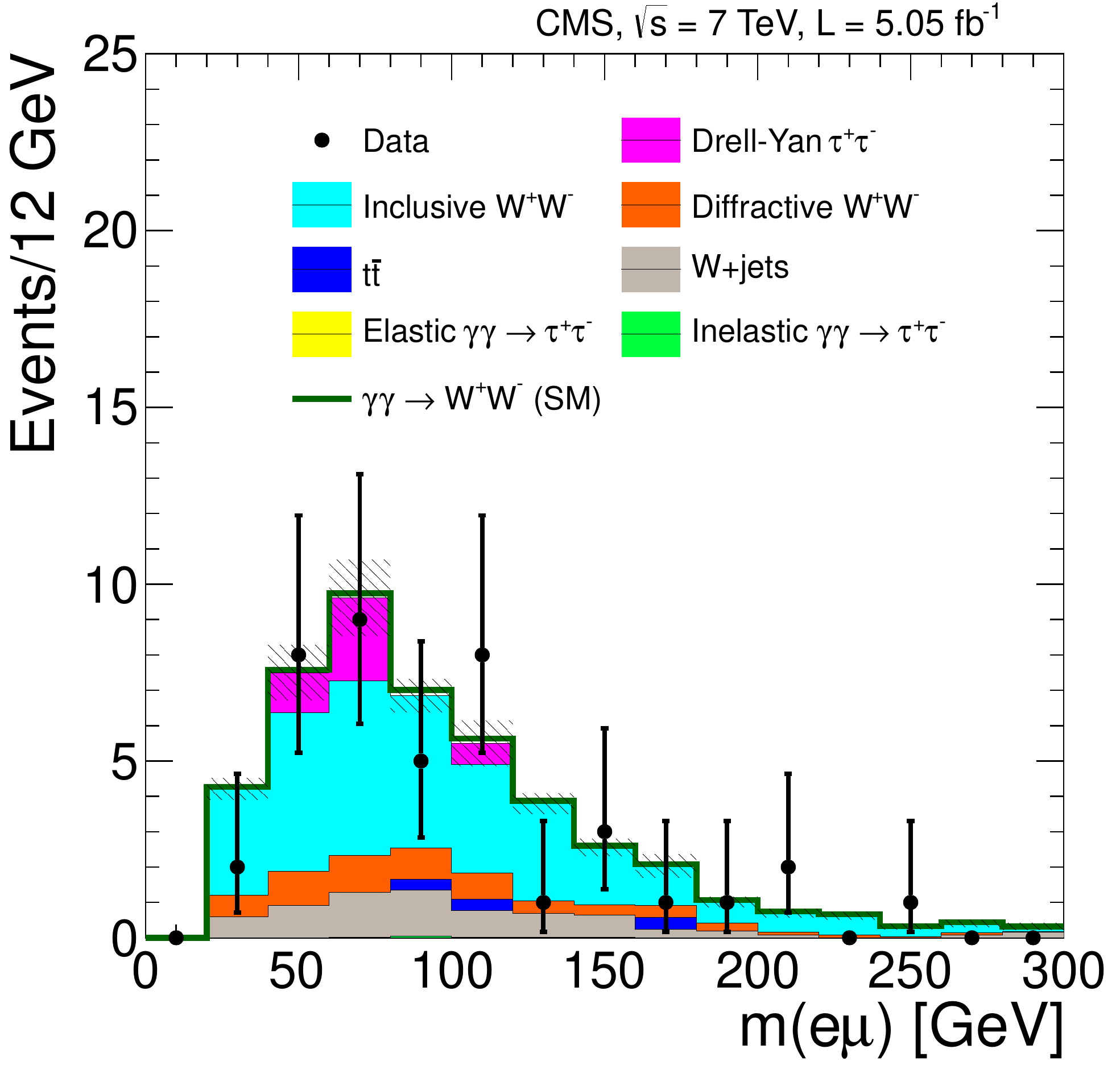}} 
\caption{\label{Fig:CMSQT2} The $e^\pm \mu^\mp$ invariant mass
  distribution for events with $1-6$ extra tracks on the vertex.
}
\end{minipage}
\end{figure*}

Since the theoretical prediction for the proton dissociation
contribution to the signal is not well known, CMS used a data-driven
method to estimate the contribution from this background source.  The
method is based on the assumption that the process $\gamma \gamma
\rightarrow \ell^+ \ell^-$ has the same probability of being produced
by elastic and inelastic $pp$ scattering as by the signal process. By
studying the $pp \rightarrow \mu^+\mu^-$ final state, which is
dominated by the $\gamma \gamma \rightarrow \mu^+ \mu^-$ process, a
correction factor $F$
using events with $m_{\mu^+\mu^-}>160$ GeV is estimated. This
correction factor is calculated as the number of observed dimuon
events corrected for the contribution from the inclusive Drell-Yan
process divided by the number of predicted events from a pure elastic
scattering process. The corresponding dimuon invariant mass
distribution is shown in Fig. \ref{Fig:CMSQT1}. A significant
undershoot of data is seen for large invariant masses, which is the
dissociation region mostly affected by rescattering effects.  The
factor $F$ is found to be $3.23\pm0.53$ and is used to scale the
predicted cross section for the $\gamma \gamma \rightarrow W^+ W^-
\rightarrow e^\pm \nu \mu^\mp \nu$ process.

The total predicted signal cross section times branching ratio is
found to be $\sigma(pp\rightarrow p W^+ W^- p \rightarrow p e^\pm \nu
\mu^\mp \nu p ) = 4.0 \pm 0.7$ fb. The associated uncertainty is mainly due limited
 statistics.  The modeling of background processes
is tested in several control regions. The dilepton invariant mass
distribution for events with $1-6$ extra tracks and $p_T(e\mu)>30\GeV$
is shown in Fig. \ref{Fig:CMSQT2}. The background in this sample is
dominated by the inclusive $W^+W^-$ production. Good agreement between
data and MC is seen.

\begin{figure*}[tb]
\begin{center}
\resizebox{0.4\textwidth}{!}{\includegraphics{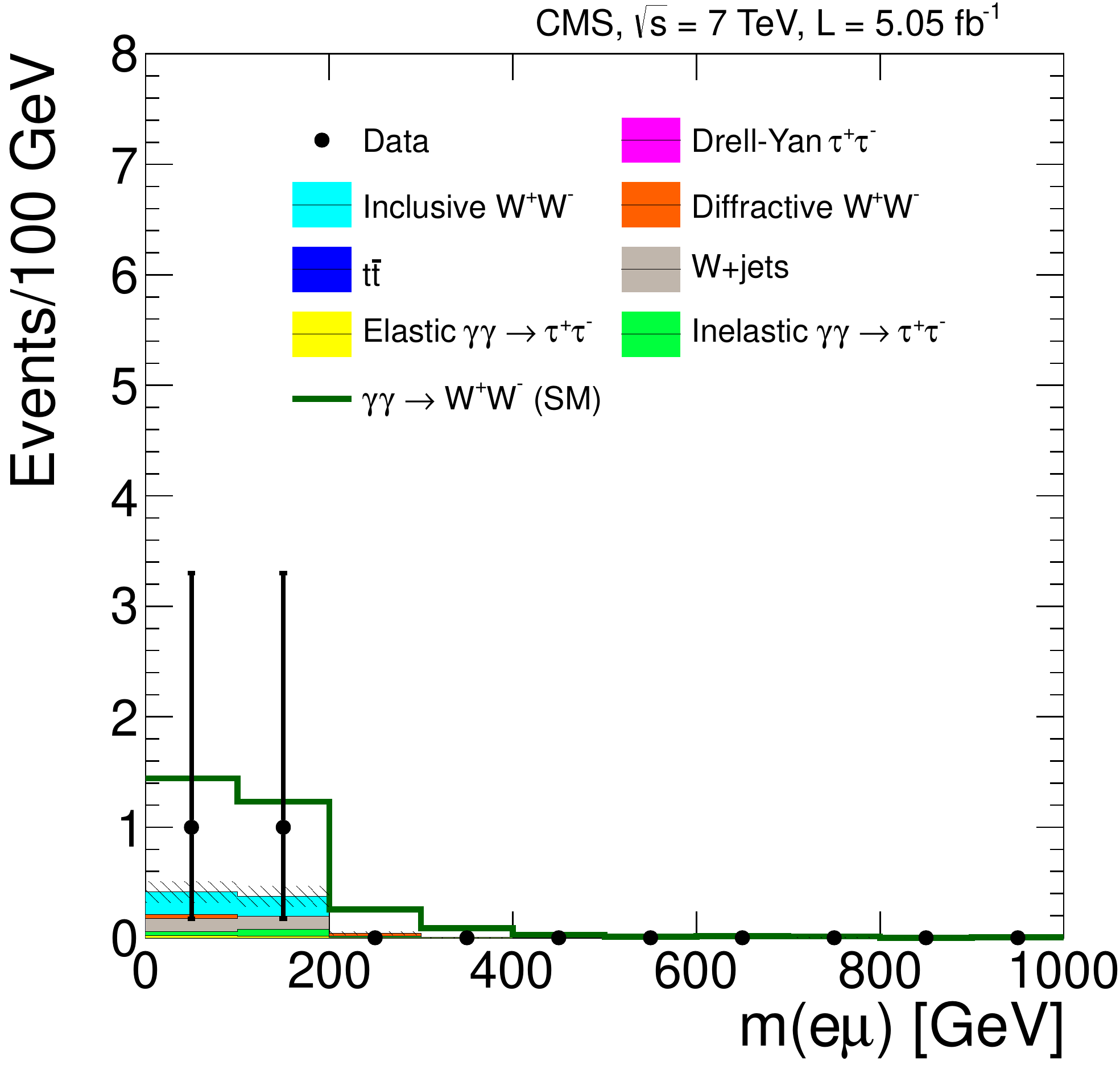}}
\resizebox{0.4\textwidth}{!}{\includegraphics{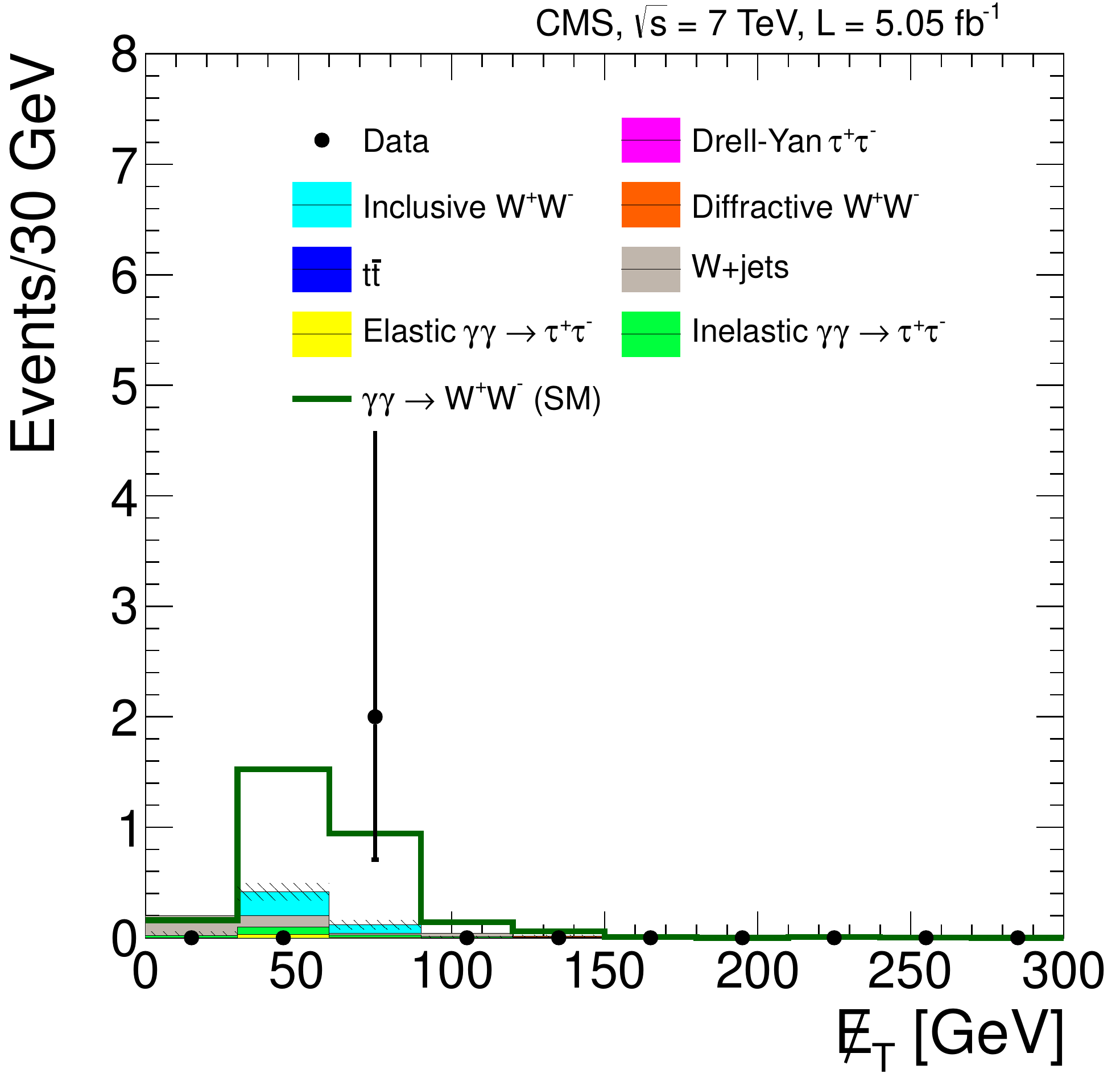}}
\caption{CMS: The $e^\pm\mu^\mp$ invariant mass (left) and \MET
  (right) distributions, for events in the signal region with zero
  extra tracks on the $e^\pm\mu^\mp$ vertex and
  $p_T(e^\pm\mu^\mp)>30\,\GeV$. The backgrounds (solid histograms) are
  stacked with statistical uncertainties indicated by the shaded
  region, the signal (open histogram) is stacked on top of the
  backgrounds.}
\label{fig:CMSppWWFinal}
\end{center}
\end{figure*}

Two events pass all selection criteria in data. The number of expected
signal events is $2.2\pm0.4$, where the largest experimental
uncertainty is due to the exclusive requirement on the tracks
associated with the common vertex.  The total background is expected
to be $0.8\pm0.2$ events.  The $e\mu$ invariant mass distribution and
the $\MET$ distribution for events in the signal region including the
predictions for signal and background are shown in
Fig. \ref{fig:CMSppWWFinal}. While the current statistics do not allow
a definite answer on the observation of QGCs at the LHC, an upper
limit on the cross section at $95\%$ CL is placed with
$\sigma(pp\rightarrow p W^+ W^- p \rightarrow p e^\pm \nu \mu^\mp \nu
p ) < 10.6$ fb.  

The larger dataset at $\sqrt{s}=8\,\TeV$ could allow
a first observation. A first evidence on vector boson scattering and the electroweak production of $W^\pm W^\pm$ with two jets in  $pp$ collisions at $\sqrt{s}=8\,\TeV$ has been recently published\cite{Aad:2014zda}. Also a search on the production of $WW\gamma$ and $WZ\gamma$ with constraints on anomalous QGCs has become available \cite{Chatrchyan:2014bza}. The studies of QGCs in general and the scattering of two heavy vector bosons in particular, will play a critical role in the upcoming years of the LHC physics program. Even though the discovery of the Higgs boson provides a unitarization scheme for the otherwise divergent longitudinally polarized vector boson scattering cross section with increasing center-of-mass energy, various scenarios of physics beyond the SM would enhance this scattering process. Last but not least, the scattering of two heavy gauge bosons has never been experimentally observed and its proof would be a further success of the predictive power of the SM.


\section{\label{sec:conclusion}Summary and Outlook}

A large variety of diboson production processes have been studied by the ATLAS and CMS experiments using data from
the LHC runs in 2010 and 2011. The diboson results of both experiments have been reviewed in
a comprehensive way for a first time in this article. 

The measured inclusive
cross sections are combined and compared with the Standard Model predictions and
a good agreement is observed. Several inclusive cross section measurements are no longer
statistically limited and the associated systematic uncertainties are
in many cases at the percent level. However, the study of differential
distributions can still profit from a larger dataset, allowing for more
precise tests of the corresponding theoretical calculations.

The good agreement between theory and experiment was used to set
limits on possible extensions of the electroweak sector, such as
 anomalous triple gauge couplings. While the limits on the anomalous $WWV$ vertices are still
dominated by the results from the LEP experiments, the most stringent
constraints on the $ZZV$- and $Z\gamma V$-vertices are already now given by both collaborations. 
In addition, the possibility to use fiducial
cross section measurements for the testing of these Standard Model extensions in a model independent way has been discussed.

\begin{figure}[h]
\begin{center}
\resizebox{0.75\textwidth}{!}{\includegraphics{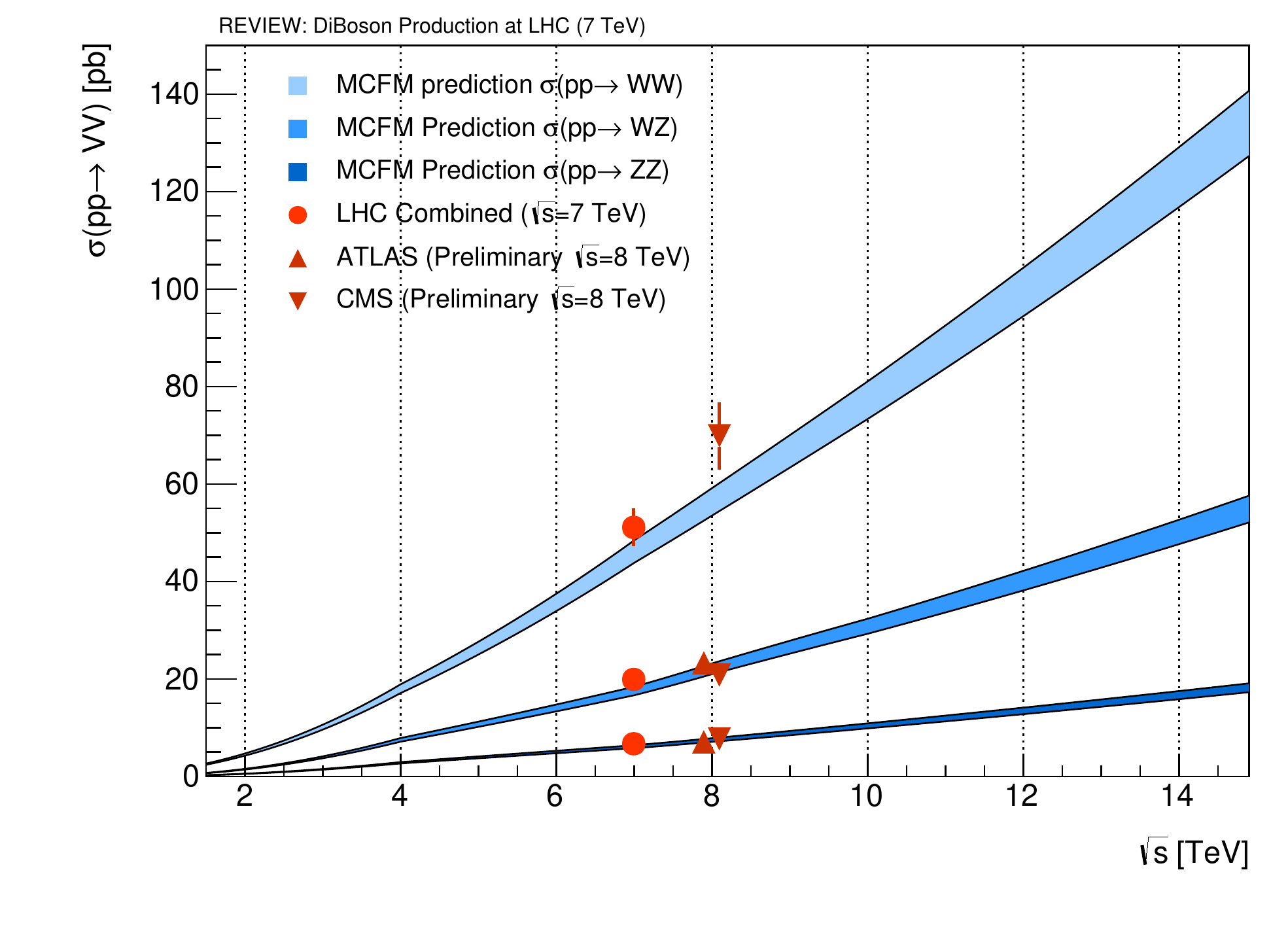}}
\caption{\label{Fig:EnergyDependence}Energy dependence of the
  predicted diboson production cross-sections and the available
  measurements.}
\end{center}
\end{figure}

Figure \ref{Fig:EnergyDependence} illustrates the dependence of the
$WW$, $WZ$, and $ZZ$ production cross sections on the center-of-mass
energy.  The available LHC measurements for $\sqrt{s}=8\,\TeV$ are also shown\cite{CMSZZ8TeV, CMSWW8TeV, CMSWZ8TeV, ATLASZZ8TeV, ATLASWZ8TeV}. While the
increase of the cross section is only $20\%$ between 
7 TeV and 8 TeV, the cross section for the second LHC run at 13-14 TeV
is expected to rise by a factor of $\approx 2.5$. This increased
energy, together with the expected integrated luminosity, will allow
for a new round of precision measurements of the electroweak sector to
gain a better understanding of the nature of the electroweak symmetry
breaking mechanism.

\section*{Acknowledgements}
It was our honor to participate during the first LHC run in study of diboson production in a new energy regime. It was a truly collaborative effort of the ATLAS and CMS experiments which have lead to a wide series of wonderful physics measurements. For this review article on the first diboson measurements at the LHC, we would like to thank Dr. Edward Diehl and Prof. Heinz-Georg Sander for their careful review of this paper. In addition, we thank the ATLAS and CMS working group conveners,  Sasha Glazov, Alessandro Tricoli, Jeffrey Berryhill and Maxime Gouzevitch for their input. The contribution from Matthias Schott was supported by the Volkswagen Foundation and the German Research Foundation (DFG). The contribution from Junjie Zhu was supported by the Department of Energy Early Career Grant under contract DE-SC0008062 .

\bibliography{DiVectorBoson7TeVJHEP}{}
\bibliographystyle{atlasnote}

\end{document}